\documentclass[12pt,a4paper,oneside]{article}

\usepackage{graphicx,amssymb,amsmath}
\usepackage{setspace} 
\usepackage[linkcolor={blue},citecolor={blue},colorlinks=true,linktocpage,urlcolor={blue}]
{hyperref}
\usepackage[capitalise]{cleveref}  
\usepackage[margin=2cm]{geometry}
\usepackage[dvipsnames]{xcolor}
\definecolor{darkblue}{rgb}{0,0,0.54}
\usepackage{tikz}
\usetikzlibrary{decorations.pathmorphing}
\usetikzlibrary{patterns}
\usetikzlibrary{calc}
\usepackage{multirow} 
\usepackage{appendix}
\usepackage{bbm}
\usepackage{import}
\usepackage{tensor}
\usepackage{pdfpages}
\usepackage{cancel} 
\usepackage{slashed} 
\usepackage{xparse} 
\usepackage{tabu} 
\usepackage{adjustbox} 
\usepackage{simpler-wick}
\usepackage[bb=boondox]{mathalfa}
\usepackage{dsfont}
\usepackage{subcaption}

\usepackage[natbib=false,maxbibnames=99,
    sorting=none,
    backend=biber,
    style=numeric,
]{biblatex}
\numberwithin{equation}{section} 
\usepackage{empheq} 

\graphicspath{{Figures/}} 

\DeclareMathOperator{\sgn}{sgn}

\DeclareMathOperator{\arcsinh}{arcsinh}

\renewcommand{\Re}{\operatorname{Re}}

\newcommand{\rdS}{\ell}
\newcommand*\dd{\mathop{}\!\mathrm{d}}

\newcommand{\propG}{\mathcal{S}}
\newcommand{\gm}{m_g} 
\newcommand{\sVol}[1]{\text{Vol}(S^{#1})} 

\usepackage{titling} 
\usepackage[affil-it]{authblk} 

\usepackage{tocloft} 
\def \beq { \begin{equation}}
\def \eeq {\end{equation}}
\def \l {\left}
\def \r {\right}

\def \th {\theta}

\def \bra {\langle}
\def \ket {\rangle}

\def \pr {\partial}
\def \ra {\rightarrow}

\renewcommand \bar \overline

\title{Out-of-time-ordered Correlators in de Sitter Revisited}
\author[1,2]{Alexey Milekhin,}
\author[3,4]{Vladimir Narovlansky,}
\author[5]{Jiuci Xu}

\affil[1]{Department of Physics and Astronomy, University of Kentucky, Lexington, KY 40506, USA} 

\affil[2]{Institute\! for\! Quantum Information\! and~Matter, California\! Institute\! of\! Technology, Pasadena, CA\! 91125,~USA}  

\affil[3]{Racah Institute of Physics, The Hebrew University of Jerusalem,
Jerusalem 91904, Israel}

\affil[4]{ORFE Department, Princeton University, Princeton, NJ 08544, USA}
    
\affil[5]{Department of Physics, University of California, Santa Barbara, CA 93106, USA}

\date{\small \texttt{
milekhin@uky.edu, narovlansky@princeton.edu, jiuci\char`_xu@ucsb.edu
}}

\makeatletter 
\let\@fnsymbol\@arabic
\makeatother

\begin{document}

\begin{titlingpage}
    \maketitle
    \begin{abstract}
	We study the 4-point out-of-time-ordered correlator (OTOC) of scalar fields using the eikonal approximation to gravity around a de Sitter background. It can be defined in a gauge-invariant way by dressing the operators to an observer.  We consider de Sitter space in any dimension and any bulk masses of the scalar fields. At tree level we find the maximal Lyapunov exponent of $2 \pi/\beta_{\rm dS}$.
	However, we describe an IR problem in this calculation at the loop level associated with the vector part of the graviton propagator. Regularizing this divergence leads to the vanishing of all odd powers in Newton's constant in the asymptotic expansion, so that the leading answer comes with twice the maximal Lyapunov exponent in simple operator configurations. We find that the perturbative OTOC in de Sitter can initially grow, which is not allowed by the quantum bound on chaos. Interestingly, we find that this growth is mediated by large diffeomorphisms in the graviton propagator. 
    \end{abstract}
\end{titlingpage}

\tableofcontents

\section{Introduction and summary}
In order to discuss dynamics in de Sitter space, it is useful to include an observer living in the space, see Fig.\ \ref{fig:worldline_dS_setup}. Indeed, in this case, we can study gauge invariant observables by dressing to the observer worldline. In such a setup, it is desirable to describe the quantum mechanics of the resulting system \cite{Chandrasekaran:2022cip,Narovlansky:2023lfz,Kolchmeyer:2024fly,Narovlansky:2025tpb}, see also \cite{Milekhin:2023bjv,Milekhin:2024vbb,Okuyama:2025hsd,Susskind:2021esx,Yang:2025lme,Geng:2025bcb}.

A particularly interesting observable is the out-of-time-order correlator (OTOC). The OTOC can be used to characterize chaotic behavior of quantum systems. In the case of gravity, particles falling to a horizon become blue-shifted, and even if initially they have a small amount of energy, they eventually get boosted and backreact on the geometry creating a shockwave. The behavior of shockwaves in de Sitter space \cite{Hotta:1992qy,Hotta:1992wb,Sfetsos:1994xa} is qualitatively different from that in anti-de Sitter and flat space. While in AdS shockwaves cause time delay, in dS they cause time advance. This effect of shockwaves on null geodesics is depicted in Fig.\ \ref{fig:shockwaves}. In certain kinematics it results in the puzzling initial \textit{growth} of the OTOC, in contrast to the initial \textit{decay} observed for AdS black holes and quantum many-body systems. This effect manifests in the ``wrong sign'' of the gravitational eikonal phase in dS compared to AdS. 

\begin{figure}[ht]
    \centering
    \begin{subfigure}[b]{0.45\textwidth}
        \centering
		\begin{tikzpicture}
                \node[fill,inner sep=0pt] (LB) at (0,0) {};
                \node[fill,inner sep=0pt] (LT) at (0,5) {};
                \node[fill,inner sep=0pt] (RB) at (5,0) {};
                \node[fill,inner sep=0pt] (RT) at (5,5) {};
                \draw[thick] (LB)--(LT);
                \draw[line width=2pt, color=blue,->] (RB)--(5,2.5);
                \draw[line width=2pt, color=blue] (RB)--(RT);
                \draw[thick,dashed] (LB)--(RT);
                \draw[thick,dashed] (LT)--(RB);
                \draw[thick] (LT)--(RT);
                \draw[thick] (RB)--(LB);
                \end{tikzpicture}
        \caption{}
        \label{fig:worldline_dS_setup}
    \end{subfigure}
    \hfill
    \begin{subfigure}[b]{0.45\textwidth}
        \centering
		\begin{tikzpicture}
                \node[fill,inner sep=0pt] (LB) at (0,0) {};
                \node[fill,inner sep=0pt] (LT) at (0,5) {};
                \node[fill,inner sep=0pt] (RB) at (5,0) {};
                \node[fill,inner sep=0pt] (RT) at (5,5) {};
                \draw[thick] (LB)--(LT) node[left,pos=0.5] {$L\,$};
                \draw[thick] (RB)--(RT) node[right,pos=0.5] {$\, R$};
                \draw[line width=2pt, color=blue,->] (RB)--(5,2.5);
                \draw[line width=2pt, color=blue] (RB)--(RT);
                \draw[line width=2pt, color=blue,->] (LB)--(0,2.5);
                \draw[line width=2pt, color=blue] (LB)--(LT);
                \draw[thick,dashed] (LB)--(RT);
                \draw[thick,dashed] (LT)--(RB);
                \draw[thick] (LT)--(RT);
                \draw[thick] (RB)--(LB);
                \end{tikzpicture}
        \caption{}
        \label{fig:two_worldlines_dS}
    \end{subfigure}
    \caption{Observers in de Sitter.}
    \label{fig:observers_dS}
\end{figure}
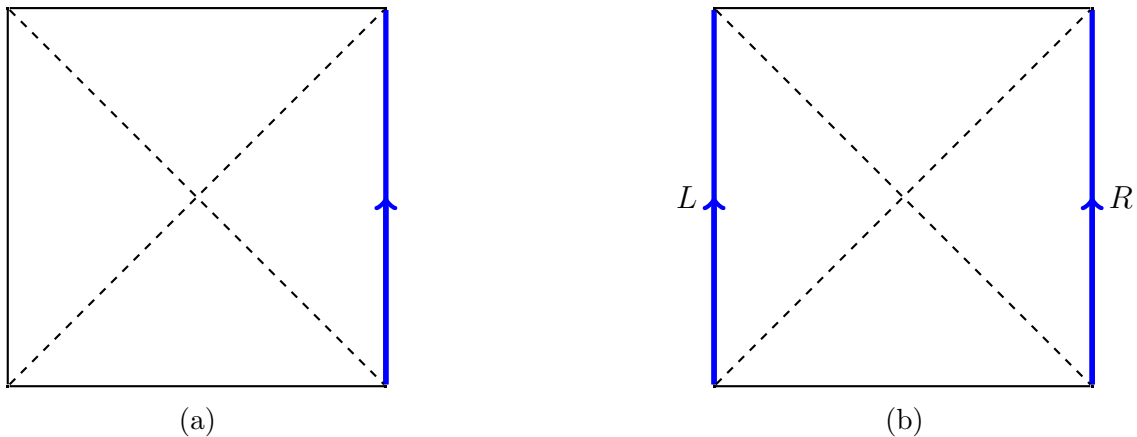
\begin{figure}[ht]
    \centering
    \begin{subfigure}[b]{0.45\textwidth}
        \centering
		\begin{tikzpicture}
                \node[fill,inner sep=0pt] (LB) at (0,0) {};
                \node[fill,inner sep=0pt] (LT) at (0,5) {};
                \node[fill,inner sep=0pt] (RB) at (5,0) {};
                \node[fill,inner sep=0pt] (RT) at (5,5) {};
                \draw[thick] (LB)--(LT);
                \draw[thick] (RB)--(RT) coordinate[pos=0.2] (geodesic_int);
                \draw[line width=2pt] (LB)--(RT);
                \draw[thick,dashed] (LT)--(RB);
                \draw[decorate,decoration={coil,aspect=0,amplitude=1pt, segment 				length=12.2pt}] (LT)--(RT) coordinate[pos=0.55] (geodesic_end);
                \draw[decorate,decoration={coil,aspect=0,amplitude=1pt, segment 				length=12.2pt}] (RB)--(LB);
                \draw[thick,dotted, color=blue] (geodesic_int)--(3,3);
                 \draw[thick,dotted, color=blue] (3.9,3.9)--(geodesic_end);
                \end{tikzpicture}
        \caption{Time delay in AdS.}
        \label{fig:AdS_shockwave}
    \end{subfigure}
    \hfill
    \begin{subfigure}[b]{0.45\textwidth}
        \centering
		\begin{tikzpicture}
                \node[fill,inner sep=0pt] (LB) at (0,0) {};
                \node[fill,inner sep=0pt] (LT) at (0,5) {};
                \node[fill,inner sep=0pt] (RB) at (5,0) {};
                \node[fill,inner sep=0pt] (RT) at (5,5) {};
                \draw[thick] (LB)--(LT) coordinate[pos=0.9] (geodesic_end);
                \draw[thick] (RB)--(RT) coordinate[pos=0.1] (geodesic_int);
                \draw[line width=2pt] (LB)--(RT);
                \draw[thick,dashed] (LT)--(RB);
                \draw[thick] (LT)--(RT);
                \draw[thick] (RB)--(LB);
                \draw[thick,dotted, color=blue] (geodesic_int)--(2.75,2.75);
                 \draw[thick,dotted, color=blue] (2.25,2.25)--(geodesic_end);
                \end{tikzpicture}
        \caption{Time advance in dS.}
        \label{fig:dS_shockwave}
    \end{subfigure}
    \caption{Shockwaves in AdS and dS. We depict the shockwaves by heavy lines. To probe them we send null geodesics shown in dotted blue.}
    \label{fig:shockwaves}
\end{figure}
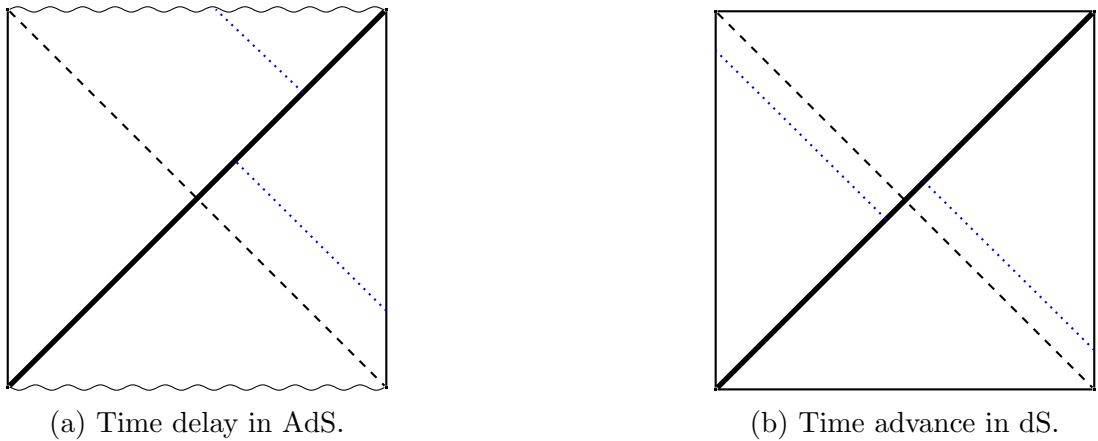

While this setup might ideally make it possible to discuss the OTOC at finite Newton's constant $G_N$, it is beneficial to start with $G_N \to 0$.
It is particularly interesting to consider an observer whose mass is much larger than the de Sitter scale, so that the observer is localized, but much smaller than the Planck scale, so that it does not affect spacetime significantly.
In this setup, the neighborhood of the observer is just de Sitter space and the observer follows a geodesic. It is then clear where the operators are located in the spacetime.
If we take simultaneously large time separations as we take $G_N \to 0$, there is still a gravitational backreaction effect.
However, the backreaction happens away from the worldline.
The observer essentially sits at a pole of the spatial sphere and we can place various operators there.
In this setup, we will essentially be interested in perturbative gravity around a de Sitter background.

We can also consider two observers, one in the north pole and one in the south pole, which are labeled by `R' and `L' in the Penrose diagram of Fig.\ \ref{fig:two_worldlines_dS}. In the limit of $G_N \to 0$, we can go from one pole to the other by shifting the times by $i\pi $ in units where the de Sitter radius is $\rdS=1$.
We assume that the observers couple to massive scalar fields in the bulk, which at leading order we take to be free. We take two species $V$ and $W$, so that we are interested in studying 4-point OTOCs of the form $\langle VW(t)VW(t)\rangle $ around a de Sitter background.
When the time $t$ is large, this OTOC corresponds to two quanta, one of the $V$ particle and one of the $W$ particle, moving very close to the two horizons. In this case we can use an eikonal approximation to calculate the correlation function. In gravity we truncate to processes where the $V$ and the $W$ particles can emit and absorb a graviton. The dominant diagrams in the eikonal approximation are `generalized ladder' diagrams of graviton exchanges resulting in a two-to-two gravitational scattering.
In fact, as we recall in more detail below, the gravitational scattering phase can be obtained by a gravitational saddle point approximation. 
We are interested mainly in three different OTOC configurations
\begin{enumerate}
\item A one-sided configuration where all operators are on a single pole of the spatial sphere, which can be written as $\langle V_R W_R(t)V_RW_R(t)\rangle $.
\item A two-sided configuration where one operator appears on the other pole, $\langle V_L W_R(t)V_RW_R(t)\rangle $.
\item The regularized configuration used in the bound on chaos \cite{Maldacena:2015waa} where we evolve by a quarter of the thermal circle between any two operators.
\end{enumerate}

The first configuration is natural to consider since it is associated with a single observer. The second configuration has a simple bulk interpretation, since in the limit of large bulk masses it is given by the length of a geodesic going between the two sides in the presence of a shockwave \cite{Shenker:2013pqa}. Here we analyze such OTOCs in de Sitter space of any dimension $d$, and for any bulk masses. We include both light and heavy masses in de Sitter units, corresponding to the complementary and principal series representations of de Sitter. The OTOC in three dimensions for conformally coupled scalars was studied in \cite{Aalsma:2020aib}.\footnote{Our analysis differs in several aspects. First, while it is argued there that the linear term in $G_N$ is absent, we believe it is because an $\arcsinh$ term was dropped. If it is kept, there will be such a term in the analysis of \cite{Aalsma:2020aib}. Second, the eikonal phase there corresponds to two antipodal (in the internal sphere) signals coming from each side of the Penrose diagram. However, an interaction exchanging particles does not have to preserve the antipodal property.}

\subsection{Summary of the results}
We start by analyzing the OTOC using the standard approach with null localized shockwaves in order to extract the eikonal phase. However, we will see that the relevant Einstein equations have no solutions because of zero modes. In order to understand better the origin of this problem we will adopt a more basic approach, which is equivalent to the shockwave computation, based on resumming Feynman diagrams representing two high-energy particle exchanging linearized gravitons. In this approach the eikonal phase is explicitly given by a certain integral of the graviton propagator. 

Using existing results for the graviton propagator in de Sitter we find that the eikonal phase is essentially IR divergent. However this divergence is associated with large diffeomorphisms residing in the vector part of the propagator. Moreover, the overall coefficient of this contribution in principle depends on the gauge fixing parameters, but it diverges for any choice of parameters. At the tree level (leading order $G_N$ contribution to the OTOC) this divergence is absent for admissible matter configurations which satisfy the gravitational constraints. However, this IR divergence becomes unavoidable at the loop level. Moreover, it is associated with internal graviton propagators in Feynman diagrams, so it is not expected to be removed by the gravitational dressing which affects only the external lines. 

At the tree level we find the usual $e^{2 \pi t/\beta_{\rm dS}}$ Lyapunov behavior ($1/\beta_{\rm dS}$ being the temperature of the cosmological horizon) and the aforementioned initial growth of the OTOC in the configurations (2) and (3). 
Interestingly, we uncover that this growth is also caused by a large diffeomorphism  in the propagator. In this case, however, its contribution is independent of the gauge fixing parameters \footnote{At least for the most general gauge fixing term which is quadratic in the fluctuations}. In contrast, for BTZ black holes, the $G_N e^{2 \pi t/\beta_{\rm AdS}}$ decay of the OTOC entirely comes from the physical traceless-transverse component of the graviton propagator.

In order to regularize it we introduce a small graviton mass $\gm$ and study the limit where $G_N, \gm \rightarrow 0$, $t \rightarrow +\infty$ but $G_N e^{2 \pi t/\beta_{\rm dS}}/\gm^2$ is held fixed. 
In writing the OTOC as an asymptotic expansion in $G_N/\gm^2$, every factor of $G_N/\gm^2$ comes with a time dependence of $e^{2\pi t/\beta_{\rm dS} } $.
We find that only even powers of these appear for simple operator configurations. In particular, the first non-trivial correction corresponds to twice the maximal Lyapunov exponent $\lambda _L=\frac{2\pi }{\beta_{\rm dS} } $. In configuration (2) above, the OTOC starts increasing at first, for times before the scrambling time, as in the geodesic approximation analysis of \cite{Aalsma:2020aib} (calculated for an isotropic shockwave). This is the time advance effect of de Sitter. We find that this is true for any bulk masses. In the configuration (3), as emphasized in \cite{Narovlansky:2025tpb}, an initially growing OTOC is not allowed by the bound on chaos \cite{Maldacena:2015waa}. Nonetheless, we find such growth for any bulk masses. This gives an interesting tension between gravity in de Sitter space and quantum mechanics. 

The paper is organized as follows. In Sec.\ \ref{sec:dS_geometry}, we review the geometry of de Sitter space and introduce the obstruction associated with the zero modes of the Einstein equations in the presence of a localized shockwave. In Sec.\ \ref{sec:diagrammatics}, we derive the gravitational eikonal approximation directly from Feynman diagrams. Using the explicit expressions obtained there, we determine the eikonal phase for scattering in de Sitter space in Sec.\ \ref{sec:eikonal}. We identify an infrared divergence in the vector sector of the graviton propagator and analyze the diffeomorphism contributions that affect the final result. Finally, in Sec.\ \ref{sec:ds_otoc}, we compute the de Sitter OTOC following the approach developed for AdS in \cite{Shenker:2014cwa}. The calculation requires two main ingredients: the  wavefunctions, obtained as Fourier transforms of 2-point functions on the space without deformations (de Sitter in our case), and the eikonal phase derived in the preceding section. We demonstrate the initial growth of the OTOC and in the case with the mass regulator, find a Lyapunov exponent equal to twice the maximal Lyapunov exponent. Additional technical details are collected in the appendices.

\textit{Note added: } While this paper was in preparation, we became aware of several other groups working on related topics.
We coordinated the arXiv submission with them.  

\begin{figure}[h]
\centering
	\begin{tikzpicture}
        \node[fill,inner sep=0pt] (LB) at (0,0) {};
        \node[fill,inner sep=0pt] (LT) at (0,5) {};
        \node[fill,inner sep=0pt] (RB) at (5,0) {};
        \node[fill,inner sep=0pt] (RT) at (5,5) {};
        \draw[thick] (LB)--(LT);
        \draw[thick] (RB)--(RT);
        \draw[thick,dashed] (LB)--(RT);
        \draw[thick,dashed] (LT)--(RB);
        \draw[thick] (LT)--(RT);
        \draw[thick] (RB)--(LB);
        \draw[thick,->] (RT) +(0.2,0.2)--+(1,1) node[right] {$v$};
        \draw[thick,->] (LT) +(-0.2,0.2)--+(-1,1) node[left] {$u$};
        \draw[->] (RB) +(-0.5,1.2) to [bend left=25] +(-0.5,3.8) node[right] {$t$};
        \draw[->] (LB) +(0.5,3.8) to [bend left=25] +(0.5,1.2) node[left] {$t$};
        \end{tikzpicture}
\caption{Penrose diagram of global de Sitter space.}
\label{fig:de_Sitter_Penrose}
\end{figure}
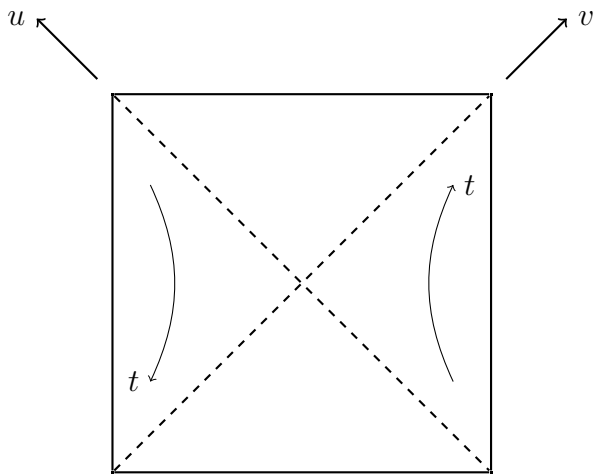

\section{de Sitter geometry} \label{sec:dS_geometry}

de Sitter space is a maximally symmetric solution to Einstein gravity with a positive cosmological constant. It can be described as a hyperboloid. Indeed, $d$-dimensional de Sitter $dS_d$ can be written using $(d+1)$-dimensional embedding coordinates $X^A$, $A=0,\cdots ,d$, satisfying
\begin{equation}
\eta _{AB} X^A X^B=\rdS^2,\qquad \eta _{AB} =\text{diag}(-1,+1,\cdots ,+1)
\end{equation}
where $\rdS$ is the de Sitter radius. The Penrose diagram of de Sitter is shown in Fig.\ \ref{fig:de_Sitter_Penrose}. Spatial slices correspond to spheres of radius $\rdS$.

Let us mention a couple of coordinate systems (see Fig.\ \ref{fig:de_Sitter_Penrose}).
A causal wedge of an observer can be written in terms of static coordinates, which do not cover the entire space. These coordinates include $t,r$, as well as coordinates for a $(d-2)$-dimensional sphere. To parametrize the latter sphere we can use $y^i$, $i=1,\cdots ,d-1$ constrained by $y^2=1$. The relation between these coordinates and the embedding coordinates is
\begin{equation}
\begin{split}
& X^0 = \sqrt{\rdS^2-r^2}\, \sinh \frac{t}{\rdS} \\
& X^d = -\sqrt{\rdS^2-r^2}\, \cosh \frac{t}{\rdS} \\
& X^i = r y^i
\end{split}
\end{equation}
The metric in terms of static coordinates is
\begin{equation}
\dd s^2 = -(1-r^2/\rdS^2) \dd t^2 + \frac{\dd r^2}{1-r^2/\rdS^2} +r^2 \dd \Omega _{d-2} ^2
\end{equation}
where $\dd \Omega _{d-2} ^2$ is the metric on the $(d-2)$-sphere.

We also have global Kruskal coordinates $u,v,y^i$ covering the entire $dS_d$. They are related to embedding coordinates by
\begin{equation}
\begin{split}
& X^0 = \rdS \frac{u+v}{1-uv} \\
& X^d = \rdS \frac{u-v}{1-uv} \\
& X^i = \rdS \frac{1+uv}{1-uv} y^i
\end{split}
\end{equation}
and the metric is given in terms of them by
\begin{equation} \label{eq:dS_Kruskal}
\dd s^2= \rdS^2 \frac{-4 \dd u \dd v+(1+uv)^2 \dd \Omega _{d-2} ^2}{(1-uv)^2} .
\end{equation}

It would be convenient to define the chord distance $Z$ between the points $X$ and $X'$ as
\beq
Z= \frac{1}{\rdS^2} X \cdot X'.
\eeq

\subsection{OTOC and shockwaves in de Sitter space} \label{sec:ds3_shockwave}

When thinking about OTOCs, it is useful to consider two-sided setups, see \cite{Shenker:2014cwa}. For AdS, this is achieved by considering a two-sided black hole with two asymptotic AdS boundaries where two CFTs live. In this case it is natural to place operators on the two CFTs. Here we are interested in de Sitter space \cite{Aalsma:2020aib}. In this case, we can imagine two observers, one living on the north pole of the spatial sphere, and the other living on the south pole. Both in AdS and in dS, one can also restrict the discussion to a single side by omitting the other. In the case of dS, we would like to include fluctuations in the geometry which makes the definition of the operators more intricate. However, we could consider the $G_N \to 0$ limit where along the observers the space is just that of de Sitter, and the backreaction only happens away from the observers.
The idea is that even when a small amount of energy is released by the operators, if it happens long enough in the past, the released perturbation gets blue-shifted when approaching the horizon and eventually backreacts on the geometry.
In such cases, the OTOC can be calculated as a 2-to-2 gravitational scattering using the eikonal approximation \cite{Shenker:2014cwa}.

In this section we briefly describe shockwaves in de Sitter and the appearance of a zero mode in this computation. The rest of the paper will be dedicated to the discussion of this zero mode.
 For simplicity let us consider for the moment three-dimensional de Sitter.
As mentioned, we can cover the entire $dS_3$ space with Kruskal coordinates given by
\begin{equation}
\dd s^2 =\rdS^2 \frac{-4 \dd u \dd v+(1+uv)^2 d\phi ^2}{(1-uv)^2} .
\end{equation}
Now the internal sphere is just a circle, parametrized by $\phi $, an angular variable of period $2\pi $.

A shockwave geometry corresponding to a source propagating at $u=0$ along the $v$ direction is generally given by
\begin{equation}
\label{eq:ds3_shock_metric}
\dd s^2 =\rdS^2 \left[ \frac{-4 \dd u \dd v+(1+uv)^2 d\phi ^2}{(1-uv)^2} + \tilde h_{uu} du^2 \right] 
\end{equation}
where we parametrize the deformation of the metric using a function $f(\phi )$
\begin{equation}
\tilde h_{uu} =- \delta (u) f(\phi ) .
\end{equation}
This is a shockwave caused by a source given by
\begin{equation}
\label{eq:Tuu3}
T_{uu} = \frac{f(\phi )+f''(\phi )}{16 \pi G_N} \delta (u)
\end{equation}
with all other energy-momentum components vanishing.
Indeed, one can check that the Einstein equations
\begin{equation}
R_{\mu \nu } -\frac{1}{2} Rg_{\mu \nu } +\rdS^{-2} g_{\mu \nu } =8\pi G_N T_{\mu \nu } 
\end{equation}
are satisfied (where $G_N$ is Newton's constant).

A simple solution is an isotropic shockwave along the internal circle. It corresponds to an energy-momentum tensor homogeneous along the circle. This gives a solution
\begin{equation}
T_{uu} = \frac{\rdS  p^v}{\pi} \delta (u), \qquad \tilde h_{uu} = - 16 G_N \rdS p^v \delta (u) 
\end{equation}
where $p^v$ is the total momentum in the $v$ direction. Note that on the horizons the geometry is flat, so we have usual special relativity.
Similarly, we could consider a symmetric shockwave going in the other direction
\begin{equation}
T_{vv} = \frac{\rdS  p^u}{\pi} \delta (v), \qquad \tilde h_{vv} = - 16G_N \rdS p^u \delta (v) .
\end{equation}

In the eikonal approximation, one is usually interested in localized shocks where the energy-momentum tensor behaves as $\delta (u)\delta (\phi -\phi _0)$.
However, here we arrive at one of the main issues we discuss extensively.
Clearly, on a circle, there is no solution to $f(\phi )+f''(\phi )=\delta (\phi-\phi_0 )$. This can be seen for instance by the fact that the LHS is annihilated on the $\pm 1$ Fourier modes, but the RHS is not, so there is no solution. This feature occurs in all dimensions: for $dS_d$ the analogue of (\ref{eq:Tuu3}) is
\beq
\label{eq:problematic_eq}
T_{uu} \propto \Delta_{S^{d-2}} f + (d-2) f .
\eeq
So the zero mode always resides in the $L=1$ (vector) angular momentum sector with respect to $SO(d-1)$. 

How should we treat this problem? Does it mean that the eikonal approximation breaks down because the matter sources cannot be localized or should we just make sure that $L=1$ sources inside the OTOC are absent, effectively solving Einstein equations with the delta-function where $L=1$ modes have been subtracted?
In order to understand this better, we now review the eikonal scattering and perform a direct Feynman diagram computation.

\section{Review of the eikonal resummation} \label{sec:diagrammatics}

In this section we derive the eikonal approximation in de Sitter by doing perturbation theory in a theory of matter coupled to perturbative gravity around de Sitter background, following a very similar discussion for anti de Sitter \cite{Cornalba:2007zb}. We will expand the matter coupled to gravity action in metric perturbations, truncating to the graviton propagator and the leading (linear in) graviton interaction with the matter. Then in the eikonal approximation where the Mandelstam $s$ variable is much larger than the Mandelstam $t$ variable and the appropriate de Sitter scale, only so-called `generalized ladder' diagrams contribute dominantly. Interestingly, these diagrams generate no UV divergences. Our conclusion is that we did not find anything specific about kinematics in de Sitter which would break the Feynman diagram analysis. In Appendix \ref{sec:saddle_point_linearized} we provide an alternative derivation based on the action principle. However, we will see a reincarnation of the shockwave zero mode we saw in the previous section in the final integral expression for the eikonal phase.

For the pure gravity part of the action, we take Einstein-Hilbert with a cosmological constant, expand around de Sitter to second order in the metric fluctuations, and add a gauge-fixing term. In terms of Feynman diagrams, this simply results in a graviton propagator.
For the matter, we consider free massive scalars. Expanding the free scalar action up to linear order in the metric fluctuations $g_{\mu \nu} \ra g_{\mu \nu} + h_{\mu \nu}$ we have 
\begin{equation}
\begin{split}
S_m&=\frac{1}{2} \int d^dx \sqrt{-g}\left( -(\nabla \phi)^2    - m^2 \phi ^2\right) \\
& +\frac{1}{2} \int d^dx \sqrt{-g} \, h_{\mu \nu } \left[ \partial ^{\mu } \phi \partial ^{\nu } \phi -\frac{1}{2} g^{\mu \nu } \left( (\nabla \phi )^2+m^2\phi ^2 \right) \right] .
\end{split}
\end{equation}
The first term gives the matter propagator in a de Sitter background. The second term is the vertex $\frac{1}{2} \int h_{\mu \nu } T^{\mu \nu } $. The hallmark of the graviton exchange is that it is quadratic in the particle momenta. We take two such scalar fields, so that the combinatorics is simpler. These are dual to $V$ and $W$ operators inside the OTOC.

\subsection{Ladder diagrams}

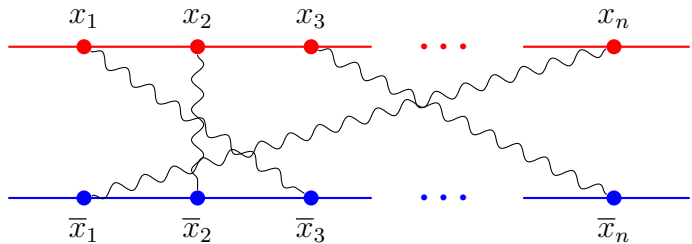
\begin{figure}[h]
\centering
	\begin{tikzpicture}

\def\yTop{2}
\def\yBot{0}

\def\xa{1}
\def\xb{2.5}
\def\xc{4}
\def\xd{8}

\def\xLeftStart{0}
\def\xLeftEnd{4.8}
\def\xRightStart{6.8}
\def\xRightEnd{9}
\def\xEllipsis{5.8}

\draw[red, thick] (\xLeftStart,\yTop) -- (\xLeftEnd,\yTop);
\draw[red, thick] (\xRightStart,\yTop) -- (\xRightEnd,\yTop);
\node[red, font=\boldmath\large] at (\xEllipsis,\yTop) {$\cdots$};

\draw[blue, thick] (\xLeftStart,\yBot) -- (\xLeftEnd,\yBot);
\draw[blue, thick] (\xRightStart,\yBot) -- (\xRightEnd,\yBot);
\node[blue, font=\boldmath\large] at (\xEllipsis,\yBot) {$\cdots$};

\node[circle, fill=red, inner sep=2pt, label=above:{$x_1$}]
  (T1) at (\xa,\yTop) {};
\node[circle, fill=red, inner sep=2pt, label=above:{$x_2$}]
  (T2) at (\xb,\yTop) {};
\node[circle, fill=red, inner sep=2pt, label=above:{$x_3$}]
  (T3) at (\xc,\yTop) {};
\node[circle, fill=red, inner sep=2pt, label=above:{$x_n$}]
  (T4) at (\xd,\yTop) {};

\node[circle, fill=blue, inner sep=2pt, label=below:{$\bar{x}_1$}]
  (B1) at (\xa,\yBot) {};
\node[circle, fill=blue, inner sep=2pt, label=below:{$\bar{x}_2$}]
  (B2) at (\xb,\yBot) {};
\node[circle, fill=blue, inner sep=2pt, label=below:{$\bar{x}_3$}]
  (B3) at (\xc,\yBot) {};
\node[circle, fill=blue, inner sep=2pt, label=below:{$\bar{x}_n$}]
  (B4) at (\xd,\yBot) {};

\draw[decorate, decoration={snake, amplitude=0.8mm, segment length=4mm}]
  (T1) -- (B3);
\draw[decorate, decoration={snake, amplitude=0.8mm, segment length=4mm}]
  (T2) -- (B2);
\draw[decorate, decoration={snake, amplitude=0.8mm, segment length=4mm}]
  (T3) -- (B4);
\draw[decorate, decoration={snake, amplitude=0.8mm, segment length=4mm}]
  (T4) -- (B1);

        \end{tikzpicture}
\caption{Generalized ladder diagrams used in the eikonal approximation. The horizontal lines are scalar propagators with heavy dots representing interaction vertices, and wiggly lines are graviton propagators. Other diagrams are suppressed in the eikonal approximation.}
\label{fig:generalized_ladder}
\end{figure}

\begin{figure}[h]
\centering
	\begin{tikzpicture}
        \node[fill,inner sep=0pt] (LB) at (0,0) {};
        \node[fill,inner sep=0pt] (LT) at (0,5) {};
        \node[fill,inner sep=0pt] (RB) at (5,0) {};
        \node[fill,inner sep=0pt] (RT) at (5,5) {};
        \draw[thick] (LB)--(LT);
        \draw[thick] (RB)--(RT);
        \draw[thick,blue] (LB)--(RT);
        \draw[thick,red] (LT)--(RB);
        \draw[thick] (LT)--(RT);
        \draw[thick] (RB)--(LB);
        \draw[decorate, decoration={snake, amplitude=0.8mm, segment length=4mm}]
  (3.5,1.5) -- (3.5,3.5);
        \end{tikzpicture}
\caption{Two particles (blue and red) follow classical null geodesic trajectories and exchange a single graviton. This is the process governing the eikonal phase.}
\label{fig:single_graviton_exchange_in_Eikonal_approx}
\end{figure}
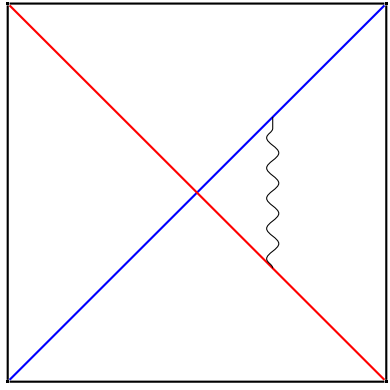

The idea of the eikonal approximation is that when the center of mass energy is large, the trajectories of the particles become classical, following null geodesics. Consider the generalized ladder diagrams of Fig.\ \ref{fig:generalized_ladder} where we allow all the possibilities of connecting with graviton propagator (``rungs'') a vertex of particle 1 (corresponding to the $V$ field) with a vertex of particle 2 (corresponding to the $W$ field).  The two sequences of propagator lines (``rails'') of the particles become the trajectories, as depicted in Fig.\ \ref{fig:single_graviton_exchange_in_Eikonal_approx}, where we show for example a single graviton exchange.

The technical statement behind the eikonal approximation is that smallness of the coupling constant can be compensated by the large null momentum factor coming from the matter ``rail'' propagators in the interaction vertex. Such momentum dependence is definitely true for graviton exchanges \footnote{In contrast, for the scalar exchange the vertices are momentum independent, so the eikonal approximation is known to break down \cite{Kabat:1992pz,Treiman,Jackiw_eikonal}.}.
Thus we want this large momentum to be preserved, resulting in graviton (or other exchange) momentum being zero along these null directions. This will also result in a drastic simplification of the side ``rail'' propagators.
In curved space-time it will require adopting the proper coordinate system where the corresponding null momentum direction is essentially flat, so that this momentum flows through the ``rails'' of the diagram.

 That is, we need a Killing vector with an associated charge, so that if we show that the particle follows a geodesic, this charge is then conserved along the geodesic. Since it is conserved, a large charge remains large.

To achieve this, we find the description of \cite{Cornalba:2007zb} useful, which we adapt here to de Sitter space. In this section we set $\rdS=1$. We use uppercase letters for dS embedding space as before.
Consider two null vectors $K_1$ and $K_2$ in embedding space. We can think of $K_1$ as the momentum of particle 1 when it is highly boosted and specifically when it moves close to the horizon (similarly for particle 2 with $K_2$).
Indeed, given a point $X $ in dS, that is $X^2=1$, such that $X \cdot K=0$ for a null $K$, such a null vector indeed generates a null geodesic in dS since $X+\lambda K$ for real $\lambda $ is also in de Sitter. However, going over all $X$ satisfying this generates only a codimension 1 surface. The construction of the null momenta away from the horizons is described momentarily.
We use the same normalization as in \cite{Cornalba:2007zb} where $-2K_1 \cdot K_2=(2\omega )^2$; we think about $\omega $ as the center of mass energy for the $K_1,K_2$ trajectories.

These two vectors define a transverse space which are all the $W $ in dS ($W^2=1$) such that $W \cdot K_1=W \cdot K_2=0$. The transverse space is a sphere.
Fixing $X_0$ in the transverse space, we get two Killing vectors of de Sitter given in embedding space coordinates by
\begin{equation}
T_i(X)=\frac{(K_i \cdot X)X_0-(X_0 \cdot X)K_i}{2\omega } ,\qquad i=1,2.
\end{equation}
This is similar to the usual way to get rotations and boosts in embedding space.
For example, in $dS_3$ choosing
\begin{equation}
\begin{split}
K_1&=\omega (1,1,0 ,0) \\
K_2&=\omega (1,-1,0 ,0) \\
X_0 &= (0,0,1 ,0)
\end{split}
\end{equation}
these Killing vectors written in dS are given in Kruskal coordinates by
\begin{equation}
\begin{split}
& T_1= \frac{1}{2} \left( -\cos\phi \, \partial _u+v^2\cos \phi \, \partial _v +\frac{2v}{1+uv} \sin \phi \, \partial _{\phi } \right)  \\
& T_2= \frac{1}{2} \left( u^2 \cos\phi \, \partial _u-\cos \phi \,  \partial _v +\frac{2u}{1+uv} \sin \phi \, \partial _{\phi } \right)  
\end{split}
\end{equation}
These are the kind of Killing vectors we want. For example, for $u=0$, $T_2$ is $-\frac{1}{2} \cos \phi \, \partial _v$ so it acts as translations along $v$.

In order to get the null momenta corresponding to null geodesics of particle 1 away from the horizon, we can use the $T_2$ Killing vector. We generally denote such null momenta by $K$ (which reduce to $K_1$ along the corresponding horizon), and they are given by $e^{\gamma T_2} K_1$  up to normalization. We fix the normalization of $K$ such that $-T_2 \cdot K=\omega $. Therefore, highly boosted trajectories of particle 1 at different spacetime points are described by tangent vectors $K$, and the charge of $T_2$ along them is fixed.
For particle 2 we denote the extension of $K_2$ using $T_1$ by $\bar K$ with the analogous normalization.

Then \cite{Cornalba:2007zb} define separate coordinates for particle 1 and particle 2 adapted to such conserved charges. In order to distinguish these coordinates from Kruskal coordinates, we use capital $U,V$. For particle 1, a coordinate system is given by
\begin{equation}
X=e^{VT_2} e^{UT_1} W
\end{equation}
so that the parameters are $W$, a transverse vector in embedding space, and $U,V$. In these coordinates a line of constant $V,W$ is then a null geodesic. Indeed, $\partial _U$ is proportional to $K$. Similarly there are coordinates adapted to particle 2, which are different (as the relevant Killing vector is different). Denoting by a bar quantities associated to particle 2, these are the $\bar U,\bar V,\bar W$ coordinates where 
\begin{equation}
\bar X=e^{\bar UT_1} e^{\bar VT_2} \bar W .
\end{equation}

The coordinates used to describe particle 1 have metric 
\begin{equation} \label{eq:metric_Penedones_coor}
\dd s^2=\dd W^2-(X_0 \cdot W)^2 \dd U \dd V+\frac{U^2}{4} (X_0 \cdot W)^2 \dd V^2
\end{equation}
where we always mean that $\dd W^2$ is the metric induced on the transverse space. The Killing vector $T_2$ in these coordinates is just $T_2=\partial _V$.  Similarly, the coordinates adapted to particle 2 have metric 
\begin{equation}
    \dd s^2 = \dd \bar W^2 - (X_0 \cdot \bar W)^2 \dd \bar U  \dd \bar V  +\frac{\bar{V}^2}{4} (X_0 \cdot \bar W )^2  \dd \bar U^2
\end{equation}
 
As mentioned, in the coordinates \eqref{eq:metric_Penedones_coor}, $K^U$ is the only non-zero component of $K$, and so $K_V$ is the only non-vanishing lower-index component. The normalization above of $K$ means that $K_V=-\omega $ (since $K \cdot T_2=K \cdot (\partial _V)=K_V=-\omega $ indeed).
Now, at the quantum level having a large charge under $T_2$ means $\partial _V \to -i\omega $ as acting on wavefunctions. This substitution with large $\omega $ is the core of the eikonal approximation. In the flat spacetime case, it is equivalent to assuming the exchanged momentum is small compared to the momentum of the particle going along an almost null trajectory and the following simplification of the matter propagators usually used.
Therefore if we only care about large $\omega $ we can replace
\begin{equation}
\partial _{\mu } \rightarrow i K_{\mu } 
\end{equation}
where $\mu $ labels coordinates on dS.
(For particle 2 we similarly have $\partial _{\mu } \rightarrow i\bar K_{\mu } $ conserved.)

Under this assumption, the propagators of the two scalar fields simplify.
Using $\partial _V=-i\omega $ and that $\omega $ is large, the Laplacian simplifies to
\begin{equation}
\nabla ^2 \approx \frac{4i\omega }{(X_0 \cdot W)^2} \partial _U
\end{equation}
and therefore looking at the Klein-Gordon equation with a source for a free scalar, we see that the scalar propagators simplify to
\begin{equation}
G _i(X_j,X_{j+1} )\approx \frac{1}{2\omega } \Theta (U_{j+1} -U_j)\delta (V_{j+1} -V_j)\delta _{S^{d-2} } (W_{j+1} ,W_j)
\end{equation}
where $\delta _{S^{d-2} }$ is the delta on the transverse sphere. Clearly acting with the Laplacian gives a delta function indeed. As claimed, this shows that particle 1 stays on the null geodesic with the same $K^{\mu } $ (or $\partial _V$).

Let us consider now the vertex $\frac{1}{2} \int h_{\mu \nu } T^{\mu \nu } $. Since particle 1 has large $K^U$, the dominant term in $T^{\mu \nu } $ is $\partial ^{\mu } \phi \partial ^{\nu } \phi $ since $g_{\mu \nu } \partial ^{\mu } \phi \partial ^{\nu } \phi $ can have at most one $K^U$ because $g_{UU} =0$. So $T^{\mu \nu } \approx \partial ^{\mu } \phi \partial ^{\nu } \phi $ which can have two factors of $K^U$, larger than the other terms.
The same discussion holds for particle 2 with the barred coordinates. 
We conclude that each graviton propagator together with the vertices gives\footnote{When replacing $\partial _{\mu } =iK_{\mu } $ on both sides of the graviton propagator, the two indices on opposite sides give $(-1)$ and there is another pair of indices giving another $(-1)$ so these factors of $i$ cancel.}
\begin{equation}
\begin{tikzpicture}[baseline=(current bounding box.center)]

\draw[red, thick] (0,2) -- (1,2);

\draw[blue, thick] (0,0) -- (1,0);

\node[circle, fill=red, inner sep=2pt, label=above:{$x$}]
  (T) at (0.5,2) {};

\node[circle, fill=blue, inner sep=2pt, label=below:{$\bar{x}$}]
  (B) at (0.5,0) {};

\draw[decorate, decoration={snake, amplitude=0.8mm, segment length=4mm}]
  (T) -- (B);

\end{tikzpicture}
=\int dX \sqrt{-g(X)}\int d\bar X \sqrt{-g(\bar X)}\, \left(  i^2 D_{\mu \nu ,\rho \sigma }(X,\bar X) K^{\mu } K^{\nu } \bar K^{\rho } \bar  K^{\sigma } \right) .
\end{equation}

In the following, we denote the determinant of the metric in the transverse directions by $g_{\perp} $. When integrating $dW$ or having a delta function in the transverse space, this implicitly includes factors of $\sqrt{g_{\perp} }$ since we write it in embedding space.

Using the simplified form of the scalar propagators and the simplification of the vertex, the sum of diagrams of the form shown in Fig. \ref{fig:generalized_ladder} is just
\begin{equation}
\begin{split}
& \int dV d\bar U \int dW d\bar W \psi _1(X_1)\psi _3(X_n) \psi _2(\bar X_1)\psi _4(\bar X_n) \\
& \int_{-\infty}^{+\infty} dU_1 \sqrt{-g(X_1)/g_{\perp} } \int^{+\infty}_{U_1} dU_2 \sqrt{-g(X_2)/g_{\perp} }\cdots \int^{+\infty} _{U_{n-1} } dU_n \sqrt{-g(X_n)/g_{\perp} }\\
&  \int_{-\infty}^{+\infty} d\bar V_1 \sqrt{-g(\bar X_1)/\bar g_{\perp} } \int^{+\infty}_{\bar V_1} d\bar V_2 \sqrt{-g(\bar X_2)/\bar g_{\perp} }\cdots \int^{+\infty}_{\bar V_{n-1} } d\bar V_n \sqrt{-g(\bar X_n)/\bar g_{\perp} }\\
& (2\omega)^{2(1-n)} \sum _{\text{perm } \sigma } \prod _{j=1} ^n \left( i^2 D_{\mu \nu ,\rho \sigma } (X_j,\bar X_{\sigma (j)} )K^{\mu } K^{\nu } \bar K^{\rho } \bar K^{\sigma } \right) 
\end{split}
\end{equation}
where the sum is over permutations of $n$ elements, and $\psi(X)$ is the propagator from the external points in the correlator to the point $X$. Since the integrand is symmetric in the different $U$'s and $\bar V$'s, we can replace $\int _{U_j \le U_{j+1} } \prod _{j=1} ^n dU_j$ by $\frac{1}{n!} \int \prod _{j=1} ^n dU_j$ with no restrictions on the relative integration variables; the same comment applies to the $\bar V$ integrations. Then all the permutations give the same contribution resulting in a factor of $n!$. Together we get $1/n!$ leading to the exponentiated  expression:
\begin{equation} \label{eq:4pf_from_diagrams}
\begin{split}
& \int dV d\bar U \int dW d\bar W \psi _1(X_1)\psi _3(X_n)\psi _2(\bar X_1)\psi _4(\bar X_n) \\
& (2\omega )^2 \exp \left[ \frac{1}{(2\omega )^2} \int dU d\bar V \sqrt{-g(X)/g_{\perp} }\sqrt{-g(\bar X)/\bar g_{\perp} } 
\left( i^2 D_{\mu \nu ,\rho \sigma } (X,\bar X )K^{\mu } K^{\nu } \bar K^{\rho } \bar K^{\sigma } \right) \right] ,
\end{split}
\end{equation}
where all integrals in the above formula are taken over the entire real axis. For notational simplicity, we leave the integration ranges implicit in this case. 
The argument of the exponent in \eqref{eq:4pf_from_diagrams} is the eikonal phase, and we denote the quantity in the exponent by $i \delta$.

This is the result of \cite{Cornalba:2007zb} adapted to de Sitter space. From the form of the metric \eqref{eq:metric_Penedones_coor} we have $\sqrt{-g/g_{\perp}} =  \frac{(X_0 \cdot W)^2}{2} $. \footnote{The relation between the notation in \cite{Cornalba:2007zb} and here is
\begin{equation}
\begin{split}
& \Pi ^{\mu \nu ,\rho \sigma } =\frac{1}{32\pi G_N} D^{\mu \nu ,\rho \sigma } ,\\
& \Pi ^{(2)} = \frac{1}{8\pi G_N} D_{\mu \nu ,\rho \sigma } K^{\mu } K^{\nu } \bar K^{\rho } \bar K^{\sigma } ,\\
& g^2 =8\pi G_N.
\end{split}
\end{equation}
(The first equation is obtained by comparing the differential equations both satisfy, and then the second equation follows from this by their definition of $\Pi ^{(2)} $. In this differential equation for a spin $j$ particle a direct calculation shows that changing AdS to dS is achieved by the familiar $j/\ell_{\text{AdS}} ^2 \to -j/\ell_{\text{dS}}^2$.)} Usually the wavepackets $\psi$ are highly localized along a null direction, effectively enforcing $V=\bar{U}=0$. This is what we will assume from now on.

In the embedding space we can choose
\begin{equation}
\begin{split}
K_1&=\omega (1,1,0,\cdots ,0) \\
K_2&=\omega (1,-1,0,\cdots ,0) \\
X_0 &= (0,0,1,0,\cdots ,0)
\end{split}
\end{equation}
and in general we will have
\begin{equation}
W=(0,0,\cos \Phi _1,\sin \Phi _1\cos \Phi _2, \cdots )
\end{equation}
(denoting the angular variables by capital letters to distinguish them from Kruskal variables in general)
so that
\begin{equation}
X_0 \cdot W = \cos \Phi _1.
\end{equation}

As mentioned, we choose for simplicity the trajectory of particle 1 to be at $V=0$ and that of particle 2 to be at $\bar U=0$. These $V,\bar U$ coordinates are of interest to us as we will see, and they also have a simple description in terms of Kruskal coordinates. Indeed, it can be checked that the relation between Kruskal coordinates and the capitalized coordinates simplifies in some cases. First
\begin{equation} \label{eq:Penedones_Kruskal_V0}
u = -\frac{U}{2} \cos\Phi _1,\quad v=0 \qquad \text{for } V=0
\end{equation}
and we can choose all $\phi_i =\Phi_i $.
In fact, it is also true that
\begin{equation} \label{eq:Penedones_Kruskal_U0}
u=0,\quad v = -\frac{V}{2} \cos\Phi _1 \qquad \text{for } U=0
\end{equation}
with the angles identified.
Similarly
\begin{equation} \label{eq:Penedones_Kruskal_Ub0}
\bar{u}=0, \quad \bar{v} = -\frac{\bar V}{2} \cos \bar \Phi_1  \qquad \text{for } \bar U=0
\end{equation}
and 
\begin{equation} \label{eq:Penedones_Kruskal_Vb0}
\bar{u} = -\frac{\bar U}{2} \cos\bar \Phi_1,\quad \bar{v}=0  \qquad \text{for } \bar V=0
\end{equation}
and in both cases $\bar \Phi _i=\phi_i $.

Using that, we can rewrite the eikonal phase in terms of the standard Kruskal coordinates $u,v$:
\beq
\label{eq:general_eikonal}
i\delta= -\frac{1}{4} \int du d\bar{v}  
 D_{uu,\bar{vv}}  p^u p^v, 
\eeq
where we used more conventional \cite{Shenker:2014cwa} notation $p^{u,v}$ for the particles momenta $K^\mu, \bar{K}^\nu$ components.
Importantly, this Green's function is Feynman's Green's function, so the integral over $u,\bar{v}$ can be rotated to the Euclidean plane of $(x=u \bar{v},\gamma)$:
\beq
\int_{-\infty}^{+\infty} du d\bar{v} \ra  -i \int_0^{2 \pi} d\gamma \int_0^{+\infty} dx = -2 \pi i  \int_0^{+\infty} dx
\eeq

\subsection{Gauge invariance and large diffeomorphisms}
\label{sec:gauge_inv}
In gauge theories, including gravity, it is not obvious that the above eikonal exponentiation of the propagator leads to the gauge-invariant answer.
In fact, as we illustrate now the eikonal phase is independent of small gauge transformations (vanishing at infinity), but it depends on large gauge transformations which do not decay at infinity.

For a warm-up, consider photon propagator in the flat space-time
in the $R_\xi$ gauge:
\beq
D_{\mu,\nu} = \bra A_\mu A_\nu \ket = \frac{1}{p^2} \l( g_{\mu \nu} + (1-\xi) \frac{p_\mu p_\nu}{p^2} \r)
\eeq
For photon exchanges, the analogue of the integrand in eq. (\ref{eq:4pf_from_diagrams})  is $D_{\mu,\nu} K^\mu \bar{K}^\nu$ so we are taking $\bra A_U A_V \ket$ part of the propagator and putting $p_U=p_V=0$. Hence the answer is $\xi$-independent.

Now, consider gravity. Eq. (\ref{eq:4pf_from_diagrams}) basically instructs us to evaluate the integral of 
\[D_{UU,\bar{V} \bar{V}}(U,V=0;\bar{U}=0,\bar{V})\]
over $U,\bar{V}$. This is a very general expression applicable for anti-de Sitter, de Sitter and flat spacetimes.
Because the relevant components are $D_{UU,\bar V\bar V}$, and the integrations are performed along the corresponding null directions $U$ and $\bar V$, an infinitesimal diffeomorphism shifts the integrand by total derivatives.
\textit{Whether such terms can be discarded, however, depends on the asymptotic behavior of the gauge transformation.} Diffeomorphisms that decay sufficiently rapidly at infinity do not contribute, whereas large diffeomorphisms that remain nontrivial at the boundary can potentially generate a finite shift in the eikonal phase. We find that for anti-de Sitter it is not the case. In de Sitter space, by contrast, certain large diffeomorphisms do contribute, as we explain in the next section. Interestingly, the $L=0$ contribution, which we will show to be responsible for the growth of the OTOC, is a large diffeomorphism.

\section{Zero modes and divergences in the eikonal phase in de Sitter}

\label{sec:eikonal}

\subsection{IR divergence in scalar exchange}

In this subsection we try to reconcile the non-existence of localized shockwave solutions and the diagrammatic derivation in the eikonal approximation. 
The core issue we are interested in happens for any spin of the exchanged particle, so let us discuss the simplest case of an exchanged scalar particle instead of a graviton. We should mention right away that for the scalar exchange the eikonal approximation is actually not valid as was discovered a long time ago \cite{Kabat:1992pz,Treiman}, but we can always analyze the analogous set of diagrams.
Studying the scalar case will eliminate potential sources of confusion in the gravitational case, such as those associated with gravitational constraints and dressing.
We conclude that the non-existence of localized shockwaves is manifested in the diagrammatic calculation as an IR divergence. 

On the one hand, following a similar analysis to \cite{Cornalba:2007zb} and generalizing to de Sitter, the integrated correlator in the exponent in \eqref{eq:4pf_from_diagrams} can be phrased in terms of an effective transverse propagator $G_T$ which is an invariant function on transverse space (invariant under rotations of $W$ and $\bar W$). This transverse propagator satisfies (which is the analogue of the shockwave equation)
\begin{equation} \label{eq:Greens_func_eq_transverse_prop}
\left( \nabla ^2_{S^{d-2} } +d-2-m^2\right) G_{m^2,T}(W,\overline{W})=-\delta _{S^{d-2} } (W, \bar W)
\end{equation}
for any spin of the exchanged particle of mass $m$.
The issue we encountered is that for $m^2=0$ there is a zero mode of the Klein-Gordon operator. More generally, we see that this is the case for discrete integer values of $m^2$ starting at
\begin{equation} \label{eq:scalar_masses_leading_to_divergent_transverse_propagator}
m^2  \le d-2 .
\end{equation}

On the other hand, we can rewrite the eikonal phase as
\begin{equation} \label{eq:scalar_case_exponent_int}
i \delta = -\frac{g^2}{4\omega ^2} (X_0 \cdot W)(X_0 \cdot \bar W) \int du d\bar{v} \, G_{m^2,F} (X \cdot \overline{X})
\end{equation}
where $g$ is the coupling constant to the exchanged scalar and $G_{m^2,F}(X \cdot \overline{X})$ is the Feynman propagator of the exchanged scalar. The Wightman \footnote{Feynman prescription corresponds to taking $Z \ra Z - i \epsilon \sgn(t-\bar{t})$, where $t,\overline{t}$ are time components of $X,\overline{X}$.} 2-point function of a free scalar in $dS_d$ is (for this formula we restore factors of $\rdS$)
\begin{equation} \label{eq:dS_2pf_general_d}
G_{m^2,W } (Z) \equiv G_{d,\mu}(Z) = \frac{\Gamma( h_{\pm} )}{\rdS^{d-2} (4\pi )^{d/2} \Gamma \left( \frac{d}{2} \right) } \, {}_2F_1 \left( h_{\pm } ;\frac{d}{2} ;\frac{1+Z}{2} \right) 
\end{equation}
where by a Gamma function with $\pm $ arguments we mean the product of the corresponding Gamma functions, and similarly two of the arguments of the hypergeometric function are the two parameters. Above
\begin{equation}
h_{\pm } =\frac{d-1}{2} \pm i\mu 
\end{equation}
and the parameter $\mu $ in general dimension is related to the mass $m$ by
\begin{equation} \label{eq:mu_parametrization_of_mass}
\mu = \sqrt{m^2\rdS^2-\left( \frac{d-1}{2} \right) ^2}.
\end{equation}
For the two trajectories we chose, the invariant chord distance is
\begin{equation}  \label{eq:Z-def}
Z = X \cdot \bar X=-2u \bar{v} + y \cdot \bar y
\end{equation}
where $y$ and $\bar y$ are the two vectors in the embedding space of the transverse sphere of the two particles (it is an invariant distance on this sphere). So the propagator and its integral over $u$ and $\bar{v}$ are invariant under transformations of the transverse sphere. Therefore \eqref{eq:scalar_case_exponent_int} agrees with the claim about the relation between $ i \delta $ and an invariant in transverse space in \cite{Cornalba:2007zb} in the scalar case.

Let us consider the region of integration where for instance we fix $u$ and integrate over $\bar{v}$. $\bar{v} \to 0$ is the region associated with the UV behavior and large $\bar{v}$ with the IR behavior. As can be seen for instance from the form of the 2-point function as a hypergeometric function \eqref{eq:dS_2pf_general_d}, it behaves for large $Z$ as $Z^{\frac{1-d}{2} \pm i\mu } $ (a combination of these two terms) where $\mu$ is given in \eqref{eq:mu_parametrization_of_mass}.  So
\begin{itemize}
\item For $m^2 > \left( \frac{d-1}{2} \right) ^2$ it behaves qualitatively like $\frac{\sin \left( \mu \log Z\right) }{Z^{\frac{d-1}{2} } } $ (we do not write the phase of the $\sin$).
\item For $0<m^2 < \left( \frac{d-1}{2} \right) ^2$, the parameter $\mu $ is purely imaginary and the 2-point function behaves as $|Z|^{|\mu |-\frac{d-1}{2} } $ with $|\mu | < \frac{d-1}{2} $.
\item If we were taking $m^2<0$, we would have $|\mu |>\frac{d-1}{2} $ and the expression would diverge for large $Z$ as seen by the previous item.
\end{itemize}

So let us integrate over large $\bar{v}$ or $Z$ (they behave the same by \eqref{eq:Z-def}). In the first case, the integral generally converges.\footnote{This is true for $d>3$. For $d=3$ strictly speaking it does not converge, but rather behaves as $\int dZ \, \frac{\sin(\mu \log Z)}{Z} $ or changing variables to $\log Z=z$ it becomes $dz \, \sin(\mu z)$ which can be regularized.} In the second case, the integral diverges starting at the solution of $|\mu |- \frac{d-1}{2} =-1$, resulting in an IR divergence for the range of $m^2 \le d-2$. This is precisely the range \eqref{eq:scalar_masses_leading_to_divergent_transverse_propagator} that covers the zero modes affecting the eikonal phase. 
The differential equation \eqref{eq:Greens_func_eq_transverse_prop} can have finite solutions for fractional values of $m^2$, which corresponds to the usual fact that the integral can be analytically continued to give a finite answer for such values.
It is curious that the integrated correlator can be analytically continued even for many integer values of $m^2$ in this range, as can be seen by the existence of solutions to the Green's function equation.

\subsection{Naive graviton exchange}

In this subsection we finally discuss the graviton exchange using the master formula (\ref{eq:4pf_from_diagrams}). 
For that we will need the full graviton propagator, which is pretty complicated. However, things simplify considerably if we focus on the ``physical'' part of the propagator, which excludes total derivative and trace terms. This part of the propagator is gauge independent. Since we are interested in the $UU,VV$ part of the propagator the trace terms never contribute. For example, in Appendix \ref{app:ads_phys} we show that in $AdS_3$ case the physical part of the propagator gives rise to the correct eikonal phase. However, as we will see momentarily, in $dS$ case this ``physical'' part in isolation produces a highly IR divergent expression. 

As we explain in Appendix \ref{app:ds_phys}, the physical part of the propagator is \cite{Morrison} 
\beq \label{eq:D-physical}
D_{\mu \nu,\mu' \nu'}(X,X') = 
16 \pi G_N (\pr_\mu \pr_{\mu'} Z \pr_\nu \pr_{\nu'} Z + \pr_\mu \pr_{\nu'} Z \pr_\nu \pr_{\mu'} Z) H_0(Z) + \text{grad} + \text{metric}
\eeq
where metric terms are proportional to $g_{\mu \nu}, g_{\mu' \nu'}$ and grad terms are derivatives in $\mu,\nu,\mu',\nu'$. Variable $Z = X \cdot X'$ is the usual chord distance. A more careful derivation of this propagator is presented in Appendix \ref{app:dS-propagator}.
The
function \(H_0(Z)\) is  the regulated massless
scalar propagator \footnote{It can also be obtained as $$\lim_{m \ra 0}  \l[ G_{m^2} - \frac{\pi^{-d/2-1/2}}{2m^2} \Gamma \l( \frac{d+1}{2} \r) \r]$$}:
\beq
\label{eq:H0_def}
H_0(Z) = \frac{\pr }{\pr m^2} (m^2 G_{m^2}(Z)) \lvert_{m^2=0}
\eeq
Scalar massive propagator is given by eq. \eqref{eq:dS_2pf_general_d}.
For the kinematics at hand $Z = - 2 u \bar{v} + y \cdot \bar{y}$ (eq. \eqref{eq:Z-def}) so $\pr_u \pr_{\bar{v}} Z = -2$, so the integrand for the eikonal phase we simply have $H_0(Z)$. The problem is that in all dimensions \footnote{For example, for $d=4$, $$
H_0 \propto -2 \log((1-Z)/2) + \frac{2}{Z-1} - \frac{14}{3}
$$}, $H_0(Z)$ behaves as $\log(Z) = \log(-2 u \bar{v})$ for large $Z$, so the integral is badly divergent. 

In this computation, we have omitted the possible contributions of large diffeomorphisms to \eqref{eq:D-physical}, even though such terms can modify the result. As we will show, their inclusion in the full graviton propagator removes the divergence found above. This cancellation, however, is accompanied by a distinct divergence associated with the $L=1$ zero mode that appears in the shockwave analysis.

\subsection{Full graviton exchange: IR divergence in the vector sector }

Before we dive into evaluating the eikonal phase for de Sitter using known formulas for the propagator, let us describe the general structure of the graviton propagator. Graviton propagator has 3 parts: transverse-traceless (TT), vector and several scalar. Specifically, the scalar part of the propagator $D_{\mu \nu, \mu' \nu'}$ has the following tensor structures:
\beq
\text{trace:} \ g_{\mu \nu} g_{\mu' \nu'}
\eeq
\beq
\text{longitudinal:} \ \nabla_{\mu} \nabla_{\nu} \nabla_{\mu'} \nabla_{\nu'}
\eeq
\beq
\text{mixed:} \ \nabla_{\mu} \nabla_{\nu} g_{\mu' \nu'} + \nabla_{\mu'} \nabla_{\nu'} g_{\mu \nu}
\eeq
TT and trace part are the only gauge-independent parts.

The trace and mixed parts above do not contribute to the eikonal phase because we are interested in the null component  $D_{uu,\bar{v} \bar{v}}$.
The longitudinal part corresponds to a diffeomorphism in the metric.   

 The whole vector contribution is a diffeomorphism because it essentially reflects a diffeomorphism transformation $g_{\mu \nu} \rightarrow g_{\mu \nu} + \nabla_\mu \xi_\nu + \nabla_\nu \xi_\mu$ on the metric. It can potentially contribute if this diffeomorphism is large, i.e. does not decay sufficiently fast at infinity. 
 TT contribution can be parametrized by a single scalar function (directly related to $H_0$ from the previous section) which is acted upon by a complicated projector onto the TT subspace. That projector is rather involved and it contains some parts which are also pure diffeomorphisms.

Now let us discuss the form of the propagator in de Sitter in more detail. We will need to fix the gauge.
The most general \footnote{It is because the most general vector linear in the metric perturbation $h_{\mu \nu}$ one can build for a maximally symmetric space like de Sitter is $\nabla^{\mu} h_{\mu \nu} - \frac{1+\beta}{\beta} \nabla_{\nu} h^\gamma_\gamma.$} quadratic gauge fixing term with two derivatives one can add to the action can be parameterized by two real numbers $(\alpha,\beta)$:
\beq 
\label{eq:gauge_fix}
16 \pi G_N S_{\text{gauge}} = -\frac{1}{2\alpha} \int d^dx \sqrt{-g} \l( \nabla^\mu h_{\mu \nu} - \frac{1+\beta}{ \beta} \nabla_{\nu} h^{\gamma}_{\gamma} \r)^2
\eeq
With this choice, the scalar part is heavily $\alpha,\beta$-dependent, whereas the vector part has simply an overall $\alpha$ factor. For $\alpha=0, \beta=-2$ one recovers de Donder gauge.
Explicit form of the propagator in different dimensions is widely available in the literature, using both Euclidean \cite{Allen:1986tt,Higuchi:2001uv,Morrison} and Lorentzian \cite{Morrison,Frob:2016hkx} approaches. These expressions and the resulting integrals are explicit but rather bulky, so we will not write down intermediate computations and instead provide the final answers for the eikonal phase.
Using these results in dimensions $d=3,4,5$ we found the following behavior of the eikonal phase integral $\int du d\bar{v} D_{uu,\bar{vv}}$:
\begin{itemize}

\item Vector part is IR divergent. For large $x=u \bar{v}$ the integrand $D_{uu,\bar{vv}}$ behaves as ($\#$ is dimension-dependent coefficient):
    \beq
    \alpha \# \frac{\cos(\theta)}{u \bar{v}}
    \eeq
This logarithmic divergence is proportional to $L=1$ mode on the transverse sphere, since $\theta$ is the equatorial angle between the transverse positions of the two particles. Moreover, it is proportional to the gauge-fixing parameter $\alpha$. One could send $\alpha$ to zero but it has to be done with care because the logarithmic divergence also formally produces an infinite constant. One can consider the limiting procedure where one introduces an IR cutoff $\Lambda_{IR}$ and then sends $\alpha \ra 0, \log(\Lambda_{IR}) \ra \infty$, keeping $\alpha \log(\Lambda_{IR})$ finite, thus having $\alpha=0$, but assigning to this $L=1$ contribution an arbitrary finite coefficient. 
In fact, this limiting procedure can be connected to the earlier shockwave discussion. The shockwave metric (\ref{eq:ds3_shock_metric}) is, in fact, in the TT gauge \footnote{It is because the perturbation $h_{uu}$ is traceless and on the horizon $D_v = \partial_v$, so $D^\mu h_{\mu u} \propto D_v h_{uu} = \partial_v (f(\phi) \delta(u))=0$.} enforced by the $\alpha=0$. Solutions to the shockwave profile equation (\ref{eq:problematic_eq}) are also defined up to adding $L=1$ zero mode.

Recall that the whole vector part is a pure diffeomorphism. However, since the integral is divergent this is a large diffeomorphism. On the tree level this divergence can be eliminated if the matter sources are orthogonal to the $L=1$ mode (as would follow from the gravitational constraints). However, on the loop level this causes a genuine divergence.

\item TT and scalar (as long as $\beta$ is not too negative) parts give a  gauge independent answer which is both IR and UV finite. After integrating over $u, \bar{v}$ one recovers precisely the answer from the shock wave computation if one projects out $L=1$ mode: 
    \beq
    \label{eq:phase_decomp}
    i \delta^{TT}_{dS}=-i 16 \pi G_N p^u p^v \sum_{L \neq 1} \frac{1}{-(L+d-3)L+(d-2)} \sum_M Y_{LM}(y) \overline{Y}_{LM}(\bar{y}),
    \eeq
    where $Y_{LM}(y)$ are normalized spherical harmonics on $(d-2)$ sphere \footnote{For example, for $d=3$: 
\beq
\frac{1}{2 \pi} \sum_{L=-\infty,|L| \neq 1}^{+\infty} \frac{e^{i \theta L}}{-L^2+1} = -\frac{1}{4 \pi} \l( \cos(\th) - (2 \pi - 2 |\th|) \sin(|\theta|) \r)
\eeq
Around $\theta=0$ it has a kink of $\frac{1}{2} \sin(|\theta|)$, giving rise to a delta-function.
For $d=4$: 
\beq
\sum_{L \neq 1} \frac{1}{-L(L+1)+2} \sum_m Y^m_{L}(\theta,\phi) Y^{-m}_L(0,\phi')
\eeq
Sum over $m$ can be simplified into $\frac{2l+1}{4 \pi} P_L(\cos(\theta))$
\beq
\sum_{L \neq 1} \frac{2l+1}{4 \pi}\frac{P_l(\cos(\theta)) }{-L(L+1)+2} = \frac{1}{32 \pi} \l( 8 + \frac{8}{3} \cos(\theta) (4 + 3 \log{ \sin^2 \l( \frac{\theta}{2}  \r)}) \r)
\eeq
For small $\theta$ it behaves as $\frac{1}{2\pi} \log(\theta)$ so it again satisfies the wave equation with a delta-function source.}.
\end{itemize}

As promised, the severe $\log(u \bar{v})$ divergence in the ``physical'' part is now absent. However, now we have a logarithmic $1/u \bar{v}$ divergence in the vector sector. In the next subsection we discuss the physical consequences of this.

\subsection{Full graviton exchange: diffeomorphism contributions}

The reason why we obtained a finite expression in the TT sector is because of the total derivative terms which come from the metric diffeomorphisms that grow at infinity and cancel the divergence in the ``physical'' part. In contrast, in the vector sector the diffeomorphisms did the opposite: they introduced a milder logarithmic divergence proportional to $L=1$ mode.

One of the most puzzling features of OTOC in de Sitter is the initial growth of the OTOC, which is related to the aforementioned time advance generated by infalling matter. Whether we get time advance or time delay is essentially determined by the sign of the eikonal phase because it is integrated metric perturbation.
It would be instructive to compare the dS answer with the AdS answer. In $AdS_d$ black hole spacetime, the shockwave profile  $f$ along the black hole horizon is determined by \cite{Shenker:2014cwa} (constant $\mu$ is related to the black hole temperature and $AdS$ radius):
\beq
\Delta_{S^{d-2}} f - \mu^2 f \propto T_{uu}, \quad \mu>0.
\eeq
leading to the eikonal phase 
\beq
i \delta_{AdS} = - i 16 \pi G_N p^u p^v \sum_{L \ge 0} \frac{1}{-(L+d-3)- \mu^2} \sum_M Y_{LM}(y) \overline{Y}_{LM}(\overline{y})
\eeq
Comparing this to the de Sitter counterpart (\ref{eq:phase_decomp}), we see that only $L=0$ mode has a different sign, indicating time advance rather than time delay. 

 Interestingly, $L=0$ contribution in the de Sitter case comes from the pure diffeomorphism. It can be traced to the following part of the propagator \cite{Allen:1986tt}:
\beq
\label{eq:s_wave_diffeo}
D_{\mu \nu,\mu' \nu'} \supset \nabla_{\mu} \nabla_{\nu} \nabla_{\mu'} \nabla_{\nu'} G^{--}_{m^2=-d}(Z)
\eeq
where $G^{--}_{m^2}(Z)$ is the propagator for a minimally coupled scalar with the singular terms for $m^2=0,-d$ have been subtracted. Specifically,
\beq
G^{--}_{m^2}(Z) = G_{m^2}(Z) - \frac{\Gamma \l( \frac{d+1}{2} \r)}{2m^2} \pi^{-d/2-1/2} - \frac{\Gamma \l(  \frac{d+3}{2} \r)}{m^2+d} Z \pi^{-1/2-d/2}
\eeq
For example:
\beq
dS_4: \quad G^{--}_{m^2=-4} \propto 
\frac{2}{1-Z} - 6 Z \log \l( \frac{1-Z}{2} \r) + \frac{1}{10}(-45-112 Z)
\eeq
On the lightcone $\nabla_{u/\bar{v}} = \pr_{u/\bar{v}}$ so the integral is easy to evaluate and it leads to a constant which is angle-independent ($L=0$ in eq. (\ref{eq:phase_decomp})). The reason why we get a finite contribution is the $Z \log(Z)$ growth which persists in all dimensions. This growth reflects the fact that \(G^{--}_{-d}\) is the regularized scalar propagator with tachyonic mass \(m^2=-d\).

It would be interesting to understand what are the physical consequences of the fact that the ``wrong sign'' of the eikonal phase in fact comes from a large diffeomorphism. One immediate consequence is the sensitivity to the integration contour choice: for general kinematics, Euclidean sphere computation could produce a different answer compared to the Lorentzian eikonal computation.

Let us briefly explain the origin of this term in the propagator and why it does not contribute in the AdS case. For that it would be convenient to adopt Euclidean approach \cite{Allen:1986tt, Higuchi:2001uv}.
One builds graviton propagator from normalized scalar $\phi^{(n,i)}$ vector $\xi_\mu^{(n,i)}$ and tensor $W_{\mu \nu}^{(n,i)}$ harmonics on a $d-$dimensional sphere. Scalar harmonics simply obey Laplace equation:
\beq
\square \phi^{(n,i)} = \lambda_n \phi^{(n,i)}, \ \qquad \lambda_n \le 0
\eeq
with index $i$ parameterizing the degeneracy.
Tensor harmonics can actually be obtained by acting on scalar harmonics with a differential operator:
\beq
W^{(n,i)}_{\mu \nu} = \mathcal{N}_{d,n}^{-1/2} \l( \nabla_\mu \nabla_\nu - \frac{1}{d} g_{\mu \nu} \square \r) \phi^{(n,i)}, \quad n \ge 2,
\eeq
where the normalization factor comes from requiring a unit norm on the sphere: 
\beq
\int \dd \Omega_{S^{d}} \  W^{(n,i)}_{\mu \nu} \overline{W}^{(n,i), \mu \nu} =1,
\eeq
resulting in the following normalization factor:
\beq
\text{dS}: \quad \mathcal{N}_{d,n} =  \lambda_n \l(  \lambda_n \frac{d-1}{d} + (d-1) \r)
\eeq
These harmonics contribute the following term to the graviton propagator
\beq
D_{\mu \nu, \mu' \nu'}(x,x') \supset \sum_{n \ge 2, i} a_n W^{(n,i)}_{\mu \nu}(x) \overline{W}^{(n,i)}_{\mu' \nu'}(x').
\eeq
Pulling aside the tensor structure part, the sum over $n,i$ essentially consists of 
\beq
\sum_{n \ge 2, i} \frac{a_n}{\lambda_n(  \lambda_n \frac{d-1}{d} + (d-1) ) } \phi^{(n,i)}(x) \overline{\phi}^{(n,i)}(x').
\eeq
Coefficient $a_n$ is gauge-dependent, but it only has a form of scalar propagator $(\lambda_n + \dots)^{-1}$.  However, it is the  denominator $1/(\lambda_n \frac{d-1}{d} + (d-1) )$ which leads to the tachyonic mode with the mass $m^2 = -d$.

AdS computation is essentially the same, except that the normalization factor has a different sign for $d-1$, resulting in physical mass $m^2=d$ \cite{Mir}:
\beq
\text{AdS}: \quad \mathcal{N}_{d,n} = \lambda_n  \l( \lambda_n \frac{d-1}{d} - (d-1) \r).
\eeq
Thus these terms are present in AdS too but they are small (rapidly decaying) diffeomorphisms.

\subsection{Regularizing the divergence: non-zero graviton mass}

We identified the problem of the eikonal approximation as an IR divergence. This suggests that we can regularize the divergence by including a small graviton mass $\gm$. In terms of the Einstein-Hilbert action expanded to second order in the metric fluctuations (denoted $S_2$) with the gauge fixing term \eqref{eq:gauge_fix}, this can be achieved by including for instance a single term such that
\begin{equation} \label{eq:quadratic_action_alpha_beta_gauge}
\begin{split}
& 16\pi G_N(S_2+S_{\text{gauge}} ) = \\
& = - \int d^d x \sqrt{- g}\, h^{\mu \nu } \bigg[ -\frac{1}{4}  g_{\mu \rho }  g_{\nu \sigma }  \nabla ^2 +\frac{1}{4}  g_{\mu \rho }  g_{\nu \sigma } \gm ^2+ \frac{1}{2\rdS^2}  g_{\mu \rho }  g_{\nu \sigma }+ \left( \frac{1}{4} -\frac{(1+\beta )^2}{2\alpha \beta ^2 } \right)   g_{\mu \nu }  g_{\rho \sigma }  \nabla ^2\\
& \qquad \qquad  + g_{\mu \nu }  g_{\rho \sigma } \frac{d-3}{4\rdS^2} 
+\left( \frac{1}{2} -\frac{1}{2\alpha } \right)  g_{\mu \rho }  \nabla _{\nu }  \nabla _{\sigma } 
+\left( -\frac{1}{2} +\frac{1}{\alpha } +\frac{1}{\alpha \beta } \right)  g_{\mu \nu }   \nabla _{\sigma }  \nabla _{\rho } 
\bigg] h^{\rho \sigma } .
\end{split}
\end{equation}
With this modification we can find a finite result for the eikonal phase:
\begin{equation}
    i \delta = 16 \pi i G_N \rdS^{6-d} p^u p^v \sum _{L,M} \frac{1}{\gm^2 \rdS^2 +L(L+d-3)-(d-2)}  Y_{LM} (\phi ) Y^*_{LM} (\bar \phi  ).
\end{equation}
We have derived this in two ways. One is based on the equivalent saddle-point description of the eikonal approximation, explained in Appendix \ref{sec:saddle_point_linearized}. In this method we solve the linearized Einstein equations resulting from \eqref{eq:quadratic_action_alpha_beta_gauge}. (For brevity we discuss explicitly $\alpha=1,\beta=-2$ but the solution is independent of this.) The second method is based on \cite{Cornalba:2007zb} and is shown in Appendix \ref{app:mg_zero}.

Now that we have added the mass regulator, we can write down a simple and finite eikonal phase.
The idea is that we are going to keep $G_N e^t / \gm ^2$ fixed in the limit $G_N \to 0$, where $t$ is the OTOC time (as discussed in the Introduction). In this limit, the eikonal phase is dominated by the $L=1$ singular mode. The corrections from the regular piece are suppressed as $\gm  \rdS \to 0$. The dominant contribution is therefore
\begin{align}
\label{eq:Eikonal_phase_result}
& i \delta = 
16 \pi i \frac{G_N}{\gm^2} \rdS^{4-d} p^u p^v \sum_{M} Y_{1M}(y) Y^*_{1M}(\bar{y})
= 16 \pi i \frac{G_N}{\gm^2} \rdS^{4-d} p^u p^v  \frac{d-1}{\sVol{d-2}} (y \cdot \bar y)
\end{align}
where we also wrote the result in the sphere embedding space parametrized by $y$.

For example, a 1-sphere is parametrized by $\phi  \in [0,2\pi )$ and in this case the angular dependence coming from $\sum_{M} Y_{1M}(y) Y^*_{1M}(\bar{y})$ is
\begin{equation}
\frac{1}{\pi} \cos(\phi -\bar \phi )
\end{equation}
and for a 2-sphere we have the usual $(\theta ,\phi )$ parametrization and
\begin{equation}
\sum _{m=-1} ^1 Y^*_{1m} (\bar \theta ,\bar \phi )Y_{1m} (\theta ,\phi )= \frac{3}{4\pi} \left( \cos \theta \cos \bar \theta +\cos(\phi -\bar \phi )\sin \theta  \sin \bar \theta \right)  .
\end{equation}

\section{OTOC in de Sitter} \label{sec:ds_otoc}

 In this section, we study out-of-time-order correlators (OTOC) of operators inserted along the observer's worldline, which create perturbations of the empty de Sitter background. We focus on the regime of large time separation, in which these perturbations become highly boosted near the static-patch horizon while the geometry away from the horizon remains approximately unaffected. The OTOC can then be evaluated as a near-horizon scattering process. This requires determining the horizon wave profiles of the initial perturbations and computing the corresponding $S$-matrix in the eikonal regime, using the eikonal phases derived in the preceding sections. We first present the wave profiles for general spacetime dimension and arbitrary insertion points, and then analyze the scattering problem in two complementary regimes.

First, we introduce the regulator $m_g$ and consider the double-scaling limit
$m_g\to 0,\, G_N\to 0$, and 
$ G_N e^{t/\ell}/m_g^2$ fixed. 
In this limit, only the $L=1$ modes remain finite, and the resulting expression resums the OTOC to all perturbative orders in $G_N e^{t/\ell}/m_g^2$. Second, we evaluate the eikonal phase without introducing the regulator. In this case, all angular-momentum modes can contribute. Because of the underlying IR issue, however, the result is reliable only at tree level. This is nevertheless sufficient to determine the leading early-time growth of the OTOC and the associated Lyapunov exponent.

In both regimes, we find an initially growing contribution to the OTOC. In the regulated double-scaling limit, the resummed answer exhibits a Lyapunov exponent
$\lambda_L=\frac{4\pi}{\beta_{\rm dS}}$, whereas the unregulated tree-level calculation gives 
$\lambda_L=\frac{2\pi}{\beta_{\rm dS}}.$
Thus the $L=1$ sector isolated by the regulated scaling limit grows with an exponent twice that obtained from the full tree-level eikonal phase.

\subsection{OTOC as a scattering problem}
In this section we recall the description of OTOCs in gravity given in \cite{Shenker:2014cwa} to set up a general master formula we will evaluate in this section.

We will parametrize the position of operators by time $t$ and spatial coordinates $x$. We are interested in the 4-point OTOC
\begin{equation}
D = \langle V_{x _1} (t_1)W_{x _2} (t_2)V_{x _3} (t_3)W_{x _4} (t_4)\rangle .
\end{equation}
Parametrizing the metric at the horizons as
\begin{equation} \label{eq:def_a0_r0}
\dd s^2=-a_0 \dd u \dd v+r_0^2 \dd x ^2,
\end{equation}
where $x$ parametrizes an internal space with a metric left implicit for the moment,
it was shown in \cite{Shenker:2014cwa} that when $t_4 \approx t_2 \approx t$ very large and positive and $t_3 \approx t_1 \approx 0$
\begin{equation} \label{eq:Stanford_Shenker_OTOC_formula}
D = \frac{a_0^4}{(4\pi )^2} \int p^v dp^v p^u dp^u r_0^{d-2} dx  \, r_0^{d-2} dx 'e^{i\delta (s,|x - x '|)} \left[ \psi _1^*(p^u,x )\psi _3(p^u,x )\right] \left[ \psi _2^*(p^v,x ')\psi _4(p^v,x ')\right] 
\end{equation}
where the wavefunctions are
\begin{equation} \label{eq:wavefunction_defs}
\begin{split}
& \psi _1(p^u,x )=\int dv \, e^{ia_0p^u v/2} \langle \phi _V(u,v,x) V _{x _1} (t_1)^{\dagger} \rangle |_{u=0} \\
& \psi _2(p^v,x )=\int du \, e^{ia_0p^v u/2} \langle \phi _W(u,v,x) W _{x _2} (t_2)^{\dagger} \rangle |_{v=0} \\
& \psi _3(p^u,x )=\int dv \, e^{ia_0p^u v/2} \langle \phi _V(u,v,x) V _{x _3} (t_3) \rangle |_{u=0} \\
& \psi _4(p^v,x )=\int du \, e^{ia_0p^v u/2} \langle \phi _W(u,v,x) W _{x _4} (t_4) \rangle |_{v=0} 
\end{split}
\end{equation}
We assume that $V$ and $W$ are Hermitian.

Let us recall the idea behind this formula. As mentioned, because of large $t$, we assume that nothing happens to the geometry away from the horizons. Therefore, we express the $W$ external quanta in terms of particles at $v=0$, and the $V$ quanta at $u=0$, treating them just as free massive quantum fields living on a rigid AdS or dS space.
Even if say $W_4$ has coordinates $x _4$, it will generally have a different $x $ when propagated to the horizon. Then, to evaluate the scattering of all the particles that we get at the two horizons, we use the eikonal approximation to gravity to get the eikonal phase $\delta $. Because of large $t$, the momenta $p^v_2\approx p^v_4$ and $p_1^u \approx p_3^u$ (since these components are large and so the rest are small) and the coordinates $x $ of the $W$ particles are preserved at the momentary interaction at the horizon, and similarly for $V$.

Let us now restrict to $dS_d$. We can get operators on the opposite pole by analytically continuing the static patch time coordinate. Now in place of the internal coordinates $x$ we have the angular variables parametrizing a unit sphere that we collectively denote by $\phi $. Because of rotational symmetry in $\phi $, starting with a localized operator at the pole will lead to a smeared wavefunction at the horizon. For the metric \eqref{eq:dS_Kruskal} we have
\begin{equation}
a_0=4 \rdS^2,\qquad r_0 = \rdS .
\end{equation}

\subsection{Wavefunctions in general dimension}

In this subsection we compute the Fourier transform along the horizon of the scalar two-point function (\ref{eq:dS_2pf_general_d}). 
Let us write the chord distance $Z$ between a point $(u,v,\phi )$ and a point at the right pole $(r=0,t)$ where we place the operators. Because of the $U(1)$ rotational symmetry, it does not depend on $\phi $. This is different from AdS, where the $\phi $ at the asymptotic (CFT) insertion point spreads only a little (in units of the AdS scale) when we propagate to the horizon. Here since the circle caps off, starting from a point source gives a completely uniform angular dependence at the horizon. Using embedding coordinates, the corresponding inner product is
\begin{equation}
Z = \frac{v \,  e^{-t/\rdS} -u \,  e^{t/\rdS} }{1-uv} .
\end{equation}
Note we use units where $u,v$ are dimensionless (but $t$ is not).

The hypergeometric function ${}_2F_1(\cdots ;z)$ has a branch cut usually taken at $ z \in (1,\infty )$. Let us concentrate on $\psi _4$. In this case $Z=-ue^{t/\rdS} $ so there is a branch cut at $u \in (-\infty ,-e^{-t/\rdS} ) $. It is shifted a little above the real axis for $u<0$ with the $i\epsilon $ prescription. For $p^v<0$ we close the contour of the $u$ integration below and get zero in \eqref{eq:wavefunction_defs}. For $p^v>0$ we close the contour above and get
\begin{equation}
\begin{split}
& \psi _4 = \frac{\Gamma (h_{\pm } )}{\rdS^{d-2} (4\pi )^{d/2} \Gamma \left( \frac{d}{2} \right) } \int _{-\infty } ^{-e^{-t/\rdS} } du \, e^{2i\rdS^2p^v u} \\
& \qquad \qquad \left[  {}_2F_1\left( h_{\pm } ;\frac{d}{2} ; \frac{1-(u-i\epsilon )e^{t/\rdS} }{2}\right)  -{}_2F_1\left( h_{\pm } ;\frac{d}{2} ;\frac{1-(u+i\epsilon )e^{t/\rdS} }{2} \right) \right]  .
\end{split}
\end{equation}
As usual in this kind of notation we mean $\epsilon  \to 0$ from above. In general, the discontinuity for $z>1$ is
\begin{equation}
\begin{split}
& {}_2F_1(a,b;c;z+i\epsilon )-{}_2F_1(a,b;c;z-i\epsilon )=\\
& \qquad = 2\pi i \frac{\Gamma (c)}{\Gamma (a)\Gamma (b)\Gamma (c-a-b+1)} (z-1)^{c-a-b} {}_2F_1(c-a,c-b;c-a-b+1;1-z).
\end{split}
\end{equation}
Using this in the integral and changing variables to $\frac{1+ue^{t/\rdS} }{2} =-\chi $ we get
\begin{equation}
\psi _4=\frac{4\pi ie^{-t/\rdS} }{\rdS^{d-2} (4\pi )^{d/2} \Gamma \left( 2-\frac{d}{2} \right) } 
\int _0^{\infty } d\chi  \, e^{-2i\rdS^2 p^v e^{-t/\rdS} (2\chi +1)} \chi ^{1-\frac{d}{2} } {}_2F_1\left( \frac{d}{2} -h_{\pm } ;2-\frac{d}{2} ;-\chi \right) .
\end{equation}
This integral can be done explicitly\footnote{For example, Mathematica gives the answer to it.} leading to
\begin{equation}
\psi _4 = \frac{4 \sqrt{\pi }ie^{-t/\rdS} }{\rdS^{d-2} (4\pi )^{d/2} } (4i\rdS^2 p^v e^{-t/\rdS} )^{\frac{d-3}{2} } K_{i\mu } (2i\rdS^2 p^v e^{-t/\rdS} ) .
\end{equation}

The other wavefunctions are analogous, and we summarize the results here
\begin{equation} \label{eq:the_four_wavefunctions_general_d}
\begin{split}
& \psi _1 = \Theta (p^u) \frac{-4 \sqrt{\pi }ie^{t^*_1/\rdS} }{\rdS^{d-2} (4\pi )^{d/2} } (-4i\rdS^2 p^u e^{t^*_1/\rdS} )^{\frac{d-3}{2} } K_{i\mu_V } (- 2i\rdS^2 p^u e^{t^*_1/\rdS} )  \\
& \psi _2 = \Theta (p^v) \frac{4 \sqrt{\pi }ie^{-t^*_2/\rdS} }{\rdS^{d-2} (4\pi )^{d/2} } (4i\rdS^2 p^v e^{-t^*_2/\rdS} )^{\frac{d-3}{2} } K_{i\mu_W } (2i\rdS^2 p^v e^{-t^*_2/\rdS} )  \\
& \psi _3 = \Theta (p^u) \frac{-4 \sqrt{\pi }ie^{t_3/\rdS} }{\rdS^{d-2} (4\pi )^{d/2} } (-4i\rdS^2 p^u e^{t_3/\rdS} )^{\frac{d-3}{2} } K_{i\mu_V } (- 2i\rdS^2 p^u e^{t_3/\rdS} )  \\
& \psi _4 = \Theta (p^v) \frac{4 \sqrt{\pi }ie^{-t_4/\rdS} }{\rdS^{d-2} (4\pi )^{d/2} } (4i\rdS^2 p^v e^{-t_4/\rdS} )^{\frac{d-3}{2} } K_{i\mu_W } (2i\rdS^2 p^v e^{-t_4/\rdS} ) 
\end{split}
\end{equation}

Similar computation can be done when the operator insertions are away from the pole. Let 
\begin{equation}
a_i\equiv 1-r_i^2/\rdS^2 .
\end{equation}
For an insertion $(u_4,v_4,\vec{y}_4)$ and a point on the $v=0$ horizon, the chord distance $Z$ is given by
\beq
Z = - 2 \frac{u v_4}{1-u_4 v_4} + \frac{1+u_4 v_4}{1-u_4 v_4} (\vec{y}_4,\vec{y}).
\eeq
Hence the relevant Fourier transform is:
\beq
{}_2F_1(h_\pm,\frac{d}{2},\frac{1}{2} - \frac{1}{2} u c_1 + \frac{c_2}{2}),
\eeq
with 
\begin{align}
   & c_1 = \frac{2 v_4}{1-u_4 v_4} = e^{t_4/\ell} \sqrt{1-r_4^2/\rdS^2}, \nonumber \\
   & c_2 = \frac{1 + u_4 v_4}{1- u_4 v_4} (\vec{y}_4,\vec{y})= \frac{r_4}{\rdS} (\vec{y}_4,\vec{y}).
\end{align}
We can shift the integral over $u$ by $c_2/c_1$ and then rescale $e^{t_4/\rdS} \rightarrow e^{t_4/\rdS} \sqrt{a_4}$ to obtain the answer from the previous formula. Thus,
\begin{align}
\psi _4 = \Theta(p^v)\exp \l( 2i p^v  \rdS \frac{r_4}{\sqrt{a_4}} e^{-t_4/\rdS} (\vec{y}_4,\vec{y}) \r) \times \nonumber \\
\times \frac{4 \sqrt{\pi }ie^{-t_4/\rdS}  }{\rdS^{d-2} (4\pi )^{d/2} \sqrt{a_4} } \l(\frac{4i\rdS^2 p^v e^{-t_4/\rdS}}{\sqrt{a_4}}  \r)^{\frac{d-3}{2} } K_{i\mu_W } \l(2i\rdS^2 p^v \frac{e^{-t_4/\rdS}}{\sqrt{a_4}} \r) .
\end{align}
For $\psi_2$ the same argument applies with $t_4,r_4,\vec y_4$ replaced by $t_2^*,r_2,\vec y_2$.

For the $V$-side wavefunctions we instead work on the $u=0$ horizon and Fourier transform in $v$. The relevant chord distance has the analogous form
\begin{equation}
Z=2\frac{v u_i}{1-u_i v_i}+\frac{1+u_i v_i}{1-u_i v_i}(\vec y_i,\vec y)
\end{equation}
with
\begin{equation}
\frac{2u_i}{1-u_i v_i}=e^{-t_i/\rdS}\sqrt{a_i},
\qquad
\frac{1+u_i v_i}{1-u_i v_i}(\vec y_i,\vec y)=r_i/\rdS \, (\vec y_i,\vec y).
\end{equation}
The shift is now by $-r_i e^{t_i/\rdS}(\vec y_i,\vec y)/(\rdS \sqrt{a_i})$, so the angular phase has the opposite sign. Summarizing, the wavefunctions for general locations are
\begin{equation} \label{eq:waveprofile-general}
\begin{split}
& \psi _1 = \Theta (p^u) \exp\l(-2i\rdS p^u \frac{r_1}{\sqrt{a_1}} e^{t^*_1/\rdS}(\vec y_1,\vec y)\r)
\frac{-4 \sqrt{\pi }ie^{t^*_1/\rdS} }{\rdS^{d-2} (4\pi )^{d/2}\sqrt{a_1} }
\l(\frac{-4i\rdS^2 p^u e^{t^*_1/\rdS}}{\sqrt{a_1}}\r)^{\frac{d-3}{2} } K_{i\mu_V } \l(\frac{- 2i\rdS^2 p^u e^{t^*_1/\rdS}}{\sqrt{a_1}}\r)  \\
& \psi _2 = \Theta (p^v) \exp\l(2i\rdS p^v \frac{r_2}{\sqrt{a_2}} e^{-t^*_2/\rdS}(\vec y_2,\vec y)\r)
\frac{4 \sqrt{\pi }ie^{-t^*_2/\rdS} }{\rdS^{d-2} (4\pi )^{d/2}\sqrt{a_2} }
\l(\frac{4i\rdS^2 p^v e^{-t^*_2/\rdS}}{\sqrt{a_2}}\r)^{\frac{d-3}{2} } K_{i\mu_W } \l(\frac{2i\rdS^2 p^v e^{-t^*_2/\rdS}}{\sqrt{a_2}}\r)  \\
& \psi _3 = \Theta (p^u) \exp\l(-2i\rdS p^u \frac{r_3}{\sqrt{a_3}} e^{t_3/\rdS}(\vec y_3,\vec y)\r)
\frac{-4 \sqrt{\pi }ie^{t_3/\rdS} }{\rdS^{d-2} (4\pi )^{d/2}\sqrt{a_3} }
\l(\frac{-4i\rdS^2 p^u e^{t_3/\rdS}}{\sqrt{a_3}}\r)^{\frac{d-3}{2} } K_{i\mu_V } \l(\frac{- 2i\rdS^2 p^u e^{t_3/\rdS}}{\sqrt{a_3}}\r)  \\
& \psi _4 = \Theta (p^v) \exp\l(2i\rdS p^v \frac{r_4}{\sqrt{a_4}} e^{-t_4/\rdS}(\vec y_4,\vec y)\r)
\frac{4 \sqrt{\pi }ie^{-t_4/\rdS} }{\rdS^{d-2} (4\pi )^{d/2}\sqrt{a_4} }
\l(\frac{4i\rdS^2 p^v e^{-t_4/\rdS}}{\sqrt{a_4}}\r)^{\frac{d-3}{2} } K_{i\mu_W } \l(\frac{2i\rdS^2 p^v e^{-t_4/\rdS}}{\sqrt{a_4}}\r) .
\end{split}
\end{equation}
The angular dependence comes from the first exponential factor in each wavefunction. It also contains the standard $p^v e^{-t_i/\rdS}$ combination on the $W$ side and $p^u e^{t_i/\rdS}$ on the $V$ side. Hence higher momentum modes are not suppressed in time.

\subsection{Calculating the OTOC: mass regularization} \label{sec:calculating_OTOC}

The two descriptions \eqref{eq:Stanford_Shenker_OTOC_formula} and \eqref{eq:4pf_from_diagrams} are the same. In the first we integrate over $V$ for particle 1 while in the second we integrate instead over $p^u$. Let us use the first description.

Putting the wavefunctions \eqref{eq:the_four_wavefunctions_general_d} together with the eikonal phase \eqref{eq:Eikonal_phase_result} into \eqref{eq:Stanford_Shenker_OTOC_formula} we get\footnote{Using that $K_a(x)^*=K_{a^*} (x^*)$ and that for $m^2 \in \mathbb{R} $ the parameter $i\mu $ is either real or purely imaginary while $K_{a} (x)=K_{-a} (x)$ we get that $K_{i\mu } (x)^*=K_{i\mu } (x^*)$.}
\begin{equation}
\begin{split}
& D = \frac{\rdS^8}{4^{2d-6} \pi ^{2d} \rdS^{2(d-2)} } \int _0 ^{\infty } p^u dp^u p^v dp^v d\phi \sqrt{g_{\perp } } d\bar \phi  \sqrt{\bar g_{\perp} } \exp\left[ 16\pi i \frac{G_N}{\gm^2} \rdS^{4-d} p^up^v \frac{d-1}{\sVol{d-2}} (y(\phi ) \cdot  y(\bar \phi ))\right] \\
& e^{-t_4/\rdS} (4i\rdS^2 p^v e^{-t_4/\rdS} )^{\frac{d-3}{2} } K_{i\mu _W} (2i\rdS^2 p^v e^{-t_4/\rdS} ) \, 
e^{-t_2/\rdS} (-4i\rdS^2 p^v e^{-t_2/\rdS} )^{\frac{d-3}{2} } K_{i\mu _W} (-2i\rdS^2 p^v e^{-t_2/\rdS} ) \\
& e^{t_3/\rdS} (-4i\rdS^2 p^u e^{t_3/\rdS} )^{\frac{d-3}{2} } K_{i\mu _V} (-2i\rdS^2 p^u e^{t_3/\rdS} ) \, 
e^{t_1/\rdS} (4i\rdS^2 p^u e^{t_1/\rdS} )^{\frac{d-3}{2} } K_{i\mu _V} (2i\rdS^2 p^u e^{t_1/\rdS} ).
\end{split}
\end{equation}
For large arguments $K_n(x) \approx e^{-x}$ so the integral is convergent for large $p$ as long as $t_3, t_4$ have a small positive imaginary part and $t_1,t_2$ have a small negative imaginary part which is the standard $i \epsilon$ prescription. For small momenta the convergence criteria are $d-1 + \Re(2 i \mu_V)>0$, $d-1 + \Re(2 i \mu_W)>0$.
We leave the notation that $\phi $ and $\bar \phi $ are $d-2$ angular variables each implicit.
Let us now change variables
\begin{equation}
p^u \to \frac{p^u e^{-t_1/\rdS} }{2i\rdS^2} ,\qquad p^v \to \frac{p^ve^{t_4/\rdS} }{2i\rdS^2} 
\end{equation}
to get
\begin{equation} \label{eq:OTOC_integral_expr}
\begin{split}
& D = \frac{1}{4^{d-1} \pi ^{2d} \rdS^{2(d-2)} } (-e^{t_{31} /\rdS} )^{\frac{d-1}{2} } (-e^{t_{42} /\rdS} )^{\frac{d-1}{2} } \int _0 ^{\infty } dp^u dp^v \int d\phi  \sqrt{g_{\perp} }d\bar \phi \sqrt{\bar g_{\perp} } \\
& \exp\left[ -4\pi i \frac{G_N}{\gm^2} \rdS^{-d} p^up^v e^{t_{41} /\rdS} \frac{d-1}{\sVol{d-2}} (y(\phi ) \cdot y(\bar \phi ))\right]  \\
& (p^up^v)^{d-2} K_{i\mu _W} (p^v)K_{i\mu _W} (-p^v e^{t_{42} /\rdS} )
K_{i\mu _V} (p^u)K_{i\mu _V} (-p^u e^{t_{31} /\rdS} ).
\end{split}
\end{equation}
Here we used the notation $t_{ij} =t_i-t_j$.

We see that the basic integral here is of the form
\begin{equation}
I_{\delta,\mu  } (t) \equiv \int _0^{\infty } dp \, p^{\delta -2} K_{i\mu } (p)K_{i\mu } (-e^{t/\rdS} p)
\end{equation}
in which case it equals
\begin{equation}
\begin{split}
I_{\delta,\mu  } (t) &= 2^{\delta -4} \Gamma \left( \frac{\delta -1}{2} \right) \\
& \cdot  \Big[
(-e^{t/\rdS} )^{i\mu } \Gamma \left( \frac{\delta -1}{2} +i\mu \right) \Gamma (-i\mu ) {}\, _2F_1\left( \frac{\delta -1}{2} ,\frac{\delta -1}{2} +i\mu ;1+i\mu ;(-e^{t/\rdS} )^2 \right) 
+(\mu  \to -\mu ) \Big].
\end{split}
\end{equation}

Now suppose we want to evaluate \eqref{eq:OTOC_integral_expr} in an asymptotic expansion in the coupling constant by expanding the exponent. At zeroth order ---  no interaction --- we just expect the product of 2-point functions. So we replace the exponent by 1 and find that the expression decouples into two terms of the form
\begin{equation}
\label{eq:p_int_relation_to_2pf}
G_{d,\mu} = \frac{1}{2^{d-1} \pi ^d \rdS^{d-2} } (-e^{t/\rdS} )^{\frac{d-1}{2} } \sVol{d-2} I_{d,\mu } (t)
\end{equation}
one with $\mu \to \mu _W$ and $t \to t_{42} $ and the other $\mu  \to \mu _V$ and $t \to t_{31} $. 

Similarly we can expand to any order since the integrals are the same. For the integrals over the transverse space we use equation \eqref{eq:transverse_integral_in_expansion} which is derived in App.\ \ref{sec:transverse_integral}. Also we write the normalized 4-point function, by dividing by the product of the 2-point functions
\begin{equation}
\begin{split}
& G_{13} =G_{d,\mu _V} (t_1,t_3) ,\\
& G_{24} =G_{d,\mu _W} (t_2,t_4) .\\
\end{split}
\end{equation}
Note that because of \eqref{eq:transverse_integral_in_expansion}, only even powers contribute, which we label by $2k$ with $k=0,1,2,\cdots $ . The expansion is then 
\begin{equation}
\begin{split}
& D / (G_{13} G_{24} ) = \sum _{k=0} ^{\infty } \frac{(-1)^k}{(2k)!} \, \frac{\Gamma \left( k+\frac{1}{2} \right) \Gamma \left( \frac{d-1}{2} \right) }{\sqrt{\pi }\,\Gamma \left( k+\frac{d-1}{2} \right) } \left( \frac{d-1}{\sVol{d-2}} \right) ^{2k} \\
& \qquad \qquad \qquad \qquad \left( 16\pi ^2 \frac{G_N^2}{\gm^4} \rdS^{-2d} \right) ^k e^{2k t_{41} /\rdS} \frac{I_{d+2k,\mu _W} (t_{42} )I_{d+2k,\mu _V} (t_{31}) }{I_{d,\mu _W} (t_{42} )I_{d,\mu _V} (t_{31} )} .
\end{split}
\end{equation}
We can also use \eqref{eq:p_int_relation_to_2pf} to write it as
\begin{equation} \label{eq:normalized_OTOC_in_terms_of_2pf}
\begin{split}
& D / (G_{13} G_{24} ) = \sum _{k=0} ^{\infty }\frac{(-1)^k}{(2k)!} \, \frac{\Gamma \left( k+\frac{1}{2} \right) \Gamma \left( k+\frac{d-1}{2} \right) }{\sqrt{\pi} \, \Gamma \left( \frac{d-1}{2} \right) } 
\left( \frac{d-1}{\sVol{d-2}} \right) ^{2k} \\
& \qquad \qquad \left( 256 \pi ^4 \frac{G_N^2}{\gm^4} \rdS^{4-2d} e^{(2t_{41} -t_{42} -t_{31} )/\rdS} \right) ^k \frac{G_{d+2k,\mu _W} (t_{42} )}{G_{d,\mu _W} (t_{42} )} \frac{G_{d+2k,\mu _V} (t_{31} )}{G_{d,\mu _V} (t_{31} )} .
\end{split}
\end{equation}

Parametrizing the times by
\begin{equation}
\begin{split}
& t_1 = i \epsilon _1 \rdS \\
& t_2 = t+i\epsilon _2 \rdS \\
& t_3 = i\epsilon _3 \rdS \\
& t_4 = t+i\epsilon _4 \rdS
\end{split}
\end{equation}
the OTOC is interesting when $t$ is very large.
We can parametrize the leading behavior in $G_N$ as 
\begin{equation}
D/(G_{13} G_{24} ) = 1 - f \cdot \left( \frac{G_N}{\gm^2} \rdS^{-d} e^{t/\rdS} \right) ^{2} +\cdots 
\end{equation}
where $f$ is a coefficient depending on the precise configuration.

We are interested in several different configurations
\begin{enumerate}
\item One-sided configuration. Here all the operators sit at the same point on the sphere. We need to regularize by small Euclidean shifts $\epsilon _1=\epsilon _2=-\epsilon _3=-\epsilon _4=-\epsilon $. Usually the leading correction to the factorized result has a purely imaginary coefficient, but this is not the case here since the leading coefficient canceled and we got instead the next term.
\item Two-sided configuration. We shift the first operator $V_1$ to the left hand side (this is the configuration that for heavy fields probes the geodesic distance between the two sides in the presence of a single shockwave). That is, we take $\epsilon _1=-\pi $ and $\epsilon _3=0$ while still regularizing $\epsilon _2,\epsilon _4$ as in the one-sided configuration. The coefficient $f$ is real and negative. 

\item The regularized OTOC used in the bound on chaos \cite{Maldacena:2015waa} is $\epsilon _1=-\pi $, $\epsilon _2=-\frac{\pi }{2} $, $\epsilon _3=0$, and $\epsilon _4=\frac{\pi }{2} $. It gives again a negative coefficient $f$ in tension with the bound on chaos as discussed in \cite{Narovlansky:2025tpb}.
\end{enumerate}

In cases (2) and (3) we got qualitatively the same behavior as leading $G_N^1$ order $s-$wave exchange discussed in the Introduction. However, it was not obvious from the start because the computation in this section is the regularized $G_N^2/m_g^4$ contribution.

\begin{figure}[ht]
    \centering
    \begin{subfigure}[b]{0.48\textwidth}
        \centering
	\includegraphics[width=1\textwidth]{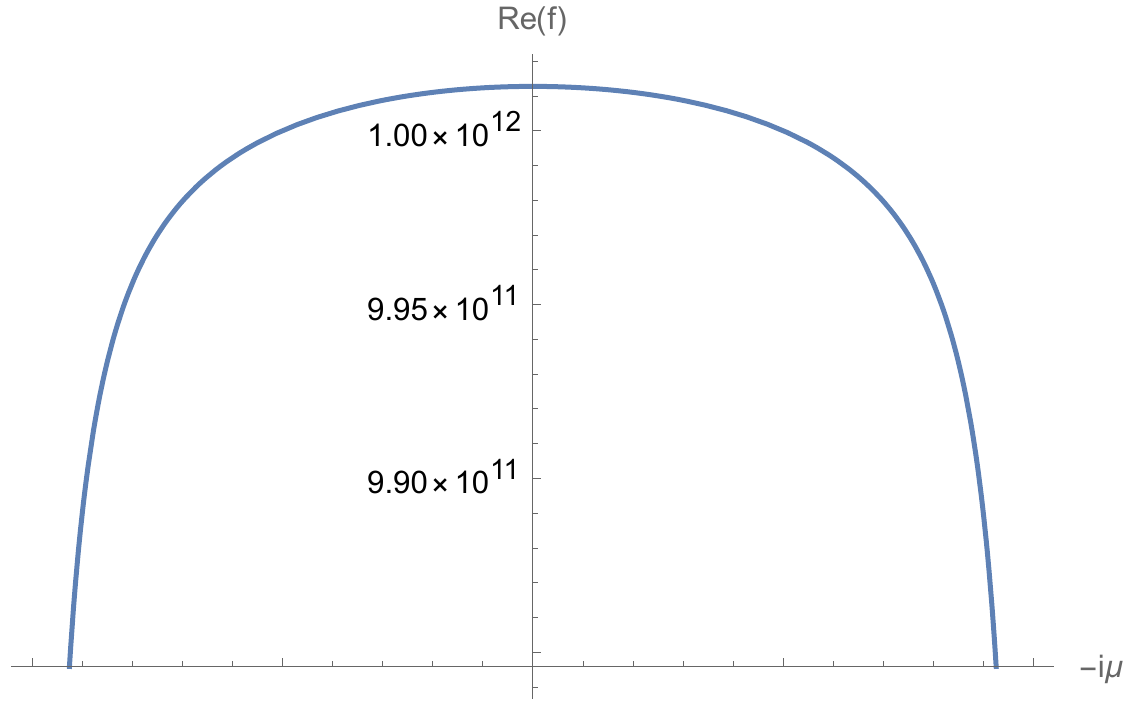}
        \caption{}
        \label{fig:f_3d_compl_one_sided}
    \end{subfigure}
    \hfill
    \begin{subfigure}[b]{0.48\textwidth}
        \centering
	\includegraphics[width=\textwidth]{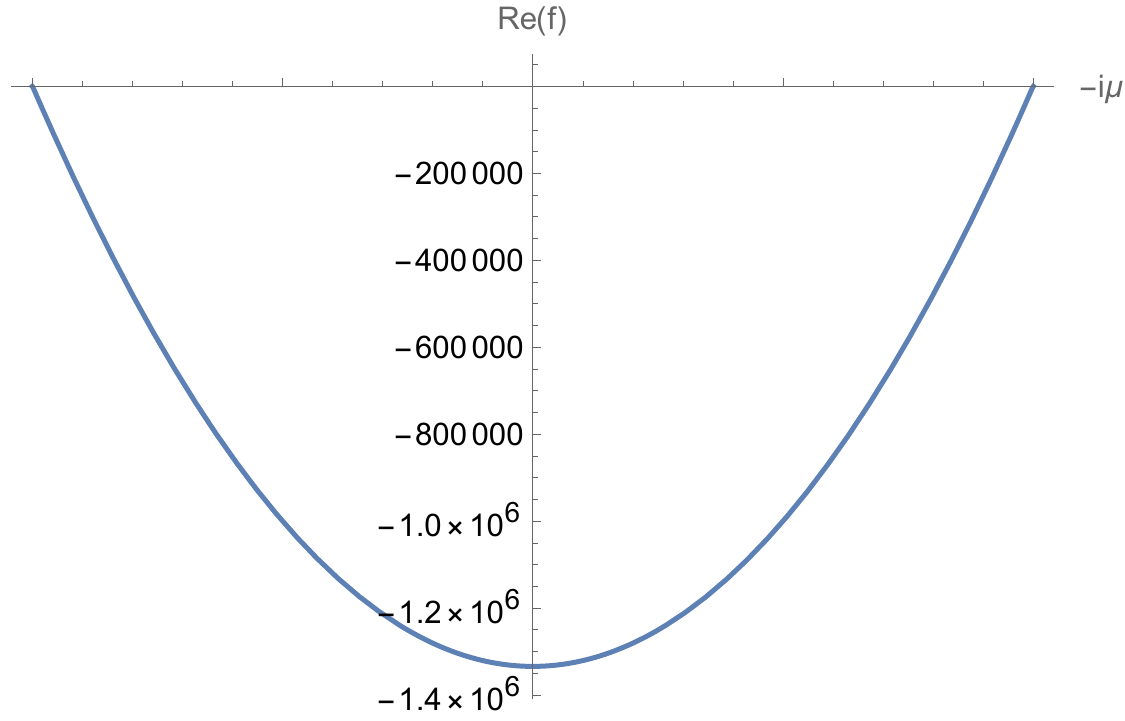}
        \caption{}
        \label{fig:f_3d_compl_two_sided}
    \end{subfigure}
        \\ \bigskip
    \begin{subfigure}[b]{0.48\textwidth}
        \centering
	\includegraphics[width=\textwidth]{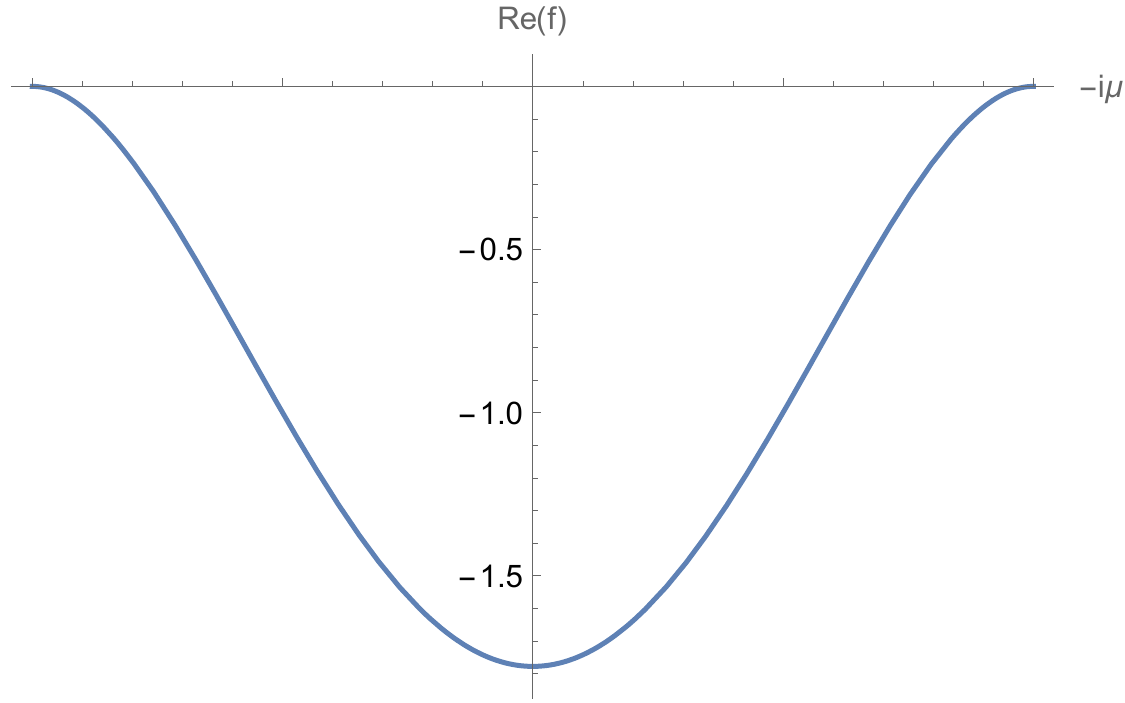}
        \caption{}
        \label{fig:f_3d_compl_boc}
    \end{subfigure}
    \caption{Plots of the coefficient $f$ as a function of $-i\mu \in (-1,1)$, parametrizing the mass of light scalars in three dimensional de Sitter in the complementary series representation. We regularize using $\epsilon =0.001$. In each case we show the real part which is the non-zero component (the other component may be non-zero because of the regulator and is small). (a) single-sided configuration, (b) two-sided configuration, (c) bound on chaos configuration.}
    \label{fig:f_3d_compl}
\end{figure}

\begin{figure}[ht]
    \centering
    \begin{subfigure}[b]{0.48\textwidth}
        \centering
	\includegraphics[width=1\textwidth]{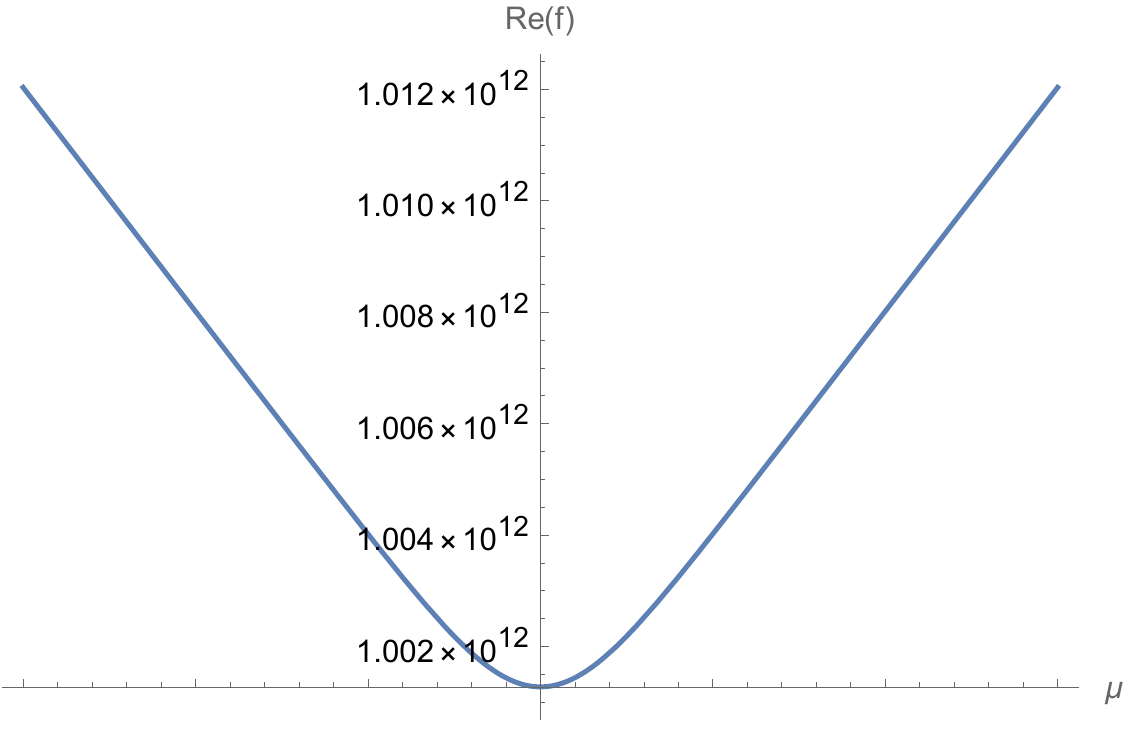}
        \caption{}
        \label{fig:f_3d_principal_one_sided}
    \end{subfigure}
    \hfill
    \begin{subfigure}[b]{0.48\textwidth}
        \centering
	\includegraphics[width=\textwidth]{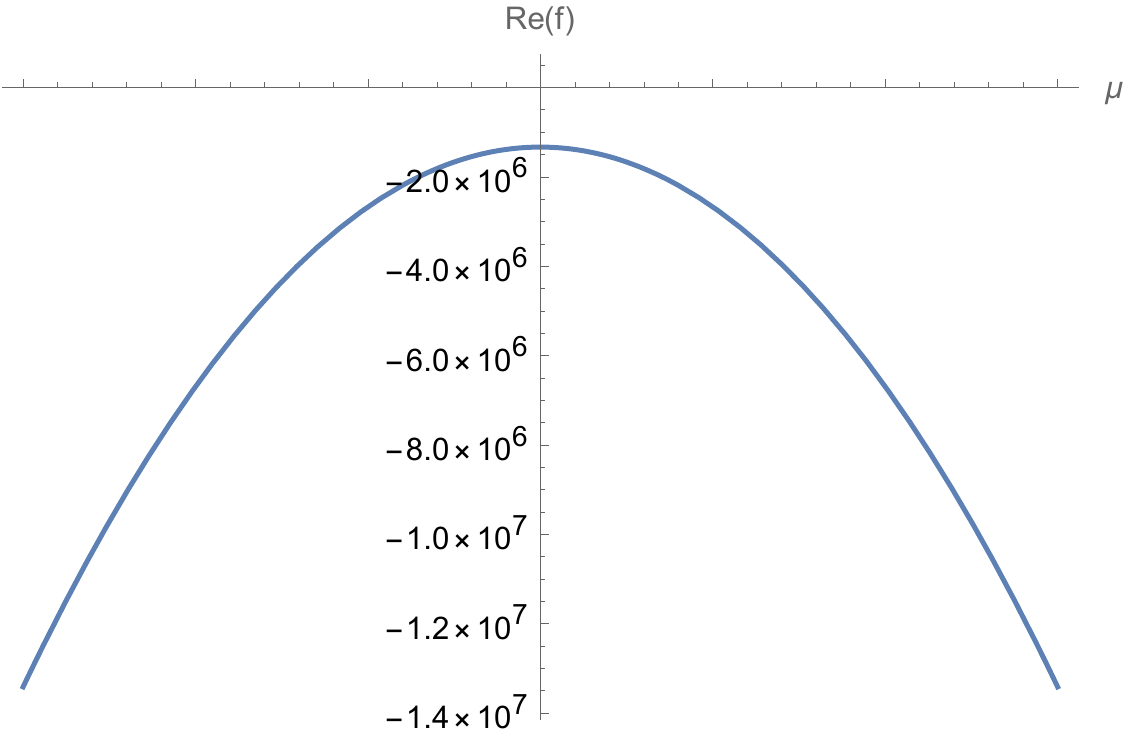}
        \caption{}
        \label{fig:f_3d_principal_two_sided}
    \end{subfigure}
        \\ \bigskip
    \begin{subfigure}[b]{0.48\textwidth}
        \centering
	\includegraphics[width=\textwidth]{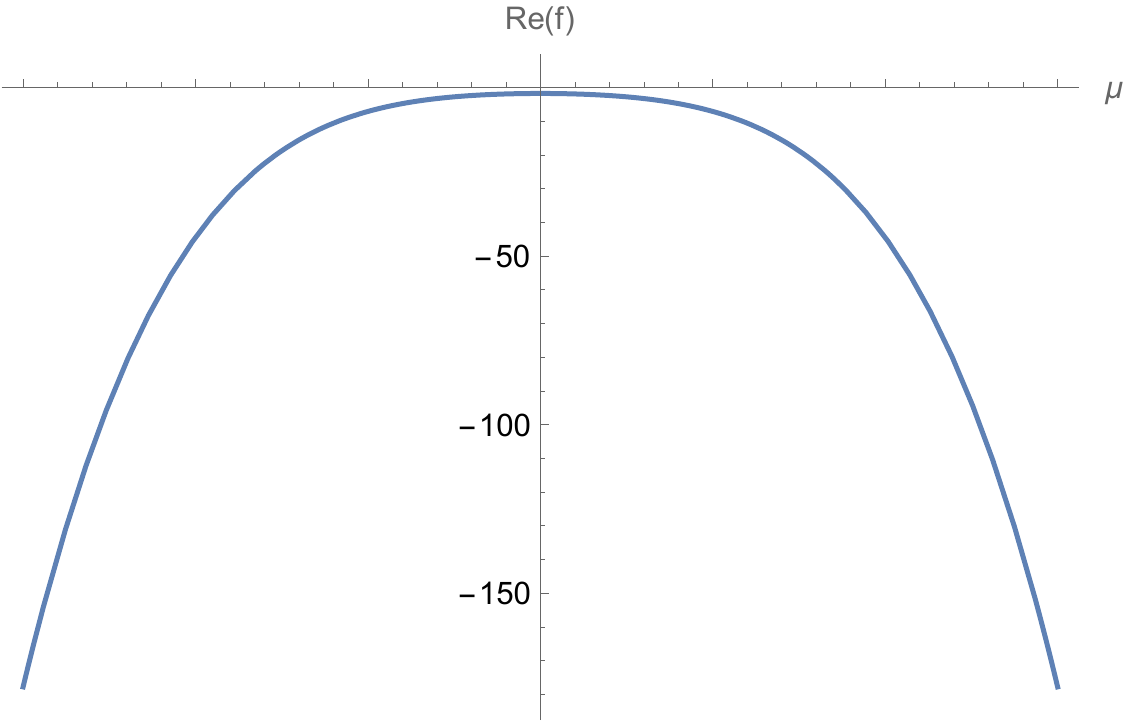}
        \caption{}
        \label{fig:f_3d_principal_boc}
    \end{subfigure}
    \caption{Plots of the coefficient $f$ as a function of $\mu \in (-3,3)$, parametrizing the mass of heavy scalars in three dimensional de Sitter in the principal series representation. We regularize using $\epsilon =0.001$. In each case we show the real part which is the non-zero component (the other component may be non-zero because of the regulator and is small). (a) single-sided configuration, (b) two-sided configuration, (c) bound on chaos configuration.}
    \label{fig:f_3d_principal}
\end{figure}

We plot the coefficient $f$ in the different configurations for $m=m_W=m_V$ in $d=3$, in Fig.\ \ref{fig:f_3d_compl} for $m^2 \rdS^2 < 1$ (that is, complementary series) and in Fig.\ \ref{fig:f_3d_principal} for $m^2 \rdS^2 >1$ (that is, principal series). In all cases, as mentioned, the coefficient $f$ is real and negative in the two-sided configuration, which is the time advance effect. Likewise it is negative in the bound on chaos configuration, which is not allowed by the bound on chaos.

\subsection{Displacing the particles away from the pole}

We now generalize the pole computation by allowing the four insertions to sit at
arbitrary static-patch positions
\[
(t_i,r_i,\vec y_i), \qquad a_i\equiv 1-r_i^2/\rdS^2,
\]
where $\vec y_i\in \mathbb{R}^{d-1}, |\vec{y}_i|=1$.  Relative to the pole configuration, the only
new ingredient is the form of the external wave packets.  The eikonal
scattering kernel itself is unchanged: the two particles still cross the two
horizons with null momenta \(p^u\) and \(p^v\), and the singular part of the
graviton exchange couples their transverse directions through the \(L=1\)
harmonic \(y\cdot \bar y\).

It is useful to absorb the trivial redshift factors into the null momenta.  We
therefore make the change of variables
\begin{equation}
p^{u}\mapsto\frac{p^{u}e^{-t_{1}/\ell}\sqrt{a_{1}}}{2i\ell^{2}},
\qquad
p^{v}\mapsto\frac{p^{v}e^{t_{4}/\ell}\sqrt{a_{4}}}{2i\ell^{2}},
\end{equation}
and then relabel the rescaled variables again as \(p^u,p^v\).  The crossed
four-point function becomes
\begin{equation}
\label{eq:OTOC_integral_general_locations}
\begin{aligned}
D_{\{r_i\}}
={}&
\frac{
\left(-e^{t_{31}/\ell}\sigma_{13}\right)^{\frac{d-1}{2}}
\left(-e^{t_{42}/\ell}\sigma_{42}\right)^{\frac{d-1}{2}}
}
{4^{d-1}\pi^{2d}\ell^{2(d-2)}}
\int_0^\infty \dd p^u \dd p^v\,(p^u p^v)^{d-2}                         \\
&\times
\int_{S^{d-2}}\dd y\sqrt{g_\perp}\,
\dd\bar y\sqrt{\bar g_\perp}\,
\exp\left[
\mathfrak{a}\, p^u p^v(y\cdot\bar y)
+p^u(\vec q_{13}\cdot y)
+p^v(\vec q_{24}\cdot\bar y)
\right]                                                                  \\
&\times
K_{i\mu_V}(p^u)
K_{i\mu_V}\!\left(-e^{t_{31}/\ell}\sigma_{13}p^u\right)
K_{i\mu_W}(p^v)
K_{i\mu_W}\!\left(-e^{t_{42}/\ell}\sigma_{42}p^v\right),
\end{aligned}
\end{equation}
where
\begin{equation}
\label{eq:eikonal_coefficient_general_locations}
\mathfrak{a}
=-\frac{4\pi iG_N}{m_g^2}\ell^{-d}e^{t_{41}/\ell}
\sqrt{a_1a_4}\,
\frac{d-1}{{\rm Vol}(S^{d-2})},
\qquad
\sigma_{ij}\equiv \sqrt{\frac{a_i}{a_j}},
\end{equation}
and 
\begin{equation}
\label{eq:q13_q24_def}
\vec q_{13}\equiv
(r_1\vec y_1-e^{t_{31}/\ell}\sigma_{13}r_3\vec y_3)/\rdS, \quad \vec q_{24}\equiv
(r_4\vec y_4-e^{t_{42}/\ell}\sigma_{42}r_2\vec y_2)/\rdS .
\end{equation}
The vectors \(\vec q_{13}\) and \(\vec q_{24}\) are the effective transverse
separations of the two wave packets in each pair after the relative
static-patch boosts have been included.  They vanish when the corresponding
insertions are placed at the pole.  Away from the pole, the external wave
profiles acquire the linear phases
\[
p^u\,\vec q_{13}\cdot y,
\qquad
p^v\,\vec q_{24}\cdot \bar y .
\]
These phases have a simple geometric interpretation: a particle emitted from a
displaced static-patch position reaches the horizon with a transverse-direction
dependent null shift.

The form \eqref{eq:OTOC_integral_general_locations} makes it possible to
organize the eikonal interaction as a differential operator.  Indeed,
\begin{equation}
\label{eq:general_location_angular_kernel_derivative}
\left(\partial_{\vec q_{13}}\cdot\partial_{\vec q_{24}}\right)
e^{p^u(\vec q_{13}\cdot y)+p^v(\vec q_{24}\cdot\bar y)}
=
p^u p^v(y\cdot\bar y)
e^{p^u(\vec q_{13}\cdot y)+p^v(\vec q_{24}\cdot\bar y)} .
\end{equation}
Thus the shockwave phase can be generated by acting with
\[
\exp\!\left[
\mathfrak{a}
\left(\partial_{\vec q_{13}}\cdot\partial_{\vec q_{24}}\right)
\right]
\]
on the product of the two disconnected angular integrals.  We may therefore
rewrite \eqref{eq:OTOC_integral_general_locations} as
\begin{equation}
\label{eq:OTOC_general_locations_diff_operator}
\begin{aligned}
D_{\{r_i\}}
={}&
\frac{
\left(-e^{t_{31}/\ell}\sigma_{13}\right)^{\frac{d-1}{2}}
\left(-e^{t_{42}/\ell}\sigma_{42}\right)^{\frac{d-1}{2}}
}
{4^{d-1}\pi^{2d}\ell^{2(d-2)}}
\int_0^\infty \dd p^u\dd p^v\,(p^u p^v)^{d-2}                          \\
&\times
\exp\!\left[
\mathfrak{a}
\left(\partial_{\vec q_{13}}\cdot\partial_{\vec q_{24}}\right)
\right]
\left[
\int_{S^{d-2}}\dd y\sqrt{g_\perp}\,
e^{p^u(\vec q_{13}\cdot y)}
\int_{S^{d-2}}\dd\bar y\sqrt{\bar g_\perp}\,
e^{p^v(\vec q_{24}\cdot\bar y)}
\right]                                                                  \\
&\times
K_{i\mu_V}(p^u)
K_{i\mu_V}\!\left(-e^{t_{31}/\ell}\sigma_{13}p^u\right)
K_{i\mu_W}(p^v)
K_{i\mu_W}\!\left(-e^{t_{42}/\ell}\sigma_{42}p^v\right).
\end{aligned}
\end{equation}

The remaining angular integrals are rotationally invariant.  For \(N=d-2\), we
define
\begin{equation}
\label{eq:sphere_exponential_integral}
\mathcal F_N(z)
\equiv
\int_{S^N}\dd y\sqrt{g_\perp}\,e^{z\,\hat q\cdot y}
=
(2\pi)^{\frac{N+1}{2}}
z^{-\frac{N-1}{2}}
I_{\frac{N-1}{2}}(z),
\end{equation}
where \(\hat q\) is a unit vector and \(I_\nu\) is the modified Bessel function.
Using this notation, the displaced four-point function becomes
\begin{equation}
\label{eq:OTOC_general_locations_F_form}
\begin{aligned}
D_{\{r_i\}}
={}&
\frac{
\left(-e^{t_{31}/\ell}\sigma_{13}\right)^{\frac{d-1}{2}}
\left(-e^{t_{42}/\ell}\sigma_{42}\right)^{\frac{d-1}{2}}
}
{4^{d-1}\pi^{2d}\ell^{2(d-2)}}
\int_0^\infty \dd p^u\dd p^v\,(p^u p^v)^{d-2}                          \\
&\times
\exp\!\left[
\mathfrak{a}
\left(\partial_{\vec q_{13}}\cdot\partial_{\vec q_{24}}\right)
\right]
\left[
\mathcal F_{d-2}\!\left(p^u|\vec q_{13}|\right)
\mathcal F_{d-2}\!\left(p^v|\vec q_{24}|\right)
\right]                                                                  \\
&\times
K_{i\mu_V}(p^u)
K_{i\mu_V}\!\left(-e^{t_{31}/\ell}\sigma_{13}p^u\right)
K_{i\mu_W}(p^v)
K_{i\mu_W}\!\left(-e^{t_{42}/\ell}\sigma_{42}p^v\right).
\end{aligned}
\end{equation}
This representation makes the factorization structure transparent: before expanding in the eikonal interaction, the two pairs $(1,3)$ and $(2,4)$ propagate independently as ordinary de Sitter two-point functions, while expanding the exponential in powers of $\mathfrak a$ inserts successive $L=1$ graviton exchanges between the two pairs. 
Each momentum integral is precisely the horizon-momentum representation of a
de Sitter two-point function:
\begin{equation}
\label{eq:general_location_2pf_momentum_representation}
G_{d,\mu}\!\left(Z(\rho,|\vec q|)\right)
=
\frac{\rho^{\frac{d-1}{2}}}{2^{d-1}\pi^d\ell^{d-2}}
\int_0^\infty\dd p\,p^{d-2}
K_{i\mu}(p)K_{i\mu}(\rho p)\mathcal F_{d-2}(p|\vec q|),
\end{equation}
with
\begin{equation}
\label{eq:Z_rho_q_def}
Z(\rho,|\vec q|)
=
\frac{1+\rho^2-|\vec q|^2}{-2\rho}.
\end{equation}
We give a detailed derivation of
\eqref{eq:general_location_2pf_momentum_representation} in
appendix \ref{app:general_location_G}.

For the \((1,3)\) pair, the relevant parameters are
\begin{equation}
\rho_{13}=-e^{t_{31}/\ell}\sigma_{13},
\qquad
\ell^2 |\vec q_{13}|^2
=
r_1^2+\rho_{13}^2 r_3^2
+2\rho_{13}r_1r_3(\vec y_1\cdot\vec y_3).
\end{equation}
Substituting these expressions into \eqref{eq:Z_rho_q_def} gives the standard
de Sitter invariant distance
\begin{equation}
Z_{13}
=
\sqrt{a_1a_3}\cosh\frac{t_{31}}{\ell}
+\frac{r_1r_3}{\ell^2}\,\vec y_1\cdot\vec y_3 .
\end{equation}
Similarly, for the \((2,4)\) pair,
\begin{equation}
\rho_{24}=-e^{t_{42}/\ell}\sigma_{42},
\end{equation}
and
\begin{equation}
Z_{24}
=
\sqrt{a_2a_4}\cosh\frac{t_{42}}{\ell}
+\frac{r_2r_4}{\ell^2}\,\vec y_2\cdot\vec y_4 .
\end{equation}
Therefore the result at general static-patch positions can be written in the
compact form
\begin{equation}
\label{eq:general_location_compact_result}
D_{\{r_i\}}
=
\exp\!\left[
\mathfrak{a}
\left(\partial_{\vec q_{13}}\cdot\partial_{\vec q_{24}}\right)
\right]
G_{d,\mu_V}(Z_{13})G_{d,\mu_W}(Z_{24}) .
\end{equation}
This formula has a transparent interpretation: the displaced answer is obtained
from the product of two ordinary de Sitter two-point functions by inserting the
eikonal graviton exchange as a transverse differential operator.  The operator
\(\partial_{\vec q_{13}}\cdot\partial_{\vec q_{24}}\) inserts one factor of the
\(L=1\) transverse harmonic \(y\cdot\bar y\), and the exponential resums an
arbitrary number of such exchanges.

Let us now examine two useful limits of \eqref{eq:general_location_compact_result}.
First, place the two $V$ insertions at the pole, which corresponds to setting
$\vec q_{13}=0$. From the definition \eqref{eq:Z_rho_q_def}, $Z_{13}$ depends
on $\vec q_{13}$ only through $|\vec q_{13}|^2$ and is therefore an even
function of $\vec q_{13}$. Consequently, all odd powers in the eikonal expansion vanish:
\begin{equation}
\left.
\left(\partial_{\vec q_{13}}\cdot\partial_{\vec q_{24}}\right)^{2k+1}
G_{d,\mu_V}(Z_{13})G_{d,\mu_W}(Z_{24})
\right|_{\vec q_{13}=0}
=0 .
\end{equation}
Physically, a rotationally invariant wave packet localized at the pole cannot
absorb an odd number of $L=1$ gravitons. The expansion therefore contains only
even powers of the eikonal coupling:
\begin{equation}
\label{eq:one_pair_at_pole_even_expansion}
\left.D_{\{r_i\}}\right|_{\vec q_{13}=0}
=
\sum_{k=0}^{\infty}
\frac{\mathfrak{a}^{2k}}{(2k)!}
\left.
\left(\partial_{\vec q_{13}}\cdot\partial_{\vec q_{24}}\right)^{2k}
G_{d,\mu_V}(Z_{13})G_{d,\mu_W}(Z_{24})
\right|_{\vec q_{13}=0}.
\end{equation}
The $W$ insertions may remain at arbitrary positions, so the even derivatives
still act nontrivially on the $Z_{24}$ dependence. Since the leading connected
contribution is quadratic in $\mathfrak a$, we conclude that whenever either
operator pair is localized at the pole, the corresponding Lyapunov exponent is
twice the maximal value $2\pi/\beta_{\rm dS}$.

We next show explicitly that the general expression
\eqref{eq:general_location_compact_result}, after normalization by
$G_{13}G_{24}$, reduces to \eqref{eq:normalized_OTOC_in_terms_of_2pf} when both
the $V$ and $W$ insertions are placed at the pole. In this limit,
\begin{equation}
\vec q_{13}=\vec q_{24}=0,
\qquad
\rho_{13}=-e^{t_{31}/\ell},
\qquad
\rho_{24}=-e^{t_{42}/\ell},
\end{equation}
and
\begin{equation}
Z_{13}=\cosh\frac{t_{31}}{\ell},
\qquad
Z_{24}=\cosh\frac{t_{42}}{\ell}.
\end{equation}
The eikonal coefficient reduces to
\begin{equation}
\mathfrak{a}_{\rm pole}
=
-\frac{4\pi iG_N}{m_g^2}\ell^{-d}e^{t_{41}/\ell}
\frac{d-1}{\operatorname{Vol}(S^{d-2})}.
\end{equation}
In this limit, the derivatives reduce to pairwise contractions of transverse
indices. Since the transverse vectors belong to $\mathbb R^{d-1}$, this gives
\begin{equation}
\label{eq:all_poles_from_displaced_limit}
D_{\rm pole}
=
\sum_{k=0}^{\infty}
\mathfrak{a}_{\rm pole}^{2k}
\frac{\left(\frac{d-1}{2}\right)_k}{k!}
\frac{
\partial_Z^kG_{d,\mu_V}\!\left(\cosh\frac{t_{31}}{\ell}\right)
\partial_Z^kG_{d,\mu_W}\!\left(\cosh\frac{t_{42}}{\ell}\right)}
{\left(\rho_{13}\rho_{24}\right)^k}.
\end{equation}
Finally, using the dimension-shifting identity
\begin{equation}
\partial_Z^kG_{d,\mu}(Z)
=
(2\pi\ell^2)^kG_{d+2k,\mu}(Z),
\end{equation}
we recover the pole result \eqref{eq:normalized_OTOC_in_terms_of_2pf}. The
agreement of the normalization follows from the identity
\begin{equation}
\frac{1}{(2k)!}
\frac{\Gamma(k+\frac12)\Gamma(k+\frac{d-1}{2})}
{\sqrt{\pi}\Gamma(\frac{d-1}{2})}
\left(256\pi^4\right)^k
=
\frac{\left(\frac{d-1}{2}\right)_k}{k!}
\left(64\pi^4\right)^k .
\end{equation}

\subsection{Calculating OTOC: leading order without mass regulator}
As we explained before, in the massless case we can only reliably compute the leading $G_N$-order correction. The simplest physical state where $L=1$ mode in the matter distribution is absent is when operators are inserted at the pole and anti-pole, producing only spherically-symmetric $L=0$ configuration. The resulting expression is
\begin{equation}
\begin{split}
& D = \frac{\rdS^8}{4^{2d-6} \pi ^{2d} \rdS^{2(d-2)} } \int _0 ^{\infty } p^u dp^u p^v dp^v d\phi \sqrt{g_{\perp } } d\bar \phi  \sqrt{\bar g_{\perp} } \left(1 + i \delta_{dS}  \right) \\
& e^{-t_4/\rdS} (4i\rdS^2 p^v e^{-t_4/\rdS} )^{\frac{d-3}{2} } K_{i\mu _W} (2i\rdS^2 p^v e^{-t_4/\rdS} ) \, 
e^{-t_2/\rdS} (-4i\rdS^2 p^v e^{-t_2/\rdS} )^{\frac{d-3}{2} } K_{i\mu _W} (-2i\rdS^2 p^v e^{-t_2/\rdS} ) \\
& e^{t_3/\rdS} (-4i\rdS^2 p^u e^{t_3/\rdS} )^{\frac{d-3}{2} } K_{i\mu _V} (-2i\rdS^2 p^u e^{t_3/\rdS} ) \, 
e^{t_1/\rdS} (4i\rdS^2 p^u e^{t_1/\rdS} )^{\frac{d-3}{2} } K_{i\mu _V} (2i\rdS^2 p^u e^{t_1/\rdS} ),
\end{split}
\end{equation}
where the eikonal phase is
\beq
i \delta_{dS} = 16 \pi i G_N \rdS^{4-d} p^u p^v {\cal G}_0(y,\bar y)
\eeq
where ($N=d-2$)
\begin{equation}
\label{eq:mG0-def}
{\cal G}_0(y,\bar y)
=
\sum_{L\neq 1}\frac{1}{L(L+N-1)-N}
\sum_M Y_{LM}(y)Y_{LM}^*(\bar y).
\end{equation}
After the rescaling
\[
p^u\mapsto \frac{p^u e^{-t_1/\rdS}}{2i\rdS^2},
\qquad
p^v\mapsto \frac{p^v e^{t_4/\rdS}}{2i\rdS^2},
\]
the leading correction is
\begin{equation}
\begin{aligned}
D_{\rm massless}^{(1)}
={}&
-4\pi iG_N \,\rdS^{-d}e^{t_{41}/\rdS}\,
{\cal P}
\int_0^\infty dp^u dp^v\,(p^up^v)^{d-1}
K_{i\mu_V}(p^u)K_{i\mu_V}(-e^{t_{31}/\rdS}p^u)
K_{i\mu_W}(p^v)K_{i\mu_W}(-e^{t_{42}/\rdS}p^v)                           \\
&\times
\int_{S^N}d\Omega\,d\bar\Omega\,{\cal G}_0(y,\bar y),
\end{aligned}
\end{equation}
where we collectively denote the prefactor as
\begin{equation}
{\cal P}
=
\frac{(-e^{t_{31}/\rdS})^{\frac{d-1}{2}}
(-e^{t_{42}/\rdS})^{\frac{d-1}{2}}}
{4^{d-1}\pi^{2d}\rdS^{2(d-2)}},
\end{equation}
Equivalently, after the rescaling the two pair wave-profile factors appearing
in the momentum integrals are
\begin{equation}
\begin{aligned}
& (p^u)^{d-1}
K_{i\mu_V}(p^u)K_{i\mu_V}(-e^{t_{31}/\rdS}p^u),\\
& (p^v)^{d-1}
K_{i\mu_W}(p^v)K_{i\mu_W}(-e^{t_{42}/\rdS}p^v).
\end{aligned}
\end{equation}
The pode--antipode configuration makes their
angular profiles constant, so the only angular dependence in the leading
correction comes from \({\cal G}_0(y,\bar y)\).

The angular integral is therefore
\begin{equation}
\begin{aligned}
\int_{S^N}d\Omega\,d\bar\Omega\,{\cal G}_0(y,\bar y)
={}&
\sum_{L\neq 1,M}
\frac{1}{L(L+N-1)-N}
\left(\int_{S^N}d\Omega\,Y_{LM}(y)\right)
\left(\int_{S^N}d\bar\Omega\,Y_{LM}^*(\bar y)\right).
\end{aligned}
\end{equation}
Since \(Y_{00}=1/\sqrt{\sVol{N}}\) and all harmonics with \(L>0\) integrate to
zero,
\begin{equation}
\int_{S^N}d\Omega\,Y_{LM}(y)
=
\sqrt{\sVol{N}}\,\delta_{L0}\delta_{M0}.
\end{equation}
Thus only the \(L=0\) harmonic contributes:
\begin{equation}
\label{eq:massless_angular_integral_result}
\int_{S^N}d\Omega\,d\bar\Omega\,{\cal G}_0(y,\bar y)
=
-\frac{\sVol{N}}{N}
=
-\frac{\sVol{d-2}}{d-2}.
\end{equation}
The exclusion of \(L=1\) is therefore automatic in this configuration.

The remaining momentum integrals are exactly of the form \(I_{\delta,\mu}\)
defined above, with \(\delta=d+1\):
\begin{equation}
\int_0^\infty dp\,p^{d-1}
K_{i\mu}(p)K_{i\mu}(-e^{t/\rdS}p)
=
I_{d+1,\mu}(t).
\end{equation}
Combining this with \eqref{eq:massless_angular_integral_result}, we obtain
\begin{equation}
\label{eq:massless_leading_OTOC_result}
D_{\rm massless}^{(1)}
=
{\cal P}\,
\frac{4\pi iG_N}{d-2}\rdS^{-d}e^{t_{41}/\rdS}\sVol{d-2}\,
I_{d+1,\mu_V}(t_{31})I_{d+1,\mu_W}(t_{42}).
\end{equation}
The zeroth-order term is
\begin{equation}
D^{(0)}
=
{\cal P}\,\sVol{d-2}^2
I_{d,\mu_V}(t_{31})I_{d,\mu_W}(t_{42})
=G_{13}G_{24}.
\end{equation}
Hence the normalized leading massless correction is
\begin{equation}
\frac{D_{\rm massless}}{G_{13}G_{24}}
=
1+
\frac{4\pi iG_N}{(d-2)\sVol{d-2}}\,
\rdS^{-d}e^{t_{41}/\rdS}
\frac{I_{d+1,\mu_V}(t_{31})}{I_{d,\mu_V}(t_{31})}
\frac{I_{d+1,\mu_W}(t_{42})}{I_{d,\mu_W}(t_{42})}
+O(G_N^2).
\end{equation}
The important point is that the sign of this term is fixed by the negative
constant mode of the transverse Green function.  Indeed, the operator whose
inverse appears in \({\cal G}_0\) has eigenvalue \(-N\) on the \(L=0\)
harmonic, and this is why the angular integral in
\eqref{eq:massless_angular_integral_result} is negative.  This negative mode
turns the sign of the linear eikonal correction into the sign opposite to the
usual decaying OTOC correction.

This is particularly transparent in the regularized configuration used for the
bound on chaos, where
\[
\epsilon_1=-\pi,\qquad
\epsilon_2=-\frac{\pi}{2},\qquad
\epsilon_3=0,\qquad
\epsilon_4=\frac{\pi}{2}.
\]
Then \(t_{31}=t_{42}=i\pi\rdS\), while
\(e^{t_{41}/\rdS}=-i e^{t/\rdS}\).  Moreover
\(-e^{t_{31}/\rdS}=-e^{t_{42}/\rdS}=1\), so the integrals
\(I_{\delta,\mu}(i\pi\rdS)\) are the positive Euclidean integrals
\(\int_0^\infty dp\,p^{\delta-2}K_{i\mu}(p)^2\).  Therefore the leading
massless correction takes the schematic form
\begin{equation}
\frac{D_{\rm massless}}{G_{13}G_{24}}
=
1+
C_{\rm massless}\,G_N\,\rdS^{-d}e^{t/\rdS}
+O(G_N^2),
\qquad
C_{\rm massless}>0 ,
\end{equation}
which grows with the maximal exponent \(1/\rdS=2\pi/\beta_{\rm dS}\) instead of
decreasing.  This linearized result should only be interpreted as the early
time behavior.  When \(G_N\rdS^{-d}e^{t/\rdS}\) becomes order one, the quadratic
term is parametrically as important as the linear term, so deciding the
late-time behavior requires including the \(O(G_N^2)\) contribution.

\section{Discussion}

In this paper we have revisited the OTOC computation in de Sitter space in the eikonal approximation. Usually the eikonal phase is easy to find using the 't Hooft--Dray  \cite{Dray:1984ha, Dray:1985yt} shockwave approach. However, in the de Sitter case the resulting equation does not have a unique solution -- a problem we aimed to address.
In order to understand the structure of the eikonal phase better, we examined its most general form (\ref{eq:general_eikonal}) instead of finding the shockwave solution. This revealed a number of unexpected things. The vector part of the propagator, which is a pure diffeomorphism, leads to a logarithmic IR divergence, which we linked to the aforementioned ambiguity in the shockwave equation. At the tree level this divergence is canceled once we integrate the eikonal phase against matter sources which satisfy the gravitational constraints. 
This leads to the standard maximal Lyapunov exponent of $2 \pi/\beta_{\rm dS}$.
However, at the loop level this divergence cannot be removed so easily. Since it is associated with graviton propagator inside the loop diagrams, it is not affected by the gravitational dressing of the external operators. Presumably, one will have to deform away from empty de Sitter in order to resolve this problem. It would be interesting to consider the simplest case of $dS_3$ with a conical defect. In this paper we instead considered massive graviton propagator and studied the regime of late times, small $G_N/m_g^2$ but $G_N e^{2 \pi t/\beta_{\rm dS}}/m_g^2$ fixed. Surprisingly, in this regime we found twice the maximal Lyapunov exponent. Other than that, the qualitative behavior of the OTOC is the same as the $s$-wave  order $G_N^1$ graviton exchange. It would be interesting to understand if this is a coincidence or not.

One of the most surprising things about OTOC in de Sitter, which was previously found in \cite{Aalsma:2020aib} and confirmed by our computation is the initial growth at early times in certain kinematical configurations. In the massless, unregularized case, this growth is due to the $s-$wave graviton exchange. We found that it is mediated by a large diffeomorphism in the propagator,  eq. (\ref{eq:s_wave_diffeo}). This part of the propagator grows at large distances, making the answer sensitive to the choice of the integration contour. In the future, we hope to further explore the physical consequences of this result, especially since this problematic growing contribution leads to the violation of the bound on chaos, as discussed in \cite{Narovlansky:2025tpb}. Previously, the role of large gauge transformations in inflationary spacetimes was discussed in \cite{Tanaka:2017nff}. 
 In contrast, in $AdS$ the eikonal phase, including $L=0$ mode, comes purely from the physical TT part of the propagator.
 Also, in our paper we did not properly discuss the gravitational dressing of the external operators, and it would be very interesting to do that, following the recent discussion in \cite{Kolchmeyer:2024fly}.

As was discussed in ~\cite{Narovlansky:2025tpb}, one microscopic way to understand growing OTOC is the following. Usually the decay of OTOC is associated with operator growth: simple operators acting on a few degrees of freedom under Heisenberg time evolution become delocalized and act on many degrees of freedom. From this perspective we should interpret growing OTOC as operators becoming simpler. This is not possible within known examples of AdS/CFT where simple operators are dual to elementary bulk fields. For example, in the original SYK model, the most basic Majorana fermion is dual to a fermionic bulk field. But perhaps we should not take this statement as something universal: it is not obvious why an elementary bulk field cannot be dual to a complicated boundary operator. This is actually realized in the recently proposed doubled SYK model with the equal energy constraint \cite{Narovlansky:2023lfz,Narovlansky:2025tpb,Goto:2026ipq}: this model is capable of reproducing matter correlation functions in $d-$dimensional de Sitter exactly. The corresponding physical operators include dressing with the equal energy projector  which is very complicated in terms of the original fermionic operators. In this model OTOC can indeed show initial growth \cite{Narovlansky:2025tpb}. It would be extremely interesting to compare our results with the results from that paper. We hope to report our results soon \cite{MNX}. However, one thing to keep in mind is that time-ordered correlation functions in de Sitter are not expected to grow and the candidate for de Sitter holographic dual must respect this. 

\section*{Acknowledgment}
We would like to thank Markus~Fr\"ob, David~Kolchmeyer,  Jonah Kudler-Flam, Juan~Maldacena, Fedor~Popov, Douglas~Stanford, Herman~Verlinde, Zhenbin~Yang, and Ying~Zhao for comments.   J.X. is supported by the U.S. Department of Energy, Office
of Science, Office of High Energy Physics, under Award Number DE-SC0011702.

\appendix

\section{Saddle point derivation of the eikonal phase} \label{sec:saddle_point_linearized}

As mentioned, a nice way to obtain the eikonal phase is by a saddle point calculation of the path integral \cite{Kabat:1992tb}. Let us consider this and relate it to the diagrammatic calculation.

The gravitational + gauge fixing action that we got is
\begin{equation} \label{eq:graviton_kinetic_term}
S_2+S_{\text{gauge}} = \frac{i}{2} \int d^dx \sqrt{-g}\, h^{\mu \nu }D^{-1} _{\mu \nu ,\rho \sigma }  h^{\rho \sigma } 
\end{equation}
and the total action also includes
\begin{equation} \label{eq:two_EM_sources}
\frac{1}{2} \int dx \sqrt{-g}\,h_{\mu \nu } T_1^{\mu \nu } +\frac{1}{2} \int dx \sqrt{-g}\,h_{\mu \nu } T_2^{\mu \nu } .
\end{equation}
The 1,2 labels are the two flavors of scalars that we have.
We can try to find a saddle point by solving
\begin{equation} \label{eq:linearized_Einstein_eq}
-2iD^{-1} _{\mu \nu ,\rho \sigma } h^{\rho \sigma } = T^{(1)} _{\mu \nu } +T^{(2)} _{\mu \nu } =T_{\mu \nu } .
\end{equation}
This equation is linear so we can also solve it separately for each source. Plugging it back we have
\begin{equation}
S = \frac{1}{4} \int dx \sqrt{-g}\,h_{\mu \nu } T^{\mu \nu } = \frac{i}{8} \int dx \sqrt{-g}\int d\bar x \sqrt{-g}\, T^{\mu \nu } (x) D_{\mu \nu ,\rho \sigma } (x,\bar x) T^{\rho \sigma } (\bar x).
\end{equation}
Since the two signals go into different directions with no propagator between them, only cross terms contribute
\begin{equation}
S= \frac{i}{4} \int dx \sqrt{-g}\int d\bar x \sqrt{-g}\, T_1^{\mu \nu } (x) D_{\mu \nu ,\rho \sigma } (x,\bar x) T_2^{\rho \sigma } (\bar x) .
\end{equation}
This is essentially what we got (in \eqref{eq:4pf_from_diagrams}) in the eikonal approximation using the diagrammatic derivation with $T_1^{\mu \nu } \sim \delta (V)\delta (W) K^{\mu } K^{\nu } $ and analogously for $T_2$. So we see that the sources are localized.

Let us make a comment. We saw that we could not solve Einstein's equations with a localized energy-momentum source. One could wonder whether at the linearized level this is possible.
Eq.\ \eqref{eq:linearized_Einstein_eq} (in the $\alpha=1,\beta=-2$ gauge) explicitly is
\begin{equation} \label{eq:linearized_Einstein_eq_explicitly}
\left( -\frac{1}{2} \delta _{\mu } ^{\rho } \delta _{\nu } ^{\sigma }   \nabla ^2+\frac{1}{\rdS^2} \delta _{\mu } ^{\rho } \delta _{\nu } ^{\sigma } +\frac{1}{4}  g_{\mu \nu  }  g^{\rho \sigma }  \nabla ^2+\frac{d-3}{2\rdS^2}  g_{\mu \nu }  g^{\rho \sigma } \right) h_{\rho \sigma }  = 8\pi G_N T_{\mu \nu } .
\end{equation}
We have not found a solution of this either for a null geodesic source localized in the transverse direction, while allowing generic metric fluctuations around de Sitter in various dimensions. There is the same issue with the $L=1$ harmonics zero modes in any dimension.

 This discussion can be generalized to the massive case. In this case the equation of motion \eqref{eq:linearized_Einstein_eq_explicitly} becomes
\begin{equation} \label{eq:linearized_Einstein_eq_explicitly_graviton_mass}
\left( -\frac{1}{2} \delta _{\mu } ^{\rho } \delta _{\nu } ^{\sigma }(   \nabla ^2-\gm^2)+\frac{1}{\rdS^2} \delta _{\mu } ^{\rho } \delta _{\nu } ^{\sigma } +\frac{1}{4}  g_{\mu \nu  }  g^{\rho \sigma }  \nabla ^2+\frac{d-3}{2\rdS^2}  g_{\mu \nu }  g^{\rho \sigma } \right) h_{\rho \sigma }  = 8\pi G_N T_{\mu \nu } .
\end{equation}

For particle 2 we anticipate an energy-momentum localized at $u=0$ and at some position in the transverse direction. In Kruskal coordinates it is given in terms of the momentum $p^v$ by (we are back to $\rdS=1$)
\begin{equation}
T_{uu} =2p^v \delta (u) \frac{\delta (\phi -\phi ^{(2)} )}{\sqrt{g_{\perp}}} 
\end{equation}
where $\phi $ is a vector of sphere coordinates and $g_{\perp}$ is the metric in the transverse space, the sphere. Similarly for particle 1 we exchange all $u \leftrightarrow v$ and $p^u \leftrightarrow p^v$ with non-zero $T_{vv} $. Note that since these particles are localized to the horizons, the kinematics is the same as in special relativity, giving this normalization.

We can anticipate a metric perturbation where only $h_{uu} $ is non-zero for particle 2. Indeed, let us try an ansatz
\begin{equation}
h_{uu} =\delta (u)F(\phi ).
\end{equation}
In this case the last two terms in round brackets in \eqref{eq:linearized_Einstein_eq_explicitly_graviton_mass} vanish. This equation is satisfied at all components except for $uu$, and this component is satisfied if
\begin{equation}
\left( \nabla ^2_{S^{d-2} } - \gm^2+d-2\right) F(\phi ) = -32 \pi G_N p^v \frac{\delta (\phi -\phi ^{(2)} )}{\sqrt{g_{\perp}}} .
\end{equation}
The singular piece of the solution here satisfies the same form of equation as \eqref{eq:Greens_func_eq_transverse_prop}. 
Therefore, the solution is
\begin{equation}
h_{uu,\text{sing}} =\delta (u) \frac{32\pi G_N p^v}{\gm^2} \frac{d-1}{\sVol{d-2}} y \cdot y^{(2)} .
\end{equation}
An analogous discussion holds for particle 1 with all $u \leftrightarrow v$. (The equation \eqref{eq:Greens_func_eq_transverse_prop} is analyzed in more detail in Appendix \ref{app:mg_zero}.)

The solution in the linearized theory is just the sum of the $h_{uu} $ solution and the $h_{vv} $ solution. The kinetic term \eqref{eq:graviton_kinetic_term}, because of the form of the graviton propagator, couples only $h^{uu} $ to $h^{vv} $. Therefore, the equation of motion of either cancels one of the terms in \eqref{eq:two_EM_sources}, and the on-shell action is just either of the terms in \eqref{eq:two_EM_sources} evaluated on-shell. The on-shell action is then
\begin{equation}
S_{\text{tot, on-shell}}  = \frac{1}{2} \int dx \sqrt{-g}\, h^{uu} T_{uu} = \frac{1}{4} \int du \, dv \int d^{d-2}\phi \sqrt{g_{\perp}}h_{vv} T_{uu} = 16\pi  \frac{G_N}{\gm^2} p^u p^v \frac{d-1}{\sVol{d-2}} y^{(1)} \cdot y^{(2)} .
\end{equation}
Remembering that $i\delta =iS$, this gives the answer \eqref{eq:Eikonal_phase_result}.

\section{Physical part of the graviton propagator}
\label{app:ds_phys}
In this Appendix we derive the physical part of the propagator which does not mix with the pure gauge (large or small) terms or pure trace terms.

Consider first the gravity part (where $S$ is the action, $G_N$ Newton's constant and $\Lambda $ the cosmological constant)
\begin{equation}
\begin{split}
16 \pi  G_N S &= \int d^d x \sqrt{-g} \left[ R-2\Lambda \right] 
\end{split}
\end{equation}
around a background $g_{\mu \nu } =\hat g_{\mu \nu } +h_{\mu \nu } $ where $\hat g_{\mu \nu } $ is dS space of radius $\rdS$ so that
\begin{equation}
\hat R= \frac{d(d-1)}{\rdS^2} ,\qquad \Lambda =\frac{(d-1)(d-2)}{2\rdS^2} .
\end{equation}
We expand the action up to second order (and integrating by parts), denoting the result by $S_2$.
Let us fix traceless transverse (TT) gauge: 
\begin{equation}
h^{\mu}_\mu = 0 , \quad \hat{\nabla}^\alpha h_{\alpha \beta}=0
\end{equation}
Then the action is simply:
\begin{equation} \label{eq:gravity_action_2nd_order}
\begin{split}
&16\pi G_N S_2  =
 - \int d^d x \sqrt{-\hat g}\, h^{\mu \nu } \bigg( -\frac{1}{4} \hat g_{\mu \rho } \hat g_{\nu \sigma } \hat \nabla ^2 + \frac{1}{2\ell^2} \hat g_{\mu \rho } \hat g_{\nu \sigma } \bigg) h^{\rho \sigma } .
\end{split}
\end{equation}
From now on we use a de Sitter background, so we omit the hat, $\hat g_{\mu \nu } \to g_{\mu \nu } $.

The propagator then basically satisfies the wave equation proportional to $\square-2/\rdS^2$:
\begin{equation} \label{eq:graviton_propagator_eq}
\begin{split}
& \frac{1}{16\pi G_N} \left( -\frac{1}{2} \delta ^{\mu } _{\rho } \delta ^{\nu } _{\sigma } \nabla _x^2+\frac{1}{\rdS^2} \delta ^{\mu } _{\rho } \delta ^{\nu } _{\sigma }  \right) \langle h^{\rho \sigma } (x)h_{\alpha \beta } (y)\rangle = \\
& \qquad \qquad = \frac{1}{2} \left( \delta ^{\mu } _{\alpha } \delta ^{\nu } _{\beta } +\delta ^{\mu } _{\beta } \delta ^{\nu } _{\alpha } \right) \frac{-i}{\sqrt{-g}} \delta (x-y).
\end{split}
\end{equation}
It is very easy to separate the part of the propagator which is not proportional to gradient terms in $\rho, \sigma, \alpha, \beta$ or pure trace terms proportional to $g_{\rho \sigma}$ or $g_{\alpha \beta}$:
\begin{equation} \label{eq:graviton_propagator}
D_{\rho \sigma ,\alpha \beta }  = \langle h_{\rho \sigma } (x)h_{\alpha \beta } (y)\rangle  = 16\pi G_N \left( \nabla_\rho \nabla_\alpha Z \nabla_{\sigma} \nabla_{\beta} Z +  \nabla_\rho \nabla_\beta Z \nabla_{\sigma} \nabla_{\alpha} Z \right) H_0 + \text{grad}  + \text{metric}
\end{equation}
Here $H_0$ is the physical, gauge invariant TT part of the propagator and the residual terms (grad and metric) have to be chosen to satisfy the TT condition. The metric terms never contribute to the eikonal phase because it is sensitive to the null components of the metric. The gradient terms play the role of total derivatives and potentially can contribute if they do not decay at infinity. For now we will drop them but then return to this question later.

We need to act on the expression \eqref{eq:graviton_propagator} with $\square$.
For a generic spacetime one cannot simply ignore the grad terms. However, for $(A)dS$ spacetimes it is true that for arbitrary $V_\sigma$:
\beq
\square \nabla_\rho V_\sigma = \text{(grad in $\rho$ or $\sigma$)} + \text{metric}.
\eeq
Hence we can simply ignore them. Finally, it is straightforward to check that \footnote{For that one needs to use $\nabla_{\alpha} \nabla_{\beta} Z = - Z g_{\alpha \beta}$.}: 
\begin{align}
\square \left( \nabla_\rho \nabla_\alpha Z \nabla_{\sigma} \nabla_{\beta} Z +  \nabla_\rho \nabla_\beta Z \nabla_{\sigma} \nabla_{\alpha} Z \right) H_0 = \nonumber \\
= \left( \nabla_\rho \nabla_\alpha Z \nabla_{\sigma} \nabla_{\beta} Z +  \nabla_\rho \nabla_\beta Z \nabla_{\sigma} \nabla_{\alpha} Z \right) (\square H_0 +\frac{2}{\ell^2} ) + \text{grad} + \text{metric}
\end{align}
Naively, one could conclude from \eqref{eq:graviton_propagator_eq} that $H_0$ has to satisfy massless wave equation:
\begin{equation}
\square H_0 = \frac{i}{\sqrt{-g}} \delta (x-y) .
\end{equation}
This would have been problematic because massless wave equation is not well-behaved in the de Sitter space.
Luckily, it is not the full story. The tensor structure  
$(\nabla_\rho \nabla_\alpha Z \nabla_{\sigma} \nabla_{\beta} Z +  \nabla_\rho \nabla_\beta Z \nabla_{\sigma} \nabla_{\alpha} Z) F$ is a gradient term or a metric term if $F$ is either a constant or linear in $Z$. So the most generic equation is
\begin{equation}
\label{eq:phys}
\square H_0 = \frac{i}{\sqrt{-g}} \delta (x-y) + a + b Z.
\end{equation}
Term $b Z$ turns out to be unimportant: it shifts the solution by a grad/metric term. But constant $a$ is important: it is fixed unambiguously by requiring that $\propG$ does not have an anti-pode singularity at $Z=-1$. 
A more careful tensor-basis analysis (adapting the AdS strategy from \cite{D_Hoker_1999}) is given in Appendix \ref{app:dS-propagator}. 
The same conclusion is reached by performing analytic continuation from the sphere \cite{Morrison}.

\section{Physical graviton exchange in BTZ}
\label{app:ads_phys}
In the case of non-rotating BTZ of radius $R$ we can perform an explicit comparison of the shock-wave answer and the perturbative computation.

Unperturbed geometry is
\beq
ds^2 = - 4 \frac{du dv}{(1+ u v)^2} + R^2 \frac{(1 - u v)^2}{(1 + u v)^2} dr^2
\eeq

Perturbing this metric by shock-wave term, $f(x) \delta(u) du^2$, the profile $f(x)$ satisfies
\beq
f'' - \mu^2 f = -\delta(x), \ \mu^2 = \frac{2 \pi R}{\beta} = R^2
\eeq
leading to $f = \frac{1}{2 R} \exp(-|r| R)$ and the eikonal phase :
\beq
\label{eq:btz_delta}
i \delta = 4 \pi i G_N R (4 p^u p^v) f(r) = 
8 \pi G_N i p^u p^v e^{-R|r|}
\eeq

In order to do the perturbative computation, it would be convenient to embed this geometry into the higher-dimensional Minkowski hyperboloid via
\beq
-X_0^2 + X_1^2 + X_2^2 - X_3^2 = -1,
\eeq
\beq
X_0 = \frac{u+v}{1+ u v}
\eeq
\beq
X_1 = \frac{1-uv}{1+ u v} \sinh(R x)
\eeq
\beq
X_2 = \frac{v-u}{1+ u v}
\eeq
\beq
X_3 = \frac{1-uv}{1+ u v} \cosh(R x)
\eeq

Denoting the chord distance by $w$:
\beq
w = - \vec{X}_1 \cdot \vec{X}_2 -1,
\eeq

The physical part of the graviton propagator can be written as
\beq
\Pi^{(phys)}_{\mu \nu,\mu' \nu'} = 16 \pi G_N
\l( \pr_\mu \pr_{\mu'} w \pr_\nu \pr_{\nu'} w  + \pr_\mu \pr_{\nu'} w
\pr_{\mu'} \pr_{\nu} w
\r) G(w)
\eeq
Function $G(w)$ is simply massless minimally coupled propagator:
\beq
G(w) = C_2 (2/w)^2 \  _2 F_1(2,3/2,3,-2/w), \ C_2 = \frac{\Gamma(3/2)}{2 (4 \pi)^{3/2}} = \frac{1}{32 \pi}
\eeq

In this case the hypergeometric function is actually a rational function and the integral over $u \bar{v}$ can easily be evaluated leading to 
\begin{align}
-\frac{1}{4}\int du d\bar{v} \Pi^{(phys)}_{u u, \bar{v} \bar{v}}(u \bar{v}, r) \vert_{v=0, \bar{u}=0} & =\nonumber
  2\pi i \frac{1}{4} \int_0^{+\infty} dx \Pi^{(TT)}_{u u, \bar{v} \bar{v}}(u \bar{v} = x , r) \vert_{v=0, \bar{u}=0}\\ 
&=  8 i \pi G_N e^{-R|r|},
\end{align}
which is exactly the phase (\ref{eq:btz_delta}).

\section{Solving for the regularized eikonal phase}
\label{app:mg_zero}
We would like to solve \eqref{eq:Greens_func_eq_transverse_prop} with a graviton mass regulator. Since as mentioned there is no solution at $\gm=0$, the solution has a piece singular in the graviton mass and a regular piece. As we explain in the main text, we will mostly be interested in the singular piece.

In order to understand this, it is natural to decompose the functions into spherical harmonics. In this appendix we use $N=d-2$ for the dimension of the sphere. For a general dimension sphere, the spectrum of $(-\nabla ^2_{S^N} )$ is $L(L+N-1)$ with $L=0,1,2,\cdots $. So for $L \neq 1$ the LHS of the equation is non-zero even for $\gm=0$ and there is no problem to solve the equation. The only issue is for $L=1$ which are the only modes for which $\nabla ^2_{S^N} +N=0$, so that if we were to take $\gm=0$, the LHS would vanish while the RHS does not. So the singular part comes precisely from the $L=1$ modes.

Consider embedding space for the sphere which we have been denoting by $y \in \mathbb{R} ^{N+1} $ with the sphere being $y^2=1$. We use $r$ as the radial direction in embedding space, and the sphere is at $r=1$. For a function $f$ defined on the sphere, we can consider an extension of it $\tilde f$ to embedding space. When acting on a function in embedding space
\begin{equation}
\nabla ^2_{\mathbb{R^{N+1} } } =\partial _r^2+\frac{N}{r} \partial _r + \frac{1}{r^2} \nabla ^2_{S^N} 
\end{equation}
where the last term acts on the angular coordinates only (that is, on the unit sphere).

Now, if a function is homogeneous of degree $L$ such that $\tilde f(\lambda y)=\lambda ^L\tilde f(y)$ it can be written as $\tilde f(y) = r^L f(y/r)$ with $f$ defined on the unit sphere. Acting on such a function we get
\begin{equation}
\nabla ^2_{\mathbb{R} ^{N+1} } \tilde f(y) = r^{L-2} \left[ \nabla ^2_{S^N} +L(L+N-1)\right] f .
\end{equation}
So the $L$ harmonics are harmonic ($\nabla ^2_{\mathbb{R} ^{N+1} } \tilde f=0$) homogeneous functions in embedding space of degree $L$.

For $L=1$ these are just the coordinate functions $y_1,\cdots y_{N+1} $ (these modes are the fundamental representation of $SO(N+1)$).
The measure on the sphere is related to the measure in embedding space simply by (we denote angular coordinates collectively by $\phi $ and the metric determinant on the sphere transverse space by $g_{\perp}$)
\begin{equation}
\int d^N \phi \sqrt{g_{\perp} } \, f = \int d^{N+1} y \, \tilde f \, \delta (r-1)
\end{equation}
and the inner product is just $(f,g) = \int d^N \phi \sqrt{g_{\perp} }\, f^* g$ which can be written as $\int d^{N+1} y \, \tilde f ^*(y) \tilde g(y) \delta (r-1)$ where again $\tilde f|_{y^2=1} =f$ and $\tilde g|_{y^2=1} =g$. Denoting the normalized spherical harmonics in general dimension by $Y_{LM} $, the $L=1$ modes extended to embedding space can be taken to be just
\begin{equation} \label{eq:L1_spherical_harmonics_in_embedding_space}
\tilde Y_{1M} = \sqrt{\frac{N+1}{\sVol{N}}} \, y_M .
\end{equation}

We also need the components in the spherical harmonic expansion of the delta function. These are simply given by
\begin{equation}
\int d\phi _1 \cdots d\phi _N \, \sqrt{g_{\perp}}\, Y_{LM} ^* \, \frac{1}{\sqrt{g_{\perp}}} \delta (\phi _1-\bar \phi_1) \cdots \delta (\phi _N-\bar \phi _N) = Y_{LM} ^*(\bar \phi ) .
\end{equation}

The equation \eqref{eq:Greens_func_eq_transverse_prop} for the $L$ modes of $G_T$ is
\begin{equation}
(\gm^2 + L(L+N-1)-N) G_{LM} =Y_{LM} ^*(\bar \phi ) .
\end{equation}
so that the solution is
\begin{equation}
\begin{split}
G_{T} (\phi ) & =\sum _{L,M} \frac{1}{\gm^2+L(L+N-1)-N} Y^*_{LM} (\bar \phi )Y_{LM} (\phi ) .
\end{split}
\end{equation}

The de Sitter version of what was shown in \cite{Cornalba:2007zb} is that for a general spin $j$ particle coupling with coupling constant $g$, the eikonal phase is
\begin{equation}
i\delta = \frac{ig^2}{2} \left[ (2\omega )^2 \frac{(1-V\bar U/4)^2}{(X_0 \cdot W) (X_0 \cdot \bar W)} \right] ^{j-1} G_T(W\cdot \bar W)
\end{equation}
where now we use the notation of embedding space in de Sitter, but the transverse space there is equivalent to embedding space for the sphere.

The wavefunctions \eqref{eq:the_four_wavefunctions_general_d} are localized on $V=\bar U=0$ for large time differences, so we use that and get
\begin{equation}
i\delta = \frac{ig^2}{2} \left[ 4p^u p^v\right]^{j-1} G_T(W \cdot \bar W) .
\end{equation}
Restricting to graviton exchanges, using Newton's coupling $G_N$, and restoring units of $\rdS$, the eikonal phase is
\begin{equation}
    i \delta = 16 \pi i G_N \rdS^{6-d} p^u p^v \sum _{L,M} \frac{1}{\gm^2 \rdS^2 +L(L+N-1)-N} Y^*_{LM} (\bar \phi )Y_{LM} (\phi )
\end{equation}
where we changed to $y,\bar y$ to signify taking angular variables on a unit sphere (rather than of radius $\rdS$).
In the $L \neq 1$ modes, we can neglect $\gm \rdS$, while the $L=1$ modes are dominated by it, and in this case we can write the angular dependence using \eqref{eq:L1_spherical_harmonics_in_embedding_space} as we do in the main text.

\section{Transverse integral of eikonal phase} \label{sec:transverse_integral}

We want to evaluate
\begin{equation}
\int d^N\phi \sqrt{g_{\perp}} \, d^N\bar \phi  \sqrt{\bar g_{\perp}} \, (y \cdot \bar y)^k
\end{equation}
where $\phi $ are angular coordinates for the $N$-sphere and $y$ are the corresponding embedding space coordinates to $\phi $. This integral is rotationally invariant, so let us choose $\bar \phi _1=0$, in which case $y \cdot \bar y = \cos \phi _1$. The integral is then simply
\begin{equation}
\sVol{N}\sVol{N-1} \int _0 ^\pi d\phi _1 \, \sin^{N-1} \phi _1 \cos ^k \phi _1 .
\end{equation}
Doing the integral we find
\begin{equation} \label{eq:transverse_integral_in_expansion}
\begin{split}
& \int d^N\phi \sqrt{g_{\perp}} \, d^N\bar \phi  \sqrt{\bar g_{\perp}} \, (y \cdot \bar y)^k = \\
& = \begin{cases}
\sVol{N} \sVol{N-1} \frac{\Gamma \left( \frac{k+1}{2} \right) \Gamma \left( \frac{N}{2} \right) }{\Gamma \left( \frac{k+N+1}{2} \right) } & k \text{ even} \\
0 & k \text{ odd} \end{cases} = \\
& = \begin{cases}
\sVol{N} ^2 \, \frac{\Gamma \left( \frac{k+1}{2} \right) \Gamma \left( \frac{N+1}{2} \right) }{\sqrt{\pi} \, \Gamma \left( \frac{k+N+1}{2} \right) } & k \text{ even} \\
0 & k \text{ odd}
\end{cases}
\end{split}
\end{equation}

\section{De Sitter invariant decomposition of graviton propagator} \label{app:dS-propagator}

In this section we repeat, for de Sitter space, the same invariant bi-tensor analysis that is often used for the AdS graviton propagator.  We work in spacetime dimension $d$ and set the de Sitter radius to $1$.  Thus
\begin{equation}
    R_{\mu\nu\rho\sigma}
    =
    g_{\mu\rho}g_{\nu\sigma}
    -
    g_{\mu\sigma}g_{\nu\rho},
    \qquad
    R_{\mu\nu}=(d-1)g_{\mu\nu},
    \qquad
    R=d(d-1).
\end{equation}

\subsection{Linearized Einstein equation}
We derive the equation satisfied by graviton propagator in this section. The trace-reversed Einstein equation following from
\((16\pi G_N)^{-1}\int\sqrt{-g}(R-2\Lambda)\) is
\begin{equation}
    R_{\mu\nu}-(d-1)g_{\mu\nu}
    =
    8\pi G_N\,\widetilde T_{\mu\nu},
    \qquad
    \widetilde T_{\mu\nu}
    \equiv
    T_{\mu\nu}
    -
    \frac{1}{d-2}g_{\mu\nu}T^\rho{}_\rho .
\end{equation}
In this subsection we use the source convention \(8\pi G_N=1\).  
Consider linearized perturbation $g_{\mu\nu}\to g_{\mu\nu}+h_{\mu\nu}$,
the linearized equation of motion for $h_{\mu\nu}$ is:
\begin{equation}
\label{eq:dS_lin_Einstein}
    -\Box h_{\mu\nu}
    -\nabla_\mu\nabla_\nu h^\rho{}_\rho
    +\nabla_\mu\nabla^\rho h_{\rho\nu}
    +\nabla_\nu\nabla^\rho h_{\mu\rho}
    +2\bigl(h_{\mu\nu}-g_{\mu\nu}h^\rho{}_\rho\bigr)
    =
    2\widetilde T_{\mu\nu}.
\end{equation}
This equation is invariant under the infinitesimal diffeomorphism:
\begin{equation}
    \delta h_{\mu\nu}=\nabla_\mu \xi_\nu+\nabla_\nu\xi_\mu .
\end{equation}
We define the graviton propagator as the inverse of the quadratic differential operator, such that for given physical source $\widetilde{T}_{\mu\nu}$, one can reverse \eqref{eq:dS_lin_Einstein} and obtain a solution of $h_{\mu\nu}$ as:
\begin{equation}
    h_{\mu\nu}(x)
    =
    \int d^d x'\,\sqrt{-g(x')}
    \,G_{\mu\nu;\mu'\nu'}(x,x')\,\widetilde{T}^{\mu'\nu'}(x').
\end{equation}
The graviton propagator must therefore satisfy
\begin{align}
\label{eq:dS_prop_eq}
    &
    -\Box G_{\mu\nu;\mu'\nu'}
    -\nabla_\mu\nabla_\nu G^\rho{}_{\rho;\mu'\nu'}
    +\nabla_\mu\nabla^\rho G_{\rho\nu;\mu'\nu'}
    +\nabla_\nu\nabla^\rho G_{\mu\rho;\mu'\nu'}
    \nonumber\\
    &\hspace{2cm}
    +2\bigl(
      G_{\mu\nu;\mu'\nu'}
      -
      g_{\mu\nu}G^\rho{}_{\rho;\mu'\nu'}
    \bigr)
    \nonumber\\
    &=
    \Bigl(
      g_{\mu\mu'}g_{\nu\nu'}
      +
      g_{\mu\nu'}g_{\nu\mu'}
    \Bigr)\delta(x,x')
    +
    \nabla_{\mu'}\Lambda_{\mu\nu;\nu'}
    +
    \nabla_{\nu'}\Lambda_{\mu\nu;\mu'} .
\end{align}
Note that the source in \eqref{eq:dS_lin_Einstein} has already been trace
reversed, the delta-function term here is the identity on symmetric two-tensor
sources rather than another trace-reversal operator.
The coefficient equations derived below are all at separated points, so this
contact-term correction does not affect the homogeneous equations for decomposition coefficients 
\(H_1,H_2,A,B,C\) defined below.
The $\Lambda$ term parameterizes the freedom to shift the propagator by a pure gauge term at the point $x'$.

\subsection{Decomposition of graviton propagator into bi-tensor basis}

We collect here the invariant variables and bi-tensors that will be used both in the differential-equation derivation and in the comparison with Higuchi's
basis \cite{Higuchi:2001uv} below.  Let $Z(x,x')$ be the standard de Sitter invariant embedding
variable, and define the de Sitter chordal distance as 
\begin{equation}
    u(x,x') \equiv 1-Z(x,x').
\end{equation}
If
\(\mu(x,x')\) is the geodesic distance, then for unit de Sitter radius
\[
    Z=\cos\mu,
\]
so that
\begin{equation}
    u=1-\cos\mu,
    \qquad
    z=\cos^2\frac{\mu}{2}=1-\frac{u}{2}.
\end{equation}
Here \(z\) is Higuchi's variable.  We will also use the abbreviations
\begin{equation}
    s(u)\equiv u(2-u),
    \qquad
    q(u)\equiv 2-u,
\end{equation}
so that \(s=\sin^2\mu\) and \(q=1+\cos\mu=2z\).

It is convenient to introduce the first and mixed second derivatives of the
chordal variable,
\begin{equation}
    U_\mu\equiv \nabla_\mu u,
    \qquad
    U_{\mu'}\equiv \nabla_{\mu'}u,
    \qquad
    M_{\mu\nu'}\equiv \nabla_\mu\nabla_{\nu'}u.
\end{equation}
In Higuchi's notation \(n_\mu=\nabla_\mu\mu\) and
\(n_{\mu'}=\nabla_{\mu'}\mu\) are the unit tangent vectors at the two endpoints
of the geodesic.  Since \(u=1-\cos\mu\), one has
\begin{equation}
    U_\mu=\sqrt{s}\,n_\mu,
    \qquad
    U_{\mu'}=\sqrt{s}\,n_{\mu'},
    \qquad
    n_\mu=\frac{U_\mu}{\sqrt{s}},
    \qquad
    n_{\mu'}=\frac{U_{\mu'}}{\sqrt{s}}.
\end{equation}

The mixed tensor \(g_{\mu\nu'}(x,x')\) is the parallel-transport tensor along
the geodesic from \(x'\) to \(x\).  If \(V^{\nu'}\) is a tangent vector at
\(x'\), then \(g^\mu{}_{\nu'}(x,x')V^{\nu'}\) is the vector obtained by
parallel transporting \(V^{\nu'}\) to the tangent space at \(x\).  Equivalently,
it is the invariant bitensor that identifies tangent indices at the two points
in a maximally symmetric way.  We use the convention that,
\begin{equation}
    g_{\mu\nu'}\nabla^\mu u=-\nabla_{\nu'}u .
\end{equation}
One has the following useful identities
\begin{equation}
    \nabla_\mu u\,\nabla^\mu u = u(2-u),
    \qquad
    \Box u = d(1-u),
    \qquad
    \nabla_\mu\nabla_\nu u = (1-u)g_{\mu\nu},
\end{equation}
and, for mixed derivatives,
\begin{equation}
\label{eq:mixed_u_identity}
    \nabla_\mu\nabla_{\nu'}u
    =
    -g_{\mu\nu'}
    -
    \frac{\nabla_\mu u\,\nabla_{\nu'}u}{2-u}.
\end{equation}
Indeed, the parallel propagator obeys
\begin{equation}
    \nabla_\mu\nabla_{\nu'}Z
    =
    g_{\mu\nu'}
    +(1-Z)n_\mu n_{\nu'},
\end{equation}
while \(\nabla_\mu u=-\nabla_\mu Z\) and
\(\nabla_\mu u\nabla_{\nu'}u/(2-u)=(1-Z)n_\mu n_{\nu'}\).  This gives
\eqref{eq:mixed_u_identity}, or in the compact notation above,
\begin{equation}
    M_{\mu\nu'}
    =
    -g_{\mu\nu'}
    -
    \frac{1}{q}U_\mu U_{\nu'}, \quad      g_{\mu\nu'}
    =
    -M_{\mu\nu'}
    -
    \frac{1}{q}U_\mu U_{\nu'}.
\end{equation}
This is the basic dictionary between the mixed parallel propagator and the
chordal-distance basis.  It is also useful to note that
\begin{equation}
    \nabla^\mu u\,\nabla_\mu\nabla_{\nu'}u
    =
    (1-u)\nabla_{\nu'}u .
\end{equation}
With these identities in hand, we now choose a maximally symmetric bi-tensor
basis.  Following the AdS analysis, we write the propagator as a sum of a
physical part and pure-gradient pieces \cite{D_Hoker_1999}, 
\begin{align}
\label{eq:dS_ansatz}
    G_{\mu\nu;\mu'\nu'}
    &=
    \Bigl(
      \nabla_\mu\nabla_{\mu'}u\,\nabla_\nu\nabla_{\nu'}u
      +
      \nabla_\mu\nabla_{\nu'}u\,\nabla_\nu\nabla_{\mu'}u
    \Bigr)H_1(u)
    +
    g_{\mu\nu}g_{\mu'\nu'}H_2(u)
    \nonumber\\
    &\quad
    +
    \nabla_{(\mu}
    \Bigl[
      \nabla_{\nu)}\nabla_{\mu'}u\,\nabla_{\nu'}u\,X(u)
    \Bigr]
    +
    \nabla_{(\mu'}
    \Bigl[
      \nabla_{\nu')}\nabla_\mu u\,\nabla_\nu u\,X(u)
    \Bigr]
    \nonumber\\
    &\quad
    +
    \nabla_{(\mu}
    \Bigl[
      \nabla_{\nu)}u\,\nabla_{\mu'}u\,\nabla_{\nu'}u\,Y(u)
    \Bigr]
    +
    \nabla_{(\mu'}
    \Bigl[
      \nabla_{\nu')}u\,\nabla_\mu u\,\nabla_\nu u\,Y(u)
    \Bigr]
    \nonumber\\
    &\quad
    +
    \nabla_\mu\bigl[\nabla_\nu u\,W(u)\bigr]g_{\mu'\nu'}
    +
    \nabla_{\mu'}\bigl[\nabla_{\nu'}u\,W(u)\bigr]g_{\mu\nu}.
\end{align}
We point out that Euclidean results from \cite{Higuchi:2001uv} can be brought into this form after some work. 
The functions \(X,Y,W\) correspond to total-derivative terms. More precisely, they are generated by derivatives acting on one or both arguments, and hence one expects their contributions vanish when the propagator is inserted between conserved gauge-invariant sources. Equivalently, they may be absorbed into the \(\Lambda\)-ambiguity on the right-hand side of \eqref{eq:dS_prop_eq}. The only tensor structures that we keep explicitly below are therefore the crossed-index structure multiplying \(H_1(u)\) and the trace structure multiplying \(H_2(u)\).

There is, however, an important subtlety in actual computations. These
total-derivative terms need not decay sufficiently rapidly at infinity, and their integrals can receive boundary-flux contributions that are sensitive to large gauge transformations. In the explicit dS\(_{3,4,5}\) computations mentioned in the text, we found that the integral over the full transverse-traceless part of the graviton propagator is finite, even though the individual non-total-derivative and total-derivative pieces can separately suffer from IR divergences.

The most general
tensor decomposition of the gauge term that can enter the coefficient equations
for the non-gradient structures,
\begin{equation}
\label{eq:dS_Lambda_ansatz}
    \Lambda_{\mu\nu;\nu'}
    =
    g_{\mu\nu}\nabla_{\nu'}u\,A(u)
    +
    \nabla_\mu u\,\nabla_\nu u\,\nabla_{\nu'}u\,C(u)
    +
    \bigl(
      \nabla_\mu\nabla_{\nu'}u\,\nabla_\nu u
      +
      \nabla_\nu\nabla_{\nu'}u\,\nabla_\mu u
    \bigr)B(u).
\end{equation}
It is useful to introduce the following independent bi-tensors:
\begin{align}
    T^{(1)}_{\mu\nu;\mu'\nu'}
    &=
    g_{\mu\nu}g_{\mu'\nu'},\\
    T^{(2)}_{\mu\nu;\mu'\nu'}
    &=  \nabla_\mu u\,\nabla_\nu u\,\nabla_{\mu'}u\,\nabla_{\nu'}u,\\
    T_{\mu\nu;\mu'\nu'}^{(3)}
    &=    \nabla_\mu\nabla_{\mu'}u\,\nabla_\nu\nabla_{\nu'}u +\nabla_\mu\nabla_{\nu'}u\,\nabla_\nu\nabla_{\mu'}u,\\
    T^{(4)}_{\mu\nu;\mu'\nu'}
    &=
    g_{\mu\nu}\,\nabla_{\mu'}u\,\nabla_{\nu'}u+g_{\mu'\nu'}\,\nabla_\mu u\,\nabla_\nu u, \\
    T^{(5)}_{\mu\nu;\mu'\nu'}
    &=\nabla_\mu\nabla_{\mu'}u\,\nabla_\nu u\,\nabla_{\nu'}u
    +\nabla_\nu\nabla_{\mu'}u\,\nabla_\mu u\,\nabla_{\nu'}u
    +
    (\mu'\leftrightarrow \nu'),
\end{align}
 Expanding
\eqref{eq:dS_ansatz} in this basis gives
\begin{equation}
\label{eq:dS_f_basis_expansion}
    G_{\mu\nu;\mu'\nu'}
    =
    \sum_{i=1}^{5}
    f_i(u)\,T^{(i)}_{\mu\nu;\mu'\nu'} .
\end{equation}
The individual pieces of \eqref{eq:dS_ansatz} contribute as
\begin{align}
    H_1\text{-term} &: \qquad H_1\,T^{(3)},\\
    H_2\text{-term} &: \qquad H_2\,T^{(1)},\\
    X\text{-terms} &: \qquad
    X\,T^{(3)}-X\,T^{(4)}+\frac12 X'\,T^{(5)},\\
    Y\text{-terms} &: \qquad
    2Y'\,T^{(2)}+(1-u)Y\,T^{(4)}+Y\,T^{(5)},\\
    W\text{-terms} &: \qquad
    2(1-u)W\,T^{(1)}+W'\,T^{(4)} .
\end{align}
Therefore the five coefficient functions in \eqref{eq:dS_f_basis_expansion} are
\begin{align}
\label{eq:dS_f_coefficients}
    f_1(u) &= H_2(u)+2(1-u)W(u),\\
    f_2(u) &= 2Y'(u),\\
    f_3(u) &= H_1(u)+X(u),\\
    f_4(u) &= -X(u)+(1-u)Y(u)+W'(u),\\
    f_5(u) &= \frac12 X'(u)+Y(u).
\end{align}
After substituting \eqref{eq:dS_ansatz} and \eqref{eq:dS_Lambda_ansatz} into
\eqref{eq:dS_prop_eq}, and using the identities above, the result may be
expanded in the same five structures.  Away from coincidence, each coefficient
must vanish separately.

\subsection{Equations determining \(H_1(u)\) and solution}

Let \({\cal D}\) denote the linear differential operator on the left-hand side
of \eqref{eq:dS_prop_eq}.  Away from the coincident point \(x=x'\), the delta
function drops out, and the propagator equation becomes
\begin{equation}
    {\cal D}G_{\mu\nu;\mu'\nu'}
    -
    \nabla_{\mu'}\Lambda_{\mu\nu;\nu'}
    -
    \nabla_{\nu'}\Lambda_{\mu\nu;\mu'}
    =
    0 .
\end{equation}
After inserting the decomposition \eqref{eq:dS_ansatz} and the gauge ansatz
\eqref{eq:dS_Lambda_ansatz}, the result can again be expanded in the five
bi-tensors introduced above:
\begin{equation}
\label{eq:dS_field_eq_basis_expansion}
    {\cal D}G_{\mu\nu;\mu'\nu'}
    -
    \nabla_{\mu'}\Lambda_{\mu\nu;\nu'}
    -
    \nabla_{\nu'}\Lambda_{\mu\nu;\mu'}
    =
    \sum_{i=1}^{5}
    {\cal F}_i(u)\,T^{(i)}_{\mu\nu;\mu'\nu'} .
\end{equation}
The five tensors \(T^{(i)}\) are independent for separated points, so the
field equation is equivalent to
\begin{equation}
    {\cal F}_i(u)=0,
    \qquad
    i=1,\ldots,5 .
\end{equation}
The coefficients \({\cal F}_i\) are differential expressions built from the
functions \(H_1,H_2,X,Y,W\) and the gauge functions \(A,B,C\).  Since \(X,Y,W\)
are total derivatives, it is more efficient to absorb their contribution into
\(\Lambda\) and use the gauge functions \(B,C\) to keep track of the relevant
pure-gauge pieces.  The equations that determine the function \(H_1(u)\)
come from the coefficients of  \(T^{(2)}\), \(T^{(5)}\), and \(T^{(3)}\).  We
write them in the order that we follow to solve for $H_1(u)$:  the
\(T^{(2)}\) equation fixes \(C'\) in terms of \(H_1''\), the \(T^{(5)}\)
equation then fixes \(B'\), and the \(T^{(3)}\) equation finally gives the
closed second-order equation for \(H_1\).

Let us spell out these three projections.  Keeping only the crossed-index
piece \(H_1(u)T^{(3)}_{\mu\nu;\mu'\nu'}\) from \eqref{eq:dS_ansatz}, the
linearized Einstein operator gives
\begin{align}
\left.
{\cal D}\!\left(H_1 T^{(3)}\right)
\right|_{T^{(2)},T^{(5)},T^{(3)}}
={}&
2H_1''\,T^{(2)}
\nonumber\\
&+
\left[
(1-u)H_1''-(d-2)H_1'
\right]T^{(5)}
\nonumber\\
&-
\left[
u(2-u)H_1''
+(d-2)(1-u)H_1'
+2(d-2)H_1
\right]T^{(3)} .
\end{align}
Let us give the intermediate steps, since this is the only place where the
crossed-index structure is differentiated nontrivially.  Write
\begin{equation}
    A\equiv 1-u,
    \qquad
    s\equiv u(2-u),
    \qquad
    h_{\mu\nu;\mu'\nu'}=H_1(u)S_{\mu\nu;\mu'\nu'},
\end{equation}
with
\begin{equation}
    S_{\mu\nu;\mu'\nu'}
    =
    M_{\mu\mu'}M_{\nu\nu'}
    +
    M_{\mu\nu'}M_{\nu\mu'} .
\end{equation}
The required derivative identities are
\begin{equation}
    \nabla_\alpha U_\beta=A g_{\alpha\beta},
    \qquad
    \nabla_\alpha M_{\beta\mu'}=-g_{\alpha\beta}U_{\mu'},
    \qquad
    U^\alpha M_{\alpha\mu'}=A U_{\mu'},
    \qquad
    \Box M_{\mu\mu'}=-M_{\mu\mu'} .
\end{equation}
They imply
\begin{align}
    S^\rho{}_{\rho;\mu'\nu'}
    &=
    2\left(g_{\mu'\nu'}-U_{\mu'}U_{\nu'}\right),
    \\
    \nabla^\rho h_{\rho\nu;\mu'\nu'}
    &=
    \left[A H_1'-(d+1)H_1\right]
    \left(
      M_{\nu\mu'}U_{\nu'}+M_{\nu\nu'}U_{\mu'}
    \right).
\end{align}
Projecting each term in \({\cal D}\) to $T^{(2)}, T^{(3)}, T^{(5)}$  then gives
\begin{align}
\left.
-\Box h
\right|_{T^{(2)},T^{(5)},T^{(3)}}
&=
2H_1'\,T^{(5)}
-
\left[sH_1''+DAH_1'-2H_1\right]T^{(3)},
\\
\left.
-\nabla_\mu\nabla_\nu h^\rho{}_{\rho}
\right|_{T^{(2)},T^{(5)},T^{(3)}}
&=
2H_1''\,T^{(2)}
+2H_1'\,T^{(5)}
+2H_1\,T^{(3)},
\\
\left.
\nabla_\mu\nabla^\rho h_{\rho\nu}
+\nabla_\nu\nabla^\rho h_{\mu\rho}
\right|_{T^{(2)},T^{(5)},T^{(3)}}
&=
\left[AH_1''-(d+2)H_1'\right]T^{(5)}
+
\left[2AH_1'-2(D+1)H_1\right]T^{(3)},
\\
\left.
2\left(h_{\mu\nu}-g_{\mu\nu}h^\rho{}_\rho\right)
\right|_{T^{(2)},T^{(5)},T^{(3)}}
&=
2H_1\,T^{(3)} .
\end{align}
Adding these four lines gives the projected action of \({\cal D}\) displayed
above.

Next, the relevant part of the gauge ambiguity is controlled only by \(B\) and
\(C\).  Projecting the gauge contribution to the left-hand side of
\eqref{eq:dS_field_eq_basis_expansion} gives
\begin{align}
\left.
\left(
-\nabla_{\mu'}\Lambda_{\mu\nu;\nu'}
-\nabla_{\nu'}\Lambda_{\mu\nu;\mu'}
\right)
\right|_{T^{(2)},T^{(5)},T^{(3)}}
={}&
-2C'\,T^{(2)}
\nonumber\\
&+
\left(-B'-C\right)T^{(5)}
-2B\,T^{(3)} .
\end{align}
The \(C'\) term comes from differentiating the scalar coefficient in
\(U_\mu U_\nu U_{\nu'}C(u)\), the \(-C\) term comes from differentiating one of
the unprimed gradients \(U_\mu,U_\nu\), the \(-B'\) term comes from
differentiating the coefficient multiplying \(M_{\mu\nu'}U_\nu+M_{\nu\nu'}U_\mu\),
and the \(-2B\) term comes from differentiating the remaining primed gradient
to produce two \(MM\) structures.  The signs of the \(C'\) and \(B'\) terms
come from the overall
\(-\nabla_{\mu'}\Lambda_{\mu\nu;\nu'}-\nabla_{\nu'}\Lambda_{\mu\nu;\mu'}\)
appearing on the left-hand side.  Therefore the relevant coefficients are
\begin{align}
\label{eq:dS_F2}
    {\cal F}_2(u)
    &=
    2H_1''(u)-2C'(u),\\
\label{eq:dS_F5}
    {\cal F}_5(u)
    &=
    (1-u)H_1''(u)-(d-2)H_1'(u)-B'(u)-C(u),\\
\label{eq:dS_F3}
    {\cal F}_3(u)
    &=
    -u(2-u)H_1''(u)-(d-2)(1-u)H_1'(u)-2(d-2)H_1(u)-2B(u).
\end{align}
Setting \({\cal F}_2={\cal F}_5={\cal F}_3=0\) gives
\begin{align}
\label{eq:dS_Ceq}
    0 &= 2H_1''(u)-2C'(u),\\
\label{eq:dS_Beqprime}
    0 &= (1-u)H_1''(u)-(d-2)H_1'(u)-B'(u)-C(u),\\
\label{eq:dS_Beq}
    0 &= -u(2-u)H_1''(u)-(d-2)(1-u)H_1'(u)-2(d-2)H_1(u)-2B(u).
\end{align}
From \eqref{eq:dS_Ceq} we obtain
\begin{equation}
\label{eq:dS_Csol}
    C(u)=H_1'(u)+c_C .
\end{equation}
In the AdS case the constant $c_C$ can be set to zero. We will keep it for now. Then \eqref{eq:dS_Beqprime} integrates to
\begin{equation}
\label{eq:dS_Bsol}
    B(u)=(1-u)H_1'(u)-(d-2)H_1(u)+c_B -u c_c,
\end{equation}

Substituting \eqref{eq:dS_Bsol} into \eqref{eq:dS_Beq} gives an uncoupled equation for $H_1(u)$,
\begin{equation}
\label{eq:dS_Geq}
    u(2-u)H_1''(u)+d(1-u)H_1'(u)= - 2 c_B -2 c_C u
\end{equation}
This is precisely the equation (\ref{eq:phys}) for the ``physical'' part of the Green's function we obtained by more heuristic means. 
Therefore the scalar coefficient
called \(H_0(Z)\) there is the same function as the coefficient \(H_1(u)\) here,
up to the overall normalization convention:
\begin{equation}
    H_1(u)
    =
    16\pi G_N\,H_0(Z), \quad Z = 1 -u.
\end{equation}
The left hand side is the wave equation for the massless minimally coupled scalar. Without coefficients $c_C, c_B$ it does not have solutions without anti-podal singularities. It turns out coefficient $c_C$ shifts the solution by $u$ which is equivalent to adding a pure gauge term. So we drop it. Coefficient $c_B$ can be adjusted to only have a singularity at $u=0$ (coincident points). Equivalently, in terms of $Z=1-u$,
\begin{equation}
\label{eq:dS_Geq_Z}
    (1-Z^2)\,H_1''(Z)-dZ\,H_1'(Z)= -2 c_B.
\end{equation}
The regulator mass term used in \eqref{eq:quadratic_action_alpha_beta_gauge} and
below, which shifts the kinetic operator as
\(-\Box h_{\mu\nu}\to -\Box h_{\mu\nu}+m_g^2h_{\mu\nu}\), only changes the
\(T^{(3)}\) equation:
\begin{equation}
    {\cal F}_3(u)\longrightarrow {\cal F}_3(u)+m_g^2H_1(u).
\end{equation}
The \(T^{(2)}\) and \(T^{(5)}\) equations still determine \(C\) and \(B\) as
above, and the eliminated equation becomes
\begin{equation} \label{eq:dS_Geq_massive}
    u(2-u)H_1''(u)+d(1-u)H_1'(u)-m_g^2H_1(u)=0.
\end{equation}
Thus the same invariant coefficient is shifted from the massless scalar
equation to the massive scalar equation, \((\Box-m_g^2)H_1=0\), away from the
coincident point.  In this appendix the de Sitter radius has been set to one;
restoring the radius replaces \(m_g^2\) in the \(u\)-equation by
\(m_g^2\rdS^2\).

For completeness, the coefficients of \(T^{(4)}\), \(T^{(1)}\), and the
trace part of the same five-basis expansion yield the remaining three
equations,
\begin{align}
\label{eq:dS_Heq1}
    0
    &=
    2H_1''(u)-2(1-u)H_1'(u)+4(d-2)H_1(u)+(d-2)H_2''(u),
    \\
\label{eq:dS_Heq2}
    0
    &=
    u(2-u)H_2''(u)
    +2(d-1)(1-u)H_2'(u)
    +2(d-1)H_2(u)
    +4H_1(u) \\
    & \quad +2(1-u)H_1'(u)
    -2(1-u)A(u),
    \\
\label{eq:dS_Aeq}
    0
    &=
    -(1-u)H_1'(u)+2dH_1(u)+A'(u).
\end{align}
Since $H_1(u)$ is already determined by \eqref{eq:dS_Geq}, this system determines the remaining scalar functions $H_2(u)$ and $A(u)$, up to the same residual gauge freedom already encoded in the $\Lambda$ term.

\section{Details of the OTOC at general insertion points}
\label{app:general_location_G}

In this appendix, we provide further details on the de Sitter OTOC with insertions at general static-patch locations. In particular, we show how the momentum-space integrals reproduce the corresponding bulk-to-bulk propagators and explain how to evaluate the derivatives with respect to $\vec q_{13}$ and $\vec q_{24}$.

\subsection{Momentum representation of de Sitter two-point function}

We first justify the momentum representation
\eqref{eq:general_location_2pf_momentum_representation}, which is the input
used below to do the \(p^u\) and \(p^v\) integrals.  Consider one pair of
insertions of the same scalar field, at static-patch data
\((t_A,r_A,\vec y_A)\) and \((t_B,r_B,\vec y_B)\).  Let
\(N=d-2\), \(M=d-1\), \(\lambda=(d-1)/2\), and
\[
    a_i=1-r_i^2/\rdS^2,\qquad
    \rho=-e^{t_{BA}/\rdS}\sqrt{\frac{a_A}{a_B}},
    \qquad
    \vec q=(r_A\vec y_A+\rho\,r_B\vec y_B)/\rdS,
\]
This is the single-sector version of the definitions used in the main text:
for the \(13\) sector one takes \((A,B)=(1,3)\), while for the \(24\) sector
one takes \((A,B)=(4,2)\).

The  standard momentum-space mode expansion of the
Bunch-Davies two-point function in planar coordinates
\[
    ds^2=\frac{\rdS^2}{\eta^2}\left(-d\eta^2+d\vec x^{\,2}\right),
\]
is given by
\begin{equation}
\label{eq:appendix_BD_mode_sum}
G_{d,\mu}(X,X')
=
\frac{\pi}{4\rdS^{d-2}}
(\eta\eta')^\lambda
\int\frac{d^M k}{(2\pi)^M}\,
e^{i\vec k\cdot(\vec x-\vec x')}\,
H^{(1)}_{i\mu}(-k\eta)\,
H^{(2)}_{i\mu}(-k\eta') , 
\end{equation}
where \(k=|\vec k|\).  Now continue to the regulated kinematics used by the horizon wavefunctions,
\begin{equation}
\label{eq:appendix_mode_sum_continuation}
    \eta=-i,\qquad
    \eta'=i\rho,\qquad
    \vec x-\vec x'=-i\vec q .
\end{equation}
Then \(e^{i\vec k\cdot(\vec x-\vec x')}=e^{\vec k\cdot\vec q}\), while the first and second Hankel functions become: 
\[
H^{(1)}_{i\mu}(ik)
=
\frac{2}{\pi i}e^{\frac{\pi\mu}{2}}K_{i\mu}(k),
\qquad
H^{(2)}_{i\mu}(-i\rho k)
=
-\frac{2}{\pi i}e^{-\frac{\pi\mu}{2}}K_{i\mu}(\rho k).
\]
 The \(\mu\)-dependent factors cancel in the product, and
\eqref{eq:appendix_BD_mode_sum} becomes
\begin{equation}
\label{eq:appendix_BD_mode_sum_K_form}
G_{d,\mu}(X,X')
=
\frac{\rho^\lambda}{2^{d-1}\pi^d\rdS^{d-2}}
\int_{\mathbb R^M}d^M k\,
e^{\vec k\cdot\vec q}
K_{i\mu}(k)K_{i\mu}(\rho k).
\end{equation}

The same expression is what one obtains by multiplying the two general-location
wave profiles at zero eikonal phase.  After the momentum rescaling used in
deriving \eqref{eq:OTOC_integral_general_locations}, the single-sector measure is
\begin{equation}
\label{eq:appendix_single_sector_measure}
d\Pi_{AB}
=
\frac{\rho^\lambda}{2^{d-1}\pi^d\rdS^{d-2}}\,
dp\,p^{d-2}\,d\phi\sqrt{g_\perp}\,
K_{i\mu}(p)K_{i\mu}(\rho p)\,
e^{p\vec q\cdot y}.
\end{equation}
The factor \(p^{d-2}\) arises from the horizon measure \(p\,dp\) together with
the two factors \(p^{(d-3)/2}\) appearing in the wave profiles. The phases in
\eqref{eq:waveprofile-general} combine into \(e^{p\vec q\cdot y}\), while all
remaining numerical factors assemble into the prefactor of
\eqref{eq:appendix_single_sector_measure}. This prefactor is inherited from the
factorized normalization appearing in
\eqref{eq:OTOC_integral_general_locations}:
\[
    \left(
    \frac{\rho_{13}^\lambda}{2^{d-1}\pi^d\rdS^{d-2}}
    \right)
    \left(
    \frac{\rho_{24}^\lambda}{2^{d-1}\pi^d\rdS^{d-2}}
    \right).
\]

Finally, writing \(d^M k=p^{M-1}dp\,d\phi\sqrt{g_\perp}\) in
\eqref{eq:appendix_BD_mode_sum_K_form} gives
\begin{equation}
\label{eq:appendix_2pf_before_angular_integral}
G_{d,\mu}(X_A,X_B)
=
\frac{\rho^\lambda}{2^{d-1}\pi^d\rdS^{d-2}}
\int_0^\infty dp\,p^{d-2}\,
K_{i\mu}(p)K_{i\mu}(\rho p)
\int_{S^N}d\phi\sqrt{g_\perp}\,e^{p\vec q\cdot y}.
\end{equation}
The angular integral is rotationally invariant in the transverse
\(\mathbb R^{N+1}\), so aligning \(\vec q\) with the polar axis gives
\begin{equation}
\int_{S^N}d\phi\sqrt{g_\perp}\,e^{p\vec q\cdot y}
=
\mathcal F_N(p |\vec q|),
\end{equation}
with \(\mathcal F_N\) defined in \eqref{eq:sphere_exponential_integral}.  This
already gives the momentum integral in
\eqref{eq:general_location_2pf_momentum_representation}.  It remains only to
identify the argument of the two-point function.  From the definition of
\(\vec q\),
\begin{equation}
\vec{q}^2
=
\frac{r_A^2+\rho^2 r_B^2
+2\rho\,r_Ar_B\,\vec y_A\cdot\vec y_B}{\ell^2} .
\end{equation}
Using \(\rho=-e^{t_{BA}/\rdS}\sqrt{a_A/a_B}\), one finds
\begin{align}
\frac{1+\rho^2-\vec{q}^2}{-2\rho}
&=
-\frac{a_A+\rho^2a_B}{2\rho}
+r_Ar_B\,\vec y_A\cdot\vec y_B/\rdS^2=
\nonumber\\
&=
\sqrt{a_Aa_B}\cosh\frac{t_{BA}}{\rdS}
+r_Ar_B\,\vec y_A\cdot\vec y_B/\rdS^2 .
\end{align}
This is precisely the de Sitter invariant \(Z(X_A,X_B)\).  Hence
\eqref{eq:appendix_2pf_before_angular_integral} becomes
\begin{equation}
G_{d,\mu}\!\left(Z(\rho,|\vec{q}|)\right)
=
\frac{\rho^\lambda}{2^{d-1}\pi^d\rdS^{d-2}}
\int_0^\infty dp\,p^{d-2}\,
K_{i\mu}(p)K_{i\mu}(\rho p)\mathcal F_N(p |\vec{q}|),
\end{equation}
which is \eqref{eq:general_location_2pf_momentum_representation}.  The formula
is first understood in the Euclidean-regulated region where the transform
converges, and the Lorentzian OTOC kinematics are reached by the same analytic
continuation used in deriving the wave profiles.

\subsection{The term linear in $\mathfrak{a}$}

Let us also spell out the linear in $\mathfrak{a}$ correction in the eikonal expansion.  From the
compact form of the general-location result,
\begin{equation}
D_{\{r_i\}}
=
\exp\!\left[
\mathfrak{a} \left(\partial_{\vec q_{13}}\cdot\partial_{\vec q_{24}}\right)
\right]
G_{d,\mu_V}(Z_{13})G_{d,\mu_W}(Z_{24}),
\end{equation}
the term linear in $\mathfrak{a}$ is
\begin{equation}
D_{\{r_i\}}^{(1)}
=
\mathfrak{a}
\left(\partial_{\vec q_{13}}\cdot\partial_{\vec q_{24}}\right)
G_{d,\mu_V}(Z_{13})G_{d,\mu_W}(Z_{24}).
\end{equation}
Here the two invariants depend on the transverse variables only through
\[
Z_{13}=Z_{13}^{(0)}+\frac{\vec q_{13}^{\,2}}{2\rho_{13}},
\qquad
Z_{24}=Z_{24}^{(0)}+\frac{\vec q_{24}^{\,2}}{2\rho_{24}},
\]
with \(\rho_{13}\) and \(\rho_{24}\) held fixed while differentiating.  Thus
\begin{equation}
\frac{\partial}{\partial q_{13}^a}G_{d,\mu_V}(Z_{13})
=
\frac{q_{13}^a}{\rho_{13}}\,
\partial_ZG_{d,\mu_V}(Z_{13}),
\qquad
\frac{\partial}{\partial q_{24}^a}G_{d,\mu_W}(Z_{24})
=
\frac{q_{24}^a}{\rho_{24}}\,
\partial_ZG_{d,\mu_W}(Z_{24}).
\end{equation}
Since the \(V\)-sector propagator is independent of \(\vec q_{24}\), and the
\(W\)-sector propagator is independent of \(\vec q_{13}\), the contracted
derivative gives
\begin{equation}
D_{\{r_i\}}^{(1)}
=
\mathfrak{a} \,
\frac{\vec q_{13}\cdot\vec q_{24}}{\rho_{13}\rho_{24}}\,
\partial_ZG_{d,\mu_V}(Z_{13})
\partial_ZG_{d,\mu_W}(Z_{24}).
\end{equation}
Using the hypergeometric form of the scalar two-point function in
\eqref{eq:dS_2pf_general_d}, with
\[
h_{V,\pm}=\frac{d-1}{2}\pm i\mu_V,
\qquad
h_{W,\pm}=\frac{d-1}{2}\pm i\mu_W,
\]
one has
\begin{equation}
\partial_ZG_{d,\mu}(Z)
=
\frac{\Gamma(h_+)\Gamma(h_-)}
{\rdS^{d-2}(4\pi)^{d/2}\Gamma(\frac d2)}
\frac{h_+h_-}{d}\,
{}_2F_1\!\left(
h_+{+}1,h_-{+}1;\frac d2{+}1;\frac{1+Z}{2}
\right).
\end{equation}
Therefore, the term linear in $\mathfrak a$ is given explicitly by
\begin{equation}
\begin{aligned}
D_{\{r_i\}}^{(1)}
={}&
\mathfrak{a} \,
\frac{\vec q_{13}\cdot\vec q_{24}}{\rho_{13}\rho_{24}}\,
\frac{
\Gamma(h_{V,+})\Gamma(h_{V,-})
\Gamma(h_{W,+})\Gamma(h_{W,-})}
{\rdS^{2d-4}(4\pi)^d\Gamma(\frac d2)^2}
\frac{
h_{V,+}h_{V,-}h_{W,+}h_{W,-}}
{d^2}
\nonumber\\
&\times
{}_2F_1\!\left(
h_{V,+}{+}1,h_{V,-}{+}1;\frac d2{+}1;\frac{1+Z_{13}}{2}
\right)
{}_2F_1\!\left(
h_{W,+}{+}1,h_{W,-}{+}1;\frac d2{+}1;\frac{1+Z_{24}}{2}
\right).
\end{aligned}
\end{equation}
The pole limit should be taken only after performing the differentiation. Setting
$\vec q_{13}=\vec q_{24}=0$ then gives
\begin{equation}
\left.D_{{r_i}}^{(1)}\right|_{\vec q_{13}=\vec q_{24}=0}=0 .
\end{equation}
This provides the explicit first-order realization of the statement that all odd powers in the eikonal expansion vanish when the insertions are at the pole. More generally, the linear term vanishes whenever either pair is placed at the pole: $\vec q_{13}=0$ or $\vec q_{24}=0$ is already sufficient to set $D_{{r_i}}^{(1)}$ to zero.

\printbibliography

@article{Frob:2016hkx,
    author = {Fr{\"o}b, Markus B. and Higuchi, Atsushi and Lima, William C. C.},
    title = "{Mode-sum construction of the covariant graviton two-point function in the Poincar{\'e} patch of de Sitter space}",
    eprint = "1603.07338",
    archivePrefix = "arXiv",
    primaryClass = "gr-qc",
    doi = "10.1103/PhysRevD.93.124006",
    journal = "Phys. Rev. D",
    volume = "93",
    number = "12",
    pages = "124006",
    year = "2016"
}

@article{Higuchi:2001uv,
    author = "Higuchi, Atsushi and Kouris, Spyros S.",
    title = "{The Covariant graviton propagator in de Sitter space-time}",
    eprint = "gr-qc/0107036",
    archivePrefix = "arXiv",
    doi = "10.1088/0264-9381/18/20/311",
    journal = "Class. Quant. Grav.",
    volume = "18",
    pages = "4317--4328",
    year = "2001"
}

@article{Chandrasekaran:2022cip,
    author = "Chandrasekaran, Venkatesa and Longo, Roberto and Penington, Geoff and Witten, Edward",
    title = "{An algebra of observables for de Sitter space}",
    eprint = "2206.10780",
    archivePrefix = "arXiv",
    primaryClass = "hep-th",
    doi = "10.1007/JHEP02(2023)082",
    journal = "JHEP",
    volume = "02",
    pages = "082",
    year = "2023"
}

@article{Kolchmeyer:2024fly,
    author = "Kolchmeyer, David K. and Liu, Hong",
    title = "{Chaos and the Emergence of the Cosmological Horizon}",
    eprint = "2411.08090",
    archivePrefix = "arXiv",
    primaryClass = "hep-th",
    reportNumber = "MIT-CTP/5805",
    month = "11",
    year = "2024"
}

@article{MNX,
    author = "Milekhin, Alexey  and Narovlansky, Vladimir and Xu, Jiuci",
    title = "{work in progress}",
    year = "2026"
}

@article{Tanaka:2017nff,
    author = "Tanaka, Takahiro and Urakawa, Yuko",
    title = "{Large gauge transformation, Soft theorem, and Infrared divergence in inflationary spacetime}",
    eprint = "1707.05485",
    archivePrefix = "arXiv",
    primaryClass = "hep-th",
    reportNumber = "YITP-17-76, KUNS-2694",
    doi = "10.1007/JHEP10(2017)127",
    journal = "JHEP",
    volume = "10",
    pages = "127",
    year = "2017"
}

@article{Goto:2026ipq,
    author = "Goto, Kanato and Milekhin, Alexey and Verlinde, Herman and Xu, Jiuci",
    title = "{Generalized Free Fields in de Sitter from 1D CFT}",
    eprint = "2605.03037",
    archivePrefix = "arXiv",
    primaryClass = "hep-th",
    reportNumber = "OU-HET-1309, RIKEN-iTHEMS-Report-26",
    month = "5",
    year = "2026"
}

@article{Allen:1986tt,
    author = "Allen, Bruce and Turyn, Michael",
    title = "{An Evaluation of the Graviton Propagator in De Sitter Space}",
    reportNumber = "TUTP-86-20",
    doi = "10.1016/0550-3213(87)90672-9",
    journal = "Nucl. Phys. B",
    volume = "292",
    pages = "813",
    year = "1987"
}

@article{Mir,
    author = "Faizal, Mir",
    title = "{Covariant Graviton Propagator in Anti-de Sitter Spacetime}",
    eprint = "1112.4369",
    archivePrefix = "arXiv",
    primaryClass = "gr-qc",
    doi = "10.1088/0264-9381/29/3/035007",
    journal = "Class. Quant. Grav.",
    volume = "29",
    pages = "035007",
    year = "2012"
}

@article{Morrison,
    author = "Morrison, Ian A.",
    title = "{On cosmic hair and ''de Sitter breaking'' in linearized quantum gravity}",
    eprint = "1302.1860",
    archivePrefix = "arXiv",
    primaryClass = "gr-qc",
    month = "2",
    year = "2013"
}

@article{Hotta:1992wb,
    author = "Hotta, M. and Tanaka, M.",
    title = "{Gravitational shock waves and quantum fields in the de Sitter space}",
    reportNumber = "TU-415",
    doi = "10.1103/PhysRevD.47.3323",
    journal = "Phys. Rev. D",
    volume = "47",
    pages = "3323--3329",
    year = "1993"
}

@article{Hotta:1992qy,
    author = "Hotta, M. and Tanaka, M.",
    title = "{Shock wave geometry with nonvanishing cosmological constant}",
    reportNumber = "TU-403",
    doi = "10.1088/0264-9381/10/2/012",
    journal = "Class. Quant. Grav.",
    volume = "10",
    pages = "307--314",
    year = "1993"
}

@article{Yang:2025lme,
    author = "Yang, Zhenbin and Zhang, Yuzhen and Zheng, Wenwen",
    title = "{Remarks on the de Sitter double cone}",
    eprint = "2505.08647",
    archivePrefix = "arXiv",
    primaryClass = "hep-th",
    reportNumber = "USTC-ICTS/PCFT-25-58",
    doi = "10.1103/jync-pmnz",
    journal = "Phys. Rev. D",
    volume = "113",
    number = "12",
    pages = "126012",
    year = "2026"
}

@article{Susskind:2021esx,
    author = "Susskind, Leonard",
    title = "{Entanglement and Chaos in De Sitter Space Holography: An SYK Example}",
    eprint = "2109.14104",
    archivePrefix = "arXiv",
    primaryClass = "hep-th",
    doi = "10.22128/jhap.2021.455.1005",
    journal = "JHAP",
    volume = "1",
    number = "1",
    pages = "1--22",
    year = "2021"
}

@article{Okuyama:2025hsd,
    author = "Okuyama, Kazumi",
    title = "{de Sitter JT gravity from double-scaled SYK}",
    eprint = "2505.08116",
    archivePrefix = "arXiv",
    primaryClass = "hep-th",
    doi = "10.1007/JHEP08(2025)181",
    journal = "JHEP",
    volume = "08",
    pages = "181",
    year = "2025"
}

@article{Milekhin:2024vbb,
    author = "Milekhin, Alexey and Xu, Jiuci",
    title = "{On scrambling, tomperature and superdiffusion in de Sitter space}",
    eprint = "2403.13915",
    archivePrefix = "arXiv",
    primaryClass = "hep-th",
    doi = "10.1007/JHEP07(2025)272",
    journal = "JHEP",
    volume = "07",
    pages = "272",
    year = "2025"
}

@article{Milekhin:2023bjv,
    author = "Milekhin, Alexey and Xu, Jiuci",
    title = "{Revisiting Brownian SYK and its possible relations to de Sitter}",
    eprint = "2312.03623",
    archivePrefix = "arXiv",
    primaryClass = "hep-th",
    doi = "10.1007/JHEP10(2024)151",
    journal = "JHEP",
    volume = "10",
    pages = "151",
    year = "2024"
}

@article{D_Hoker_1999,
   title={Graviton and gauge boson propagators in AdS+1},
   volume={562},
   ISSN={0550-3213},
   url={http://dx.doi.org/10.1016/S0550-3213(99)00524-6},
   DOI={10.1016/s0550-3213(99)00524-6},
   number={1-2},
   journal={Nuclear Physics B},
   publisher={Elsevier BV},
   author={D’Hoker, Eric and Freedman, Daniel Z. and Mathur, Samir D. and Matusis, Alec and Rastelli, Leonardo},
   year={1999},
   month=Nov, pages={330–352} }

@article{Kabat:1992tb,
    author = "Kabat, Daniel N. and Ortiz, Miguel",
    title = "{Eikonal quantum gravity and Planckian scattering}",
    eprint = "hep-th/9203082",
    archivePrefix = "arXiv",
    reportNumber = "MIT-CTP-2069",
    doi = "10.1016/0550-3213(92)90627-N",
    journal = "Nucl. Phys. B",
    volume = "388",
    pages = "570--592",
    year = "1992"
}

@article{Jackiw_eikonal,
  title = {Failure of the Eikonal Approximation for the Vertex Function in a Boson Field Theory},
  author = {Eichten, E. and Jackiw, R.},
  journal = {Phys. Rev. D},
  volume = {4},
  issue = {2},
  pages = {439--443},
  numpages = {0},
  year = {1971},
  month = {Jul},
  publisher = {American Physical Society},
  doi = {10.1103/PhysRevD.4.439},
  url = {https://link.aps.org/doi/10.1103/PhysRevD.4.439}
}

@article{Treiman,
  title = {Relativistic Eikonal Approximation},
  author = {Tiktopoulos, George and Treiman, S. B.},
  journal = {Phys. Rev. D},
  volume = {3},
  issue = {4},
  pages = {1037--1040},
  numpages = {0},
  year = {1971},
  month = {Feb},
  publisher = {American Physical Society},
  doi = {10.1103/PhysRevD.3.1037},
  url = {https://link.aps.org/doi/10.1103/PhysRevD.3.1037}
}

@article{Kabat:1992pz,
    author = "Kabat, Daniel N.",
    title = "{Validity of the Eikonal approximation}",
    eprint = "hep-th/9204103",
    archivePrefix = "arXiv",
    reportNumber = "MIT-CTP-2095",
    journal = "Comments Nucl. Part. Phys.",
    volume = "20",
    number = "6",
    pages = "325--335",
    year = "1992"
}

@article{Geng:2025bcb,
    author = "Geng, Hao and Jiang, Yikun and Xu, Jiuci",
    title = "{Algebras, entanglement islands, and observers}",
    eprint = "2506.12127",
    archivePrefix = "arXiv",
    primaryClass = "hep-th",
    doi = "10.1088/1361-6382/ae811e",
    journal = "Class. Quant. Grav.",
    volume = "43",
    number = "13",
    pages = "135012",
    year = "2026"
}

@article{Narovlansky:2023lfz,
    author = "Narovlansky, Vladimir and Verlinde, Herman",
    title = "{Double-scaled SYK and de Sitter Holography}",
    eprint = "2310.16994",
    archivePrefix = "arXiv",
    primaryClass = "hep-th",
    month = "10",
    year = "2023"
}

@article{Narovlansky:2025tpb,
    author = "Narovlansky, Vladimir",
    title = "{Towards a microscopic description of de Sitter dynamics}",
    eprint = "2506.02109",
    archivePrefix = "arXiv",
    primaryClass = "hep-th",
    month = "6",
    year = "2025"
}

@article{Aalsma:2020aib,
    author = "Aalsma, Lars and Shiu, Gary",
    title = "{Chaos and complementarity in de Sitter space}",
    eprint = "2002.01326",
    archivePrefix = "arXiv",
    primaryClass = "hep-th",
    reportNumber = "MAD-TH-20-01",
    doi = "10.1007/JHEP05(2020)152",
    journal = "JHEP",
    volume = "05",
    pages = "152",
    year = "2020"
}

@article{Shenker:2013pqa,
	title        = {{Black holes and the butterfly effect}},
	author       = {Shenker, Stephen H. and Stanford, Douglas},
	year         = 2014,
	journal      = {JHEP},
	volume       = {03},
	pages        = {067},
	doi          = {10.1007/JHEP03(2014)067},
	eprint       = {1306.0622},
	archiveprefix = {arXiv},
	primaryclass = {hep-th},
	reportnumber = {SU-ITP-13-08},
	slaccitation = {%%CITATION = ARXIV:1306.0622;%%}
}

@article{Shenker:2014cwa,
	title        = {{Stringy effects in scrambling}},
	author       = {Shenker, Stephen H. and Stanford, Douglas},
	year         = 2015,
	journal      = {JHEP},
	volume       = {05},
	pages        = 132,
	doi          = {10.1007/JHEP05(2015)132},
	eprint       = {1412.6087},
	archiveprefix = {arXiv},
	primaryclass = {hep-th},
	slaccitation = {%%CITATION = ARXIV:1412.6087;%%}
}

@article{Maldacena:2015waa,
	title        = {{A bound on chaos}},
	author       = {Maldacena, Juan and Shenker, Stephen H. and Stanford, Douglas},
	year         = 2015,
	eprint       = {1503.01409},
	archiveprefix = {arXiv},
	primaryclass = {hep-th},
	slaccitation = {%%CITATION = ARXIV:1503.01409;%%}
}

@article{Dray:1984ha,
	title        = {{The Gravitational Shock Wave of a Massless Particle}},
	author       = {Dray, Tevian and 't Hooft, Gerard},
	year         = 1985,
	journal      = {Nucl. Phys.},
	volume       = {B253},
	pages        = {173--188},
	doi          = {10.1016/0550-3213(85)90525-5},
	reportnumber = {Print-84-0773 (UTRECHT)},
	slaccitation = {%%CITATION = NUPHA,B253,173;%%}
}

@article{Dray:1985yt,
	title        = {{The Effect of Spherical Shells of Matter on the Schwarzschild Black Hole}},
	author       = {Dray, Tevian and 't Hooft, Gerard},
	year         = 1985,
	journal      = {Commun. Math. Phys.},
	volume       = 99,
	pages        = {613--625},
	doi          = {10.1007/BF01215912},
	reportnumber = {Print-85-0107 (IAS,PRINCETON)},
	slaccitation = {%%CITATION = CMPHA,99,613;%%}
}

@article{Sfetsos:1994xa,
	title        = {{On gravitational shock waves in curved space-times}},
	author       = {Sfetsos, Konstadinos},
	year         = 1995,
	journal      = {Nucl. Phys.},
	volume       = {B436},
	pages        = {721--746},
	doi          = {10.1016/0550-3213(94)00573-W},
	eprint       = {hep-th/9408169},
	archiveprefix = {arXiv},
	primaryclass = {hep-th},
	reportnumber = {THU-94-13},
	slaccitation = {%%CITATION = HEP-TH/9408169;%%}
}

@article{Cornalba:2007zb,
	title        = {{Eikonal approximation in AdS/CFT: Resumming the gravitational loop expansion}},
	author       = {Cornalba, Lorenzo and Costa, Miguel S. and Penedones, Joao},
	year         = 2007,
	journal      = {JHEP},
	volume       = {09},
	pages        = {037},
	doi          = {10.1088/1126-6708/2007/09/037},
	eprint       = {0707.0120},
	archiveprefix = {arXiv},
	primaryclass = {hep-th},
	reportnumber = {ROM2F-2007-11, LPTENS-07-27},
	slaccitation = {%%CITATION = ARXIV:0707.0120;%%}
}
\end{document}